\documentclass[preprint]{aastex63}

\usepackage{longtable}

\newcommand{\tcondslope}{\ensuremath{\frac{\partial[\mathrm{X/Fe}]}{\partial T_{\rm cond}}}}
\usepackage{xcolor}

\definecolor{my_color}{HTML}{3a18b1}

\newcommand{\teff}{\mbox{$T_{\rm eff}$}}

\newcommand{\feh}{\mbox{$\rm [Fe/H]$}}
\newcommand{\mg}{\mbox{$\rm [Mg/Fe]$}}

\newcommand{\alphafe}{\mbox{$\rm [\alpha/Fe]$}}
\newcommand{\aia}{\mbox{A$_{\rm Ia}$}}
\newcommand{\acc}{\mbox{A$_{\rm cc}$}}
\newcommand{\kpmalpha}{\mbox{[A$_{\rm cc}$/A$_{\rm Ia}$]}}
\newcommand{\fcc}{\mbox{f$_{\rm cc}$}}

\newcommand{\logg}{\mbox{$\log g$}}

\newcommand{\TSun}{\ifmmode {T_{\odot}}\else${T_{\odot}}$\fi}
\newcommand{\LSun}{\ifmmode {\logg_{\odot}}\else${\logg_{\odot}}$\fi}
\newcommand{\FSun}{\ifmmode {\feh_{\odot}}\else${\feh_{\odot}}$\fi}
\newcommand{\mSun}{\ifmmode {M_{\odot}}\else${M_{\odot}}$\fi}

\newcommand{\rbirth}{\mbox{$R_{\rm birth}$}}

\newcommand{\tcond}{\mbox{$T_{\rm cond}$}}
\newcommand{\rsq}{\mbox{$R^2$}}

\usepackage{multirow}
\usepackage{appendix}
\usepackage{amsmath}
\usepackage{mathtools}
\defcitealias{Rampalli24}{R24}

\shorttitle{GCE explains the Sun's Chemistry}
\shortauthors{Rampalli et al.}

\begin{document}

\title{A Galactic Perspective on the (Unremarkable) Relative Refractory Depletion Observed in the Sun}

\correspondingauthor{Rayna Rampalli}
\email{raynarampalli@gmail.com}

\author[0000-0001-7337-5936]{Rayna Rampalli}
\altaffiliation{NSF GRFP Fellow} 
\affiliation{Department of Physics and Astronomy, Dartmouth College, Hanover, NH 03755, USA}

\author[0000-0002-6534-8783]{James W. Johnson}
\affiliation{Carnegie Science Observatories, 813 Santa Barbara St., Pasadena, CA 91101, USA}

\author[0000-0001-5082-6693]{Melissa K. Ness}
\affiliation{Research School of Astronomy \& Astrophysics, Australian National University, Canberra, ACT 2611, Australia}

\author[0000-0002-3285-4858]{Graham H. Edwards}
\affiliation{Department of Earth \& Environmental Geosciences, Trinity University, San Antonio, TX 78212, USA}

\author[0000-0003-4150-841X]{Elisabeth R. Newton}
\affiliation{Department of Physics and Astronomy, Dartmouth College, Hanover, NH 03755, USA}

\author[0000-0001-9345-9977]{Emily J. Griffith}
\altaffiliation{NSF Astronomy and Astrophysics Postdoctoral Fellow}
\affiliation{Center for Astrophysics and Space Astronomy, Department of Astrophysical and Planetary Sciences, University  of Colorado, 389~UCB, Boulder,~CO 80309-0389, USA}

\author[0000-0001-9907-7742]{Megan Bedell}
\affiliation{Center for Computational Astrophysics, Flatiron Institute, 162 Fifth Avenue, New York, NY 10010, USA}

\author[0000-0002-5879-3360]{Kaile Wang}
\affiliation{Department of Physics, University of Texas at Austin, 2515 Speedway, Stop C1400, Austin, TX 78712-1205, USA}

\accepted{March 26, 2026}
\submitjournal{ApJ}




\begin{abstract}

Over the last two decades, the Sun has been observed to be depleted in refractory elements as a function of elemental condensation temperature (\tcond) relative to $\sim$ 80\% of its counterparts. We assess the impact of Galactic chemical evolution (GCE) on refractory element-\tcond\ trends for 109,500 unique solar analogs from the GALAH, APOGEE, Gaia RVS, and \cite{bedell18} surveys. We find that a star's \feh\ and \alphafe\ are predictive of its \tcond\ slope (\rsq\ =  $15 \pm 5$, $23 \pm 10$\% respectively) while \teff\ and \logg\ contribute more weakly (\rsq\ = $9 \pm 5$, $13 \pm 16$\%). The Sun's abundance pattern resembles that of more metal-rich (0.1 dex) and $\alpha$-depleted stars ($-0.02$ dex), suggesting a connection to broader GCE trends. To more accurately model stars' nucleosynthetic signatures, we apply the K-process model from \cite{Griffith24}, which casts each star’s abundance pattern as a linear combination of core-collapse and Type Ia supernovae contributions. We find the Sun appears chemically ordinary in this framework, {consistent with the intrinsic population scatter expected from stellar nucleosynthesis.} 
We show that refractory element–\tcond\ trends arise because elements with higher \tcond\ have higher contributions from core-collapse supernovae. Refractory element depletion trends primarily reflect nucleosynthetic enrichment patterns shaped by GCE and local ISM inhomogeneities, with these processes accounting for $> 90$\% of the observed variation within 2$\sigma$. This work highlights how abundance diversity due to local and global chemical enrichment {complicates the interpretation of population-scale planet-related chemical signatures in current datasets.}

\end{abstract}


\section{Introduction}\label{sec:intro}
The Sun appears chemically anomalous compared to other Sun-like stars. Specifically, it shows a relative depletion in refractory\footnote{We define an element as “refractory” if its 50\% condensation temperature from \cite{Lodders03} exceeds 900 K, following \cite{Flores23}.} element abundances (Na, Mn, Cr, Si, Fe, Ni, Mg, V, Ca, Ti, Al)\footnote{These are the refractory elements examined in this work.}, as a function of their condensation temperatures (\tcond). The slope of the Sun's \tcond\ trend is $\sim 10^{-4}$ dex/K, consistently placing it below $\sim$80\% of its counterparts across various datasets, \citep[e.g.,][]{Melendez09,Ramirez09,bedell18,Rampalli24}. The origin of this depletion remains an open question. While both planetary and Galactic explanations have been proposed, none have fully accounted for the observed trends. In this work, we show that \tcond\ correlates with the element's relative enrichment from Type Ia and core-collapse supernovae. We recontextualize the Sun’s refractory–\tcond\ gradient as a consequence of nucleosynthetic differences between stars driven by the chemical evolution of the Galaxy and the local interstellar medium.

Two main scenarios rooted in planet formation have been proposed to explain the Sun’s refractory depletion. In one, rocky planets sequester refractory-enriched material within their interiors \citep{Melendez09,Ramirez09}, while in the other, giant planets create dust traps that prevent the dust's accretion onto the Sun \citep{Booth20}. In our previous work, \citet[][hereafter referred to as \citetalias{Rampalli24}]{Rampalli24}, we showed that the Sun remains relatively refractory depleted compared to other Sun-like planet hosts using medium-resolution spectra from the Radial Velocity Spectrometer (RVS) instrument on Gaia \citep{GaiaMission,recio-blanco}. However, we did not detect trends based on planet-host type, suggesting that planet-driven scenarios are unlikely to be the primary explanation for this chemical signature. Other work, using high-resolution spectra, similarly shows that the Sun’s refractory depletion better aligns with broader Galactic chemical evolution trends rather than with the presence or absence of planets \citep[e.g.,][]{Sun24,Carlos25,Sun25}.

Alternative explanations for the Sun’s apparent refractory depletion invoke chemical enrichment processes tied to Galactic chemical evolution (GCE), the local interstellar medium (ISM), or a combination of both \citep[e.g.,][]{Gaidos15}. The age-abundance trends observed for the solar twin population \citep[e.g.,][]{Nissen15,Nissen16,Spina18} show that the Galaxy's chemical evolution over time can shape stars' abundances. These relations have thus been widely used to account for GCE effects in subsequent solar twin studies \citep[e.g.,][]{bedell18,Sun24,Martos25,Sun25}. \cite{bedell18} corrected for GCE effects using linear age-abundance trends and still found a refractory depletion trend with \tcond\ in the Sun relative to other solar twins (see their Figure 3). However, \cite{Cowley22} later showed that assuming a quadratic age-abundance trend removed the depletion trend entirely. This discrepancy highlights the sensitivity of such corrections to various GCE assumptions. 

\cite{Adibekyan14} found tentative evidence that the Sun's abundance pattern is correlated with its current Galactocentric radius. Ultimately, stellar ages can remain imprecise for individual, main-sequence field stars \citep[e.g.,][]{soderblom}, and present-day Galactocentric radius is not always a reliable proxy for birth location since stars can radially migrate from their original birthplace due to interactions with Milky Way spiral arm or bar resonances \citep[e.g.,][]{SellwoodBinney,Roskar08,Minchev10,DiMatteo13}. 

GCE has traditionally been characterized using observationally accessible, but physically entangled, descriptors like age, \feh, and \alphafe. We can now instead describe stars in a more physically motivated framework, defined by their nucleosynthetic signatures and birth environments. Data-driven approaches collapse these observables into physically interpretable parameters, such as a star’s Galactic birth radius (\rbirth, e.g., \citealt[][]{Minchev18,Lu22,Wang2024}) or the fractional contributions from Type Ia and core-collapse supernovae to a star's composition \citep[e.g.,][]{Griffith19,Weinberg19,Griffith22,Weinberg22,Griffith24}. These descriptors move beyond proxies and can circumvent degeneracies between individual stellar parameters. In this work, we recast stars' elemental abundances as relative contributions from core-collapse and Type Ia supernovae to explicitly connect stellar chemistry to large-scale Galactic chemical evolution and localized interstellar enrichment. By leveraging the stars' full suite of elemental abundances to infer supernova contribution patterns, we replace simplified, representative abundance ratios traditionally used to trace GCE and high-dimensional chemical vectors with a more physically grounded picture of how stars inherit their compositions locally and across the Galaxy’s history.

Here, we show how data-driven descriptors of GCE enable us to investigate the physical processes driving the apparent chemical differences in the solar analog population. Throughout this paper, we use the term ``Sun-like" to refer broadly to stars with similar fundamental properties to the Sun. Within this group, following \cite{Berke23}, we define a ``solar analog"  more specifically as a star whose \teff, \logg, and \feh\ fall within a certain threshold of the solar values. Finally, we define a ``solar twin" as a star with \teff, \logg, and \feh\ consistent with solar values to within their 1$\sigma$ uncertainties. We compile abundances for 109,500 unique solar analogs from four different catalogs. We examine correlations in various combinations of stellar parameters (\teff, \logg, \feh, \alphafe) and physically-motivated data-driven parameters (\acc, \aia, and \fcc) with \tcond\ trends to contextualize the origin of the Sun's refractory depletion. In Section \ref{sec:data}, we discuss the four catalogs used to make up our solar analog dataset. We outline how we measure \tcond-trends, determine Type Ia and core-collapse supernovae enrichment amounts for stars, and characterize their effect on the \tcond\ trends in Section \ref{sec:methods}. In Sections \ref{sec:results} and \ref{sec:kpm_results}, we discuss how these quantities correlate with \tcond\ trends. We discuss the implications of these results and their relation to planet-related scenarios in Section \ref{sec:discussion}. We conclude in Section \ref{sec:conclusion}.

\section{Data} \label{sec:data}

\begin{figure*}
    \centering
    \includegraphics[width=\textwidth]{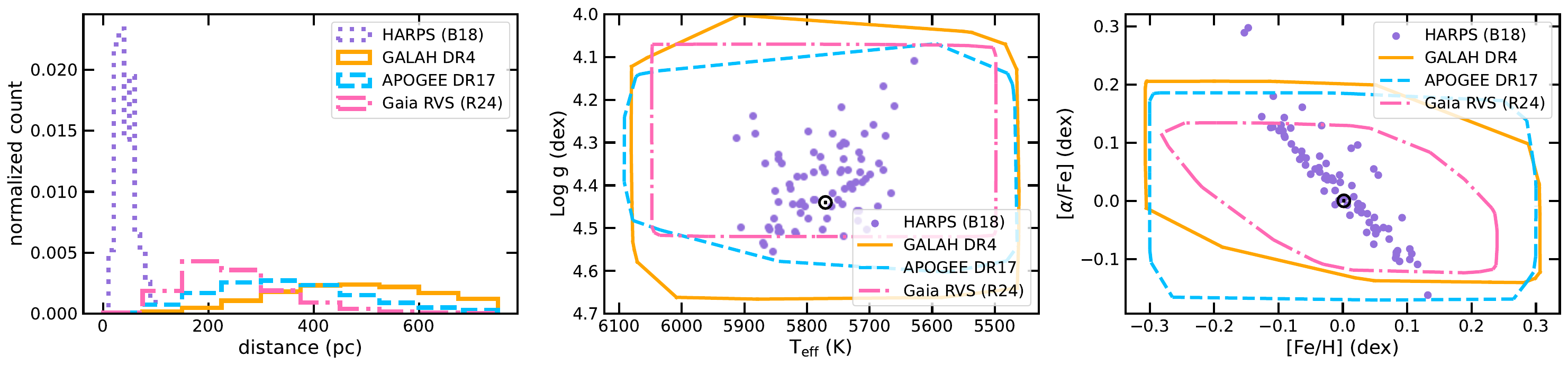}
    \caption{Summary of stellar parameters for solar analogs from \cite{bedell18} in purple, GALAH DR4 \citep{GALAHDR4} in orange, APOGEE DR17 \citep{apogeedr17} in blue, and Gaia RVS \citep{Rampalli24} in pink. \textit{Left}: Histograms of distances probed by each survey. \cite{bedell18} surveyed stars nearby in the solar neighborhood while our samples from Gaia, APOGEE, and GALAH span further distances. \textit{Middle}: \teff-\logg\ plane of solar analogs in each sample. \textit{Right}: \feh-\alphafe\ plane of solar analogs in each sample. GALAH, APOGEE, and Gaia shown as $3\sigma$ contour regions rather than individual scatter point distributions for clarity in middle and right panels. }
    \label{fig:cat_params}
\end{figure*}

To build our dataset, we compile solar analogs from the following surveys: \cite{bedell18}, GALAH DR4 \citep{GALAHDR4}, APOGEE DR17 \citep{apogeedr17}, and Gaia RVS (solar analogs curated by \citetalias{Rampalli24}). Further details on these surveys are provided below.

We select stars according to the following criteria defined by \cite{Berke23}: \teff: $\TSun\pm 300$ K, \logg: $\LSun\pm 0.3$, \feh: $\FSun\pm 0.3$ dex, where \TSun: $5772$ K, \LSun: $4.44$, and \FSun: $0.0$ dex \citep{Prvsa16}. Note that all stars in \cite{bedell18} and \citetalias{Rampalli24} already meet these criteria. We find 109,500 unique solar analogs, of which 144 are identified planet hosts in the NASA Exoplanet Archive \citep{ps}.  We cross-match these stars with the third data release (DR3) from Gaia for stellar kinematics and the Gaia DR3 source ID to identify duplicates \citep{GaiaDr3}.

All stars in this sample have abundance measurements for O, as well as 10 refractory elements in addition to Fe: Na, Mn, Cr, Si, Ni, Mg, V, Ca, Ti, and Al. These are the common set of elements available across all surveys. This set is largely set by Gaia RVS, which has the most narrow wavelength coverage and subsequently the most limited abundance coverage. 
We also calculate \alphafe\ abundance and its associated uncertainty for all stars by averaging the abundances of the \alphafe\ elements with respect to Fe (O, Mg, Si, Ca, and Ti). The \alphafe\ uncertainty is reported as the standard error of the mean of these abundance measurements: the similar individual element uncertainties are added in quadrature, the square root is taken, and the result is divided by the number of elements. 

\citet{bedell18} conducted a high-precision spectroscopic study of 79 Sun-like stars within 100 parsecs using the HARPS spectrograph (R=115,000) on the ESO 3.6-meter telescope. They measured abundances using a differential, line-by-line equivalent width measurements, and individual spectral lines were analyzed relative to the Sun's spectrum using MOOG and Kurucz stellar models \citep{MOOG,Kurucz}. 
This sample achieved the smallest average uncertainties of the four catalogs, with precisions of 4 K for \teff, 0.01 dex for \logg, and $< 0.02$ dex for [X/H]. In addition to the abundances reported above, they also measured abundances for S, Sc, Co, Cu, Zn, Sr, Y, Zr, Ba, La, Ce, Pr, Nd, Sm, Eu, Gd, and Dy, though we do not focus on these elements in this paper. 

GALAH is a high-resolution spectroscopic survey (R = 28,000) that has observed 917,588 stars across the Galaxy in its fourth data release \citep{GALAHDR4} for Galactic archaeology applications. The survey is conducted with the HERMES instrument on the Anglo-Australian Telescope \citep{galah_mission}. Stellar parameters are determined using 1D MARCS model atmospheres \citep{MARCSatmospheres} and the spectral synthesis code Spectroscopy Made Easy \citep{SME1,SME2}. Up to 32 elements were measured for each star including Li, K, Sc, Cu, Zn, Rb, Sr, Y, Zr, Mo, Ba, La, Ce, Nd, Ru, Sm, Eu, beyond what we list above and use. 
We follow catalog guidelines\footnote{The GALAH catalog guidelines can be found at \href{https://www.galah-survey.org/dr4/overview/}{https://www.galah-survey.org/dr4/overview/}.} to filter out stars with poor spectroscopic quality flags, abundance error flags, and low signal-to-noise ratios ($< 30$). Of the 917,588 observed stars, 72,229 are solar analogs that meet these quality criteria. The average parameter uncertainties in this study are 71 K for \teff, 0.1 dex for \logg, and 0.02–0.2 dex for [X/Fe].

APOGEE is another high-resolution (R = 22,500) spectroscopic survey designed for galactic archaeology, primarily targeting red giants across the Galaxy \citep{apogee}. The 733,901 stars in DR17 \citep{apogeedr17} have stellar parameters derived using the APOGEE Stellar Parameter and Chemical Abundances Pipeline (ASPCAP), which fits spectra with synthetic stellar atmosphere models \citep{aspcap}. In addition to the elements we use in this work, APOGEE reports elemental abundances for S, K, Co, and Ce. We remove stars flagged as “bad” and retain 27,569 solar analogs. The average parameter uncertainties from this catalog are 32 K for \teff, 0.03 dex for \logg, and 0.01–0.2 dex for [X/Fe].

Finally, \citetalias{Rampalli24} inferred abundances for 17,412 solar analogs using Gaia RVS spectra ($R=11,200$) with the Cannon, a data-driven method \citep{cannon}. This analysis yielded abundances with average uncertainties of 61 K for \teff, 0.09 dex for logg, and 0.04--0.1 dex for [X/H]. 

Following previous work, we use [X/Fe] rather than [X/H] to calculate \tcond\ slopes, so we convert the refractory abundances and respective uncertainties from \cite{bedell18} and \citetalias{Rampalli24} to [X/Fe] using the \feh\ values and uncertainties. We show the distribution of distances in each of these surveys in the left hand panel, the stars' positions in the \teff-\logg\ plane in the middle panel, and their positions in the \feh-\alphafe\ plane in the right panel of Figure \ref{fig:cat_params}. Given the high precision of HARPS, the \cite{bedell18} sample is the most Sun-like and surveys close-by stars. The other surveys span a similar range of distances and occupy a similar region of \teff-\logg\ and \feh-\alphafe\ parameter space.

\section{Methods} \label{sec:methods}

\subsection{\tcond\ Slope Determination and the Role of Element Choice and Survey Precision}\label{sec:tcond_slope}
\begin{figure*}
    \centering
    \includegraphics[width=\textwidth]{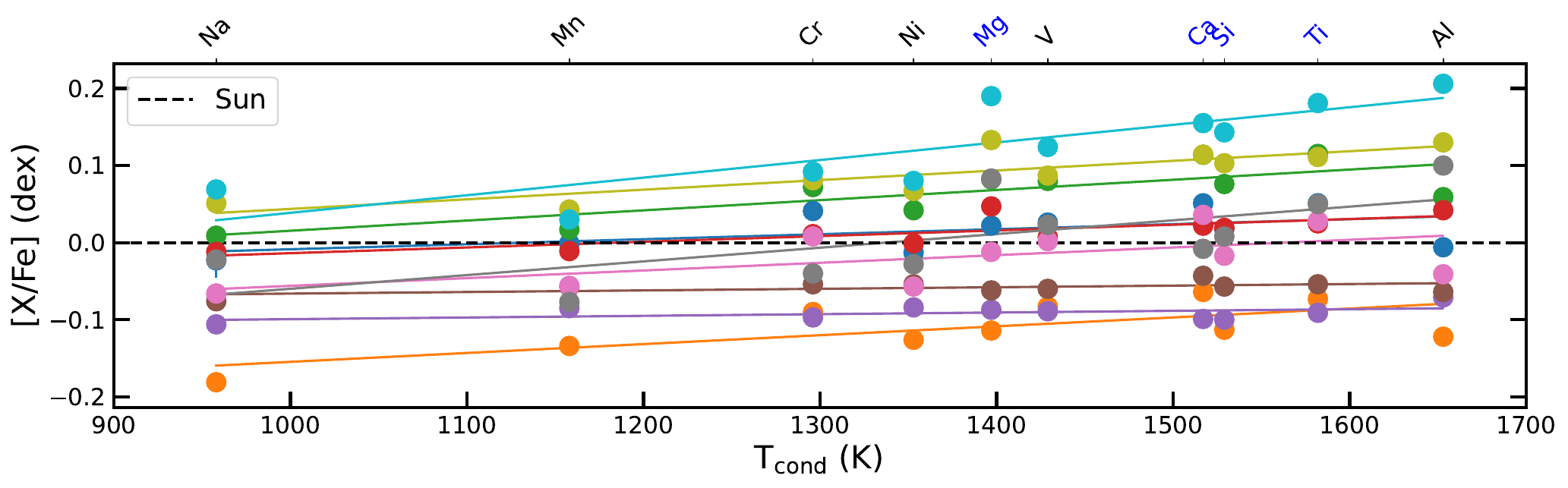}
    \caption{\tcond\ linear fits for 10 random stars' refractory abundances as a function of their 50\% condensation temperature as reported in \cite{Lodders03}. Each color represents a different star's elemental abundances, and $\alpha$-elements are noted by blue text labels.}
    \label{fig:line-fit}
\end{figure*}

The Sun's relative refractory depletion has historically been quantified by a linear trend of refractory element abundances ([X/Fe]) as a function of the elements' 50\% condensation temperatures (\tcond), as defined in \cite{Lodders03}. We continue to use this metric and measure \tcond\ slopes for all of the solar analogs in our sample. We fit a line to the abundance measurements and uncertainties for Na, Mn, Cr, Ni, Mg, V, Ca, Si, Ti, and Al and their respective \tcond\ values (958, 1158, 1296, 1353, 1397, 1429, 1517, 1529, 1582, and 1653 K) using \texttt{numpy}'s \texttt{polyfit} function to calculate a \tcond\ slope and associated uncertainty. Therefore, we mathematically define
\begin{equation}\label{eqn:tcond}
\tcond\ \text{slope} = \frac{\partial[\mathrm{X/Fe}]}{\partial T_{\rm cond}}.
\end{equation}
We show example fits for 10 random stars from the sample in Figure~\ref{fig:line-fit}.


Given that we are using data from four different surveys, systematic offsets in abundance trends are to be expected \citep{Jofre19}. {We determine if these offsets are significant by examining the 188 stars present in at least three of the four catalogs, enabling a direct star-by-star comparison of survey-to-survey differences.} We compare the observed spread in abundance measurements across surveys to the expected spread derived from 1000 bootstrap resamplings of the reported measurement uncertainties. On average, for each element, the survey-to-survey spread in abundances is consistent with the expected bootstrapped uncertainties, suggesting that observed differences between samples are primarily due to measurement uncertainty rather than systematic offsets. We apply the same resampling procedure to the \tcond\ slopes and their uncertainties and likewise find that the variations across surveys are not statistically significant. We emphasize that although this work draws on multiple spectroscopic surveys, all abundance modeling, \tcond\ slope measurements, and KPM inference are performed on a star-by-star basis within each survey. At no point are abundance measurements combined or jointly fit across surveys. {Differences in abundance analysis techniques can introduce element-dependent systematics that affect not only abundance zero points but also the detailed \tcond\ slope distributions within each survey \citep{Hinkel16,Jofre19,Sun25}. Our approach is therefore to treat each survey independently and assess whether the same qualitative enrichment–\tcond\ relationships emerge across datasets, rather than interpreting the combined sample as a single homogeneous distribution. Comparisons between surveys are used to assess consistency, to quantify the impact of measurement uncertainties and systematics, and to identify trends that are robust across independently analyzed datasets. In later sections, multiple surveys are shown together primarily for visual comparison; these figures do not reflect any joint fitting or cross-survey calibration. The consistency of these trends across independently analyzed surveys, as well as in the homogeneous HARPS solar-twin sample, supports their astrophysical origin despite survey-specific systematics.}

\tcond\ slope measurements can shift depending on which elements, and how many, are included in the fit. To test this sensitivity, we iteratively remove the $\alpha$-process elements (Mg, Ca, Si, and Ti) one at a time and recalculate \tcond\ slopes. We find that Ti, Ca, Si, and Mg have the largest impact (in that order). The effect is more pronounced in the three lower-precision catalogs (average shifts of $-0.9 \times 10^{-5}$, $-9 \times 10^{-5}$, and $5 \times 10^{-5}$ dex/K in GALAH DR4, APOGEE, and \citetalias{Rampalli24}) compared to the \cite{bedell18} catalog ($9 \times 10^{-6}$ dex/K). This significant element-by-element variability suggests that \tcond\ slope is a sensitive and incomplete metric to describe stellar chemistries. 

To avoid circular reasoning when we analyze trends in \tcond\ slope as a function of \alphafe\ in Section~\ref{sec:tcondtrends}, we also calculate and report \tcond\ slopes excluding $\alpha$-process elements entirely (leaving six elements). For consistency, we use these 6-element slopes throughout the rest of the paper. However, we note that our results are qualitatively unchanged when using the full 10-element slopes, and all major trends are robust to the choice of elemental subset. We report both types of \tcond\ slopes for each star in Table \ref{t:sas}.

\begin{figure}
    \centering
    \includegraphics[width=0.5\linewidth]{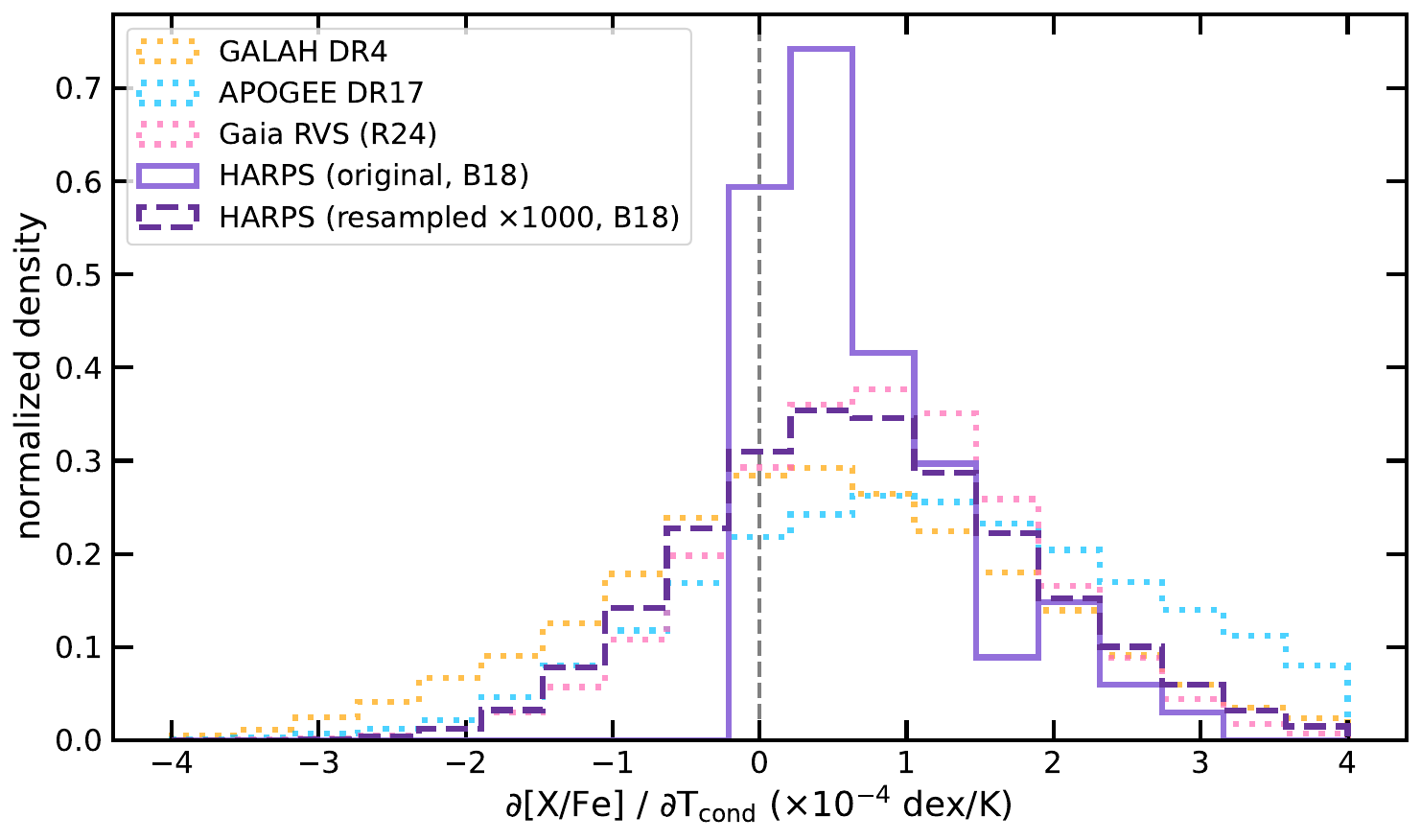}
    \caption{Distributions of \tcond\ slopes for solar analogs across different surveys. The HARPS sample from \cite{bedell18} is shown in solid purple, with resampled slopes using mean abundance uncertainties from GALAH, APOGEE, and Gaia RVS shown in dashed purple. GALAH, APOGEE, and Gaia RVS distributions are shown in orange, blue, and pink, respectively. Injecting larger uncertainties into the HARPS sample broadens the slope distribution and increases the fraction of stars with \tcond\ slopes $< 0$, making the Sun appear less anomalous (though still relatively refractory depleted) and more consistent with the three lower resolution surveys.}
    \label{fig:tcond_slope_dists}
\end{figure}

The differences in reported element abundance measurements and precisions as propagated from the different spectral resolutions, signal to noise, and wavelength coverage, between the sample from \cite{bedell18} (R $> 100,000$) and the remainder of the samples (R $< 22,000$) impacts the distributions of \tcond\ slopes. In the \cite{bedell18} sample, 92\%\footnote{This percentage differs from that reported in \cite{bedell18} due to the argument discussed earlier: the number of and choice of elements changes the \tcond\ slope.} of stars have \tcond\ slope $ > 0$, showing a heavy skew towards stars that are relatively refractory enriched compared to the Sun. This contrasts with the samples from GALAH DR4, APOGEE, and \citetalias{Rampalli24}, where more stars appear relatively refractory depleted (like the Sun), with 60–80\% showing \tcond\ slopes $> 0$. We resample the abundances from \cite{bedell18} with larger measurement uncertainties that reflect the mean abundance uncertainties from the lower resolution surveys. After injecting these uncertainties 1000 times and remeasuring \tcond\ slopes, the fraction of stars with \tcond\ slopes $< 0$ increases from 6.2\% to 26.9\%, and the Sun’s percentile rank shifts from the 6th to the 27th percentile. The broader slope distribution makes the Sun appear less anomalous and more in line with the distributions from the other surveys (Figure~\ref{fig:tcond_slope_dists}). Although the Sun’s exact percentile varies, it consistently remains below the mean of all four samples, reinforcing its classification as relatively refractory depleted. 

\subsection{Determining Stars' Nucleosynthetic fingerprints with the K-Process Model}\label{sec:kpm}
We calculate the contribution of Type Ia and core-collapse supernovae to each star's abundance profile by applying the K-process model (KPM) from \cite{Griffith24,zenodo_kpm}. The code used in \cite{Griffith24} and what we implement can be found in their Github repository\footnote{\href{https://github.com/13emilygriffith/KProcessModel}{https://github.com/13emilygriffith/KProcessModel}}. This framework uses a data-driven approach to model the relative contributions of nucleosynthetic processes that drive stars' abundance patterns in [X/H] (compared to the \tcond\ slopes measured in [X/Fe]). In agreement with previous work using similar methods \citep{Griffith19,Weinberg19,Griffith22,Weinberg22}, they show that stars' chemistries can be modeled reasonably well with a K=2 model, where the two processes are a prompt process (astrophysically corresponding to massive stars and core-collapse supernovae) and delayed process (corresponding to Type Ia supernovae).

Each elemental abundance for star $i$ is described as a linear combination of the two processes:
\begin{equation} \label{eqn:xh}
    \text{[X/H]}_{i} = A_{i}^{cc}*q_{cc}^{X} + A_{i}^{Ia}*q_{Ia}^{X}, 
\end{equation}
where the process amplitudes, $A_{i}^{k}$, describe the relative contributions of prompt and delayed processes ($k =$ CC, Ia) to each star and the process vector components ($q_{k}^{Z}$) describe the relative yields of each element, $Z$ (represented as $X$ for [X/H] in Equation \ref{eqn:xh}) produced by the two nucleosynthetic processes for each element as a function of metallicity.

The fractional contribution of the prompt process to element $Z$, in the $i$th star can be written as:

\begin{equation} \label{eqn:fcc}
    f_{i,Z}^{cc} = \frac{A_{i}^{cc} q^{Z}_{cc}}{A_{i}^{cc} q^{Z}_{cc} + A_{i}^{Ia} q^{Z}_{Ia}}.
\end{equation}

 We retain the assumption from \cite{Griffith24} that $q^{Mg}_{cc} = 1$, $q^{Mg}_{Ia} = 0$, $q^{Fe}_{cc} = 0.4$, and $q^{Fe}_{Ia} = 0.6$, substantiated by theoretical yields \citep[e.g.,][]{Woosley95,Arnett96,Thielemann02,Andrews17,Rybizki17,Anderson19}. The KPM is constrained by the 12 abundances available for all stars. The KPM then uses $\chi^2$ minimization to find the best-fitting values for $q$ and $A$. While conceptually there are 12 $q_{\mathrm{CC}}$ and 12 $q_{\mathrm{Ia}}$ (one for each element), in practice each star has its own metallicity-dependent $q$ values, so the dataset contains $100{,}000+$ inferred $q_{\rm CC}$ and $q_{\rm Ia}$ values in total. There are also $100{,}000+$ $A_{\mathrm{CC}}$ and $A_{\mathrm{Ia}}$ (one set for each star). We then assess the KPM's accuracy by comparing the measured abundances to those calculated by the KPM using Equation~\ref{eqn:xh}.

In Table~\ref{tab:resid_errors_condT}, we report the average residuals (the standard deviation between KPM-inferred and input, literature abundances, $\sigma_{\mathrm{KPM}}$), mean observational uncertainties ($\langle \sigma_{\mathrm{survey}} \rangle$), and inferred intrinsic dispersion ($\sigma_{\mathrm{int}}$, computed as \(\sqrt{\sigma_{\mathrm{KPM}}^2 - \langle \sigma_{\mathrm{survey}} \rangle^2}\)) for each element, [X/H], and survey. Most elements are reasonably matched, with residuals comparable to or smaller than the mean observational uncertainties, especially for HARPS, GALAH, and Gaia RVS data. For Gaia RVS, the reported abundance uncertainties are derived from a data‑driven model and represent upper limits. This likely contributes to the consistently smaller KPM residuals relative to the quoted uncertainties for this survey. In contrast, several elements from APOGEE, particularly Cr, Ti, V, and Na, exhibit significantly larger residuals (e.g., \(\sigma_{\mathrm{KPM}} = 0.42\)~dex for Na). We attribute this to known limitations in certain APOGEE abundance determinations, consistent with issues documented in \citet{Jonsson2020} and \citet{Griffith21}, rather than a breakdown of the KPM model. 

\begin{longtable*}{llccc}
\caption{Residual scatter from the KPM fits (\(\sigma_{\mathrm{KPM}}\)) and mean observational uncertainties (\(\langle \sigma_{\mathrm{survey}} \rangle\)) for each element, [X/H], and survey. The intrinsic dispersion column is computed as \(\sqrt{\sigma_{\mathrm{KPM}}^2 - \langle \sigma_{\mathrm{survey}} \rangle^2}\).}
\label{tab:resid_errors_condT} \\

\hline
Element & Survey & \(\sigma_{\mathrm{KPM}}\) (dex) & \(\langle \sigma_{\mathrm{survey}} \rangle\) (dex) & $\sigma_{\rm int}$ (dex) \\
\hline
\endfirsthead

\multicolumn{5}{c}{{\tablename\ \thetable{} -- Continued}} \\
\hline
Element & Survey & \(\sigma_{\mathrm{KPM}}\) (dex) & \(\langle \sigma_{\mathrm{survey}} \rangle\) (dex) & $\sigma_{\rm int}$ (dex) \\
\hline
\endhead

\hline
\multicolumn{5}{r}{{Continued on next page}} \\
\endfoot

\hline
\endlastfoot

\multirow{4}{*}{Al} 
& APOGEE DR17     & 0.08 & 0.02 & 0.08 \\
& GALAH DR4       & 0.09 & 0.07 & 0.07 \\
& Gaia RVS (R24)  & 0.04 & 0.07 & -- \\
& HARPS (B18)     & 0.03 & 0.01 & 0.03 \\
\hline
\multirow{4}{*}{Ca} 
& APOGEE DR17     & 0.05 & 0.02 & 0.05 \\
& GALAH DR4       & 0.05 & 0.04 & 0.04 \\
& Gaia RVS (R24)  & 0.03 & 0.07 & -- \\
& HARPS (B18)     & 0.02 & 0.01 & 0.02 \\
\hline
\multirow{4}{*}{Cr} 
& APOGEE DR17     & 0.22 & 0.07 & 0.21 \\
& GALAH DR4       & 0.03 & 0.02 & 0.02 \\
& Gaia RVS (R24)  & 0.02 & 0.06 & -- \\
& HARPS (B18)     & 0.02 & 0.01 & 0.02 \\
\hline
\multirow{4}{*}{Fe} 
& APOGEE DR17     & 0.01 & 0.01 & 0.00 \\
& GALAH DR4       & 0.02 & 0.06 & -- \\
& Gaia RVS (R24)  & 0.02 & 0.04 & -- \\
& HARPS (B18)     & 0.05 & 0.00 & 0.05 \\
\hline
\multirow{4}{*}{Mg} 
& APOGEE DR17     & 0.03 & 0.01 & 0.03 \\
& GALAH DR4       & 0.05 & 0.03 & 0.04 \\
& Gaia RVS (R24)  & 0.03 & 0.06 & -- \\
& HARPS (B18)     & 0.01 & 0.01 & 0.01 \\
\hline
\multirow{4}{*}{Mn} 
& APOGEE DR17     & 0.04 & 0.02 & 0.04 \\
& GALAH DR4       & 0.04 & 0.03 & 0.03 \\
& Gaia RVS (R24)  & 0.06 & 0.11 & -- \\
& HARPS (B18)     & 0.02 & 0.01 & 0.01 \\
\hline
\multirow{4}{*}{Na} 
& APOGEE DR17     & 0.42 & 0.15 & 0.39 \\
& GALAH DR4       & 0.06 & 0.03 & 0.05 \\
& Gaia RVS (R24)  & 0.05 & 0.09 & -- \\
& HARPS (B18)     & 0.03 & 0.01 & 0.03 \\
\hline
\multirow{4}{*}{Ni} 
& APOGEE DR17     & 0.03 & 0.01 & 0.03 \\
& GALAH DR4       & 0.03 & 0.02 & 0.02 \\
& Gaia RVS (R24)  & 0.02 & 0.06 & -- \\
& HARPS (B18)     & 0.01 & 0.01 & 0.01 \\
\hline
\multirow{4}{*}{O} 
& APOGEE DR17     & 0.11 & 0.08 & 0.08 \\
& GALAH DR4       & 0.12 & 0.07 & 0.09 \\
& Gaia RVS (R24)  & 0.06 & 0.09 & -- \\
& HARPS (B18)     & 0.04 & 0.01 & 0.03 \\
\hline
\multirow{4}{*}{Si} 
& APOGEE DR17     & 0.03 & 0.02 & 0.03 \\
& GALAH DR4       & 0.03 & 0.02 & 0.02 \\
& Gaia RVS (R24)  & 0.02 & 0.06 & -- \\
& HARPS (B18)     & 0.01 & 0.00 & 0.01 \\
\hline
\multirow{4}{*}{Ti} 
& APOGEE DR17     & 0.22 & 0.08 & 0.21 \\
& GALAH DR4       & 0.04 & 0.03 & 0.02 \\
& Gaia RVS (R24)  & 0.03 & 0.06 & -- \\
& HARPS (B18)     & 0.02 & 0.01 & 0.02 \\
\hline
\multirow{4}{*}{V} 
& APOGEE DR17     & 0.21 & 0.15 & 0.15 \\
& GALAH DR4       & 0.05 & 0.05 & 0.02 \\
& Gaia RVS (R24)  & 0.04 & 0.07 & -- \\
& HARPS (B18)     & 0.02 & 0.01 & 0.02 \\
\hline

\end{longtable*}


 Throughout the rest of the manuscript, we refer to the star-specific process amplitudes $A_{i}^{cc}$ and $A_{i}^{Ia}$ as \acc\ and \aia, and the element-specific process yields $q^{Z}_{\rm CC}$ and $q^{Z}_{\rm Ia}$ as $q_{\rm cc}$ and $q_{\rm Ia}$, respectively. The fractional contribution of the prompt process, $f_{i,Z}^{cc}$, for each element for each star is referred to as \fcc. {We apply the KPM to each of the four samples independently and report \acc, \aia, $q_{\rm cc}$, $q_{\rm Ia}$, and \fcc\ values for each survey separately. For ease of comparison across figures, we show the Sun’s mean values across surveys. Importantly, all stellar samples are analyzed independently within each survey, and the Sun’s per-survey values are reported explicitly in Table~\ref{t:suns}. }

\subsection{Testing Predictive Power of Stellar Parameters For \tcond\ slopes with Linear Regressions}

To understand which stellar parameters and chemical abundances have the most predictive power for a star's \tcond\ slope, we determine which parameters are most correlated with \tcond\ slope using a series of linear regressions. Using the \texttt{sklearn} package in Python \citep{sklearn}, we model \tcond\ slope as a single or linear combination of the stellar parameters of interest (e.g., \feh, \alphafe) in the following manner: 
\begin{equation} \label{eqn:lr}
    \tcondslope\ = A \cdot \feh\ + B \cdot \alphafe\ + C,
\end{equation}
where $A$ and $B$ are the regression coefficients and $C$ is the intercept. For a given star, we remove it from the dataset and perform a regression of \tcond\ slope as a function of a chosen combination of stellar parameters (\feh\ and \alphafe\ from Equation \ref{eqn:lr}) using the remaining stars. Using the resulting regression coefficients, we compute the expected \tcond\ slope for the removed star from Equation~\ref{eqn:lr}. This process is repeated for each star and each parameter combination. The predicted \tcond\ slope for star $i$, $\left( \frac{\partial[\mathrm{X/Fe}]}{\partial T_{\rm cond}} \right)_i^{\mathrm{mod}}$ is then compared to the observed value, $\left( \frac{\partial[\mathrm{X/Fe}]}{\partial T_{\rm cond}} \right)_i^{\mathrm{obs}}$, and their difference defines the residual. 

We assess model performance using the \rsq\ metric, which compares the variance in the residuals to the variance in the observed \tcond\ slopes:

\begin{equation}
R^2 = 1 - \frac{\sum \left( \left( \frac{\partial[\mathrm{X/Fe}]}{\partial T_{\rm cond}} \right)_i^{\mathrm{obs}} - \left( \frac{\partial[\mathrm{X/Fe}]}{\partial T_{\rm cond}} \right)_i^{\mathrm{pred}} \right)^2}{\sum \left( \left( \frac{\partial[\mathrm{X/Fe}]}{\partial T_{\rm cond}} \right)_i^{\mathrm{obs}} - \left\langle \frac{\partial[\mathrm{X/Fe}]}{\partial T_{\rm cond}} \right\rangle^{\mathrm{obs}} \right)^2},
\end{equation}

where \( \left( \frac{\partial[\mathrm{X/Fe}]}{\partial T_{\rm cond}} \right)_i^{\mathrm{obs}} \) and
\( \left( \frac{\partial[\mathrm{X/Fe}]}{\partial T_{\rm cond}} \right)_i^{\mathrm{pred}} \) are defined above
and \( \left\langle \frac{\partial[\mathrm{X/Fe}]}{\partial T_{\rm cond}} \right\rangle^{\mathrm{obs}} \) is the mean observed \tcond\ slope across the dataset. A higher \rsq\ indicates that the chosen stellar parameters applied in the regression better explain variance in the \tcond\ slope.

We run a series of linear regressions using combinations of \teff, \logg, \feh, \alphafe, \acc, \aia, [Ce/Fe], and [Y/Fe] (the latter two are presented in Appendix~\ref{sec:dimensionality}). We test a representative set of parameter combinations to see what drives \tcond\ slope variation such as the impact of individual parameters, how proxies compare (e.g., \alphafe\ vs. \acc/\aia), and how different combinations perform together in Sections \ref{sec:tcondtrends}, \ref{sec:acc_aia_results}, and Appendix \ref{sec:dimensionality}. The tested combinations are noted in each section. Polynomial regressions yield minimal \rsq\ improvement, and are not reported.

\section{Empirical Trends in Measured Stellar Properties}\label{sec:results}
To contextualize the observed chemical trends, we compile a set of stellar parameters and inferred nucleosynthetic properties for each solar analog in Table~\ref{t:sas}. We report:
\begin{itemize}
    \item each solar analog's original survey identifiers
    \item survey source 
    \item status as a confirmed planet host 
    \item  kinematics (Gaia DR3 RA, Dec, and parallax)
    \item fundamental stellar parameters and their uncertainties (\teff, \logg, \feh, and \alphafe)

    \item $\text{[X/Fe]}$ for 11 elements and their associated uncertainties
    \item measured \tcond\ slope and intercept with uncertainties (with and without $\alpha$-elements)
    \item inferred material contributions from core-collapse and Type Ia supernovae (\acc\ and \aia) from the KPM following \cite{Griffith24}
    \item inferred relative yields for each element ($q_{\rm cc}$, $q_{\rm Ia}$) from the KPM following \cite{Griffith24}
    \item inferred fraction of each stars' elements originating from core-collapse supernovae (\fcc) from the KPM following \cite{Griffith24}.
\end{itemize} 

These parameters form the basis for our subsequent analysis of how stellar parameters and inferred enrichment histories shape abundance patterns in solar analogs.

\begin{figure*}
    \centering
\includegraphics[width=0.47\textwidth]{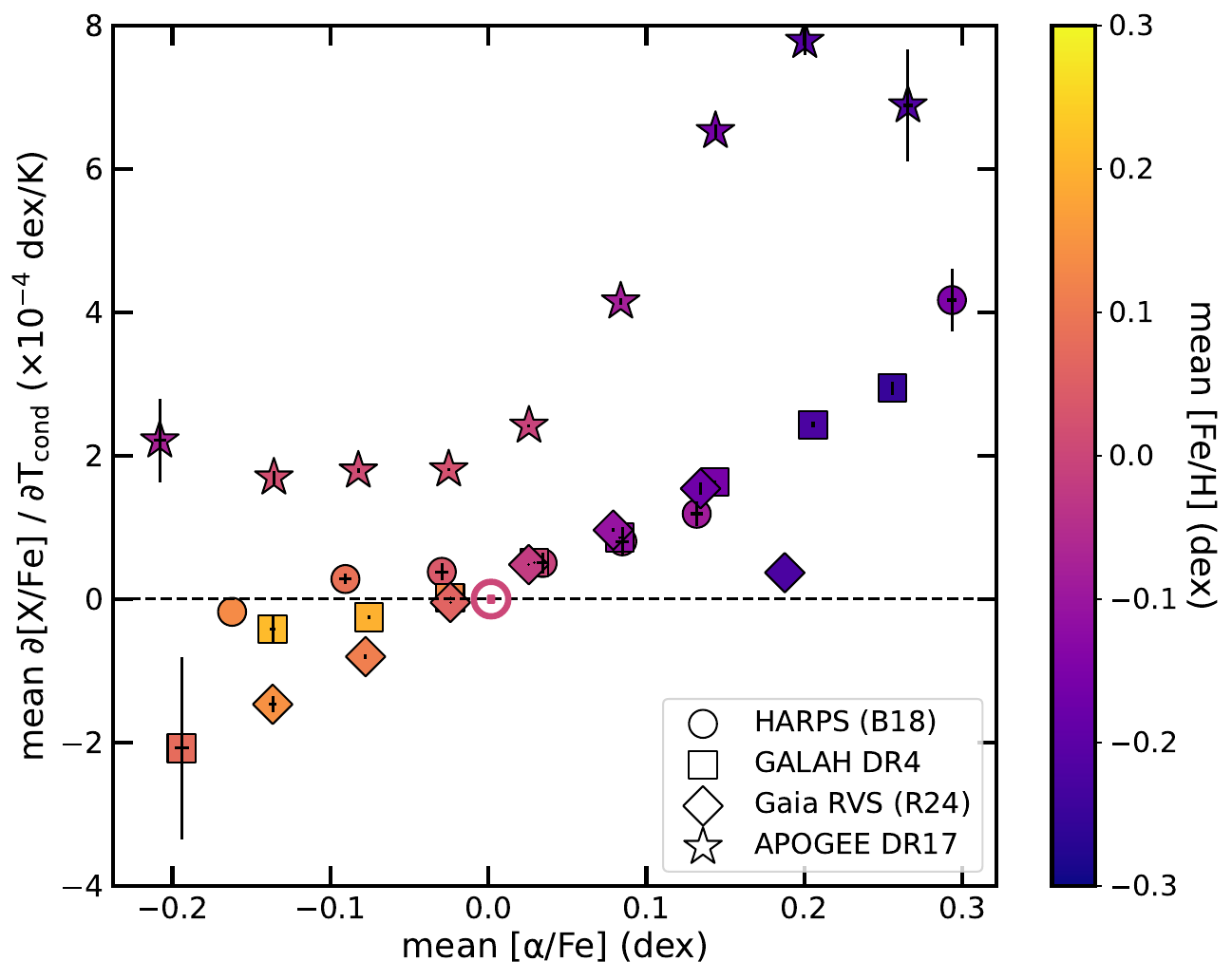}
\includegraphics[width=0.49\textwidth]{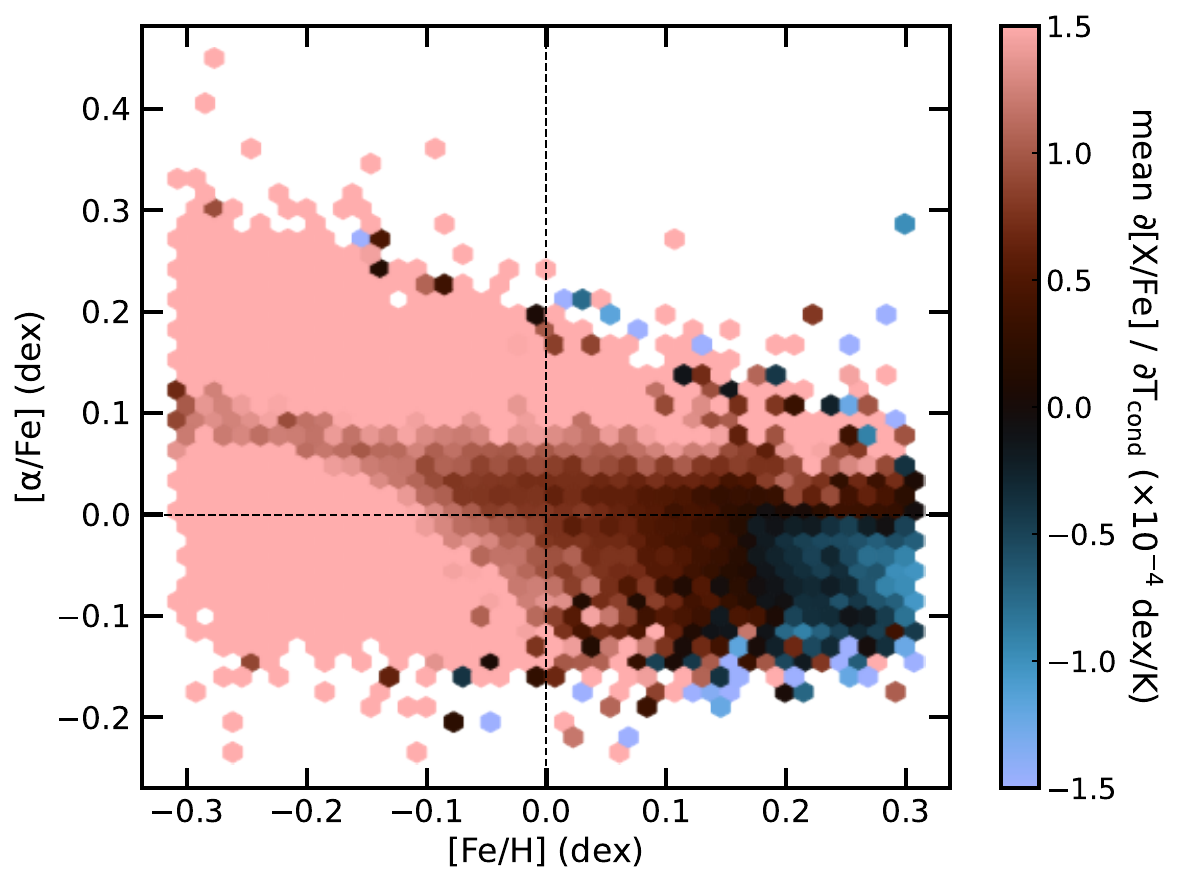}
    \caption{\textit{Left:} Mean \tcond\ slope (\tcondslope) as a function of mean \alphafe\ and colored by mean \feh\ with uncertainties on the mean for solar analogs with \cite{bedell18} as circles, GALAH as squares, \citetalias{Rampalli24} as diamonds, and APOGEE as stars. There is a positive gradient in \tcond\ slope with \alphafe\ (10$\pm4$E-4 K$^{-1}$) and a similarly strong negative gradient with \feh\ as well (-4$\pm1$E-4 K$^{-1}$). Metal-rich, $\alpha$-enriched stars tend to have \tcondslope\ = 0, consistent with the Sun. The APOGEE stars are systematically offset from the rest of the surveys given their difference in kinematic selection. \textit{Right:} \feh-\alphafe\ distribution of stars colored by their \tcond\ slope. We confirm what was seen on the left panel: the majority of stars with a \tcond\ slope consistent with zero are metal-rich and depleted in \alphafe\ compared to the Sun. Hexbin cells with mean \tcond\ slopes consistent with zero are centered at \feh\ = 0.1 dex and \alphafe\ = $-0.02$ dex.} 
    \label{fig:fehalphafe}
\end{figure*}

\subsection{\tcond\ Slope Trends in \feh-\alphafe\ and \teff-\logg\ Space} \label{sec:tcondtrends}

We examine how \tcond\ slopes vary across the \feh--\alphafe\ and \teff--\logg\ planes in Figures~\ref{fig:fehalphafe} and \ref{fig:tefflogg}. In the left panel of Figure~\ref{fig:fehalphafe}, we bin stars by \alphafe\ and plot their mean \tcond\ slope, color-coded by mean \feh. With increasing \alphafe, \tcond\ slopes increase while \feh\ decreases, producing a range of \tcond\ slopes from $-2-8 \times 10^{-4}$ dex/K in our sample. By applying linear regressions across samples, we measure gradients of $10\pm 4 \times 10^{-4}$ K$^{-1}$ and $-4 \pm 1 \times 10^{-4}$ K$^{-1}$ in \tcond\ slope as a function of \alphafe\ and \feh\ respectively. The corresponding $R^2$ values are $23 \pm 10\%$ and $15 \pm 5\%$, indicating that \alphafe\ is a better predictor of \tcond\ slope than \feh.

By definition, the Sun has a \tcond\ slope of 0 (represented by the dashed line in Figure \ref{fig:fehalphafe}) and is best matched by stars that are more metal-rich and $\alpha$-depleted. We highlight this in the right hand panel of Figure \ref{fig:fehalphafe} by combining all the samples and showing a binned \feh-\alphafe\ plane colored by the mean \tcond\ slope. Hexbin cells with mean \tcond\ slopes consistent with zero are centered at a mean \feh\ $= 0.1$ dex and \alphafe\ $= -0.02$ dex while the hexbin cell at (0,0) shows an average \tcond\ slope of $\sim 1 \times 10^{-4}$ dex/K, consistent with the solar twin \tcond\ slope average reported in \cite{bedell18}.

\begin{figure}
    \centering
    \includegraphics[width=0.5\linewidth]{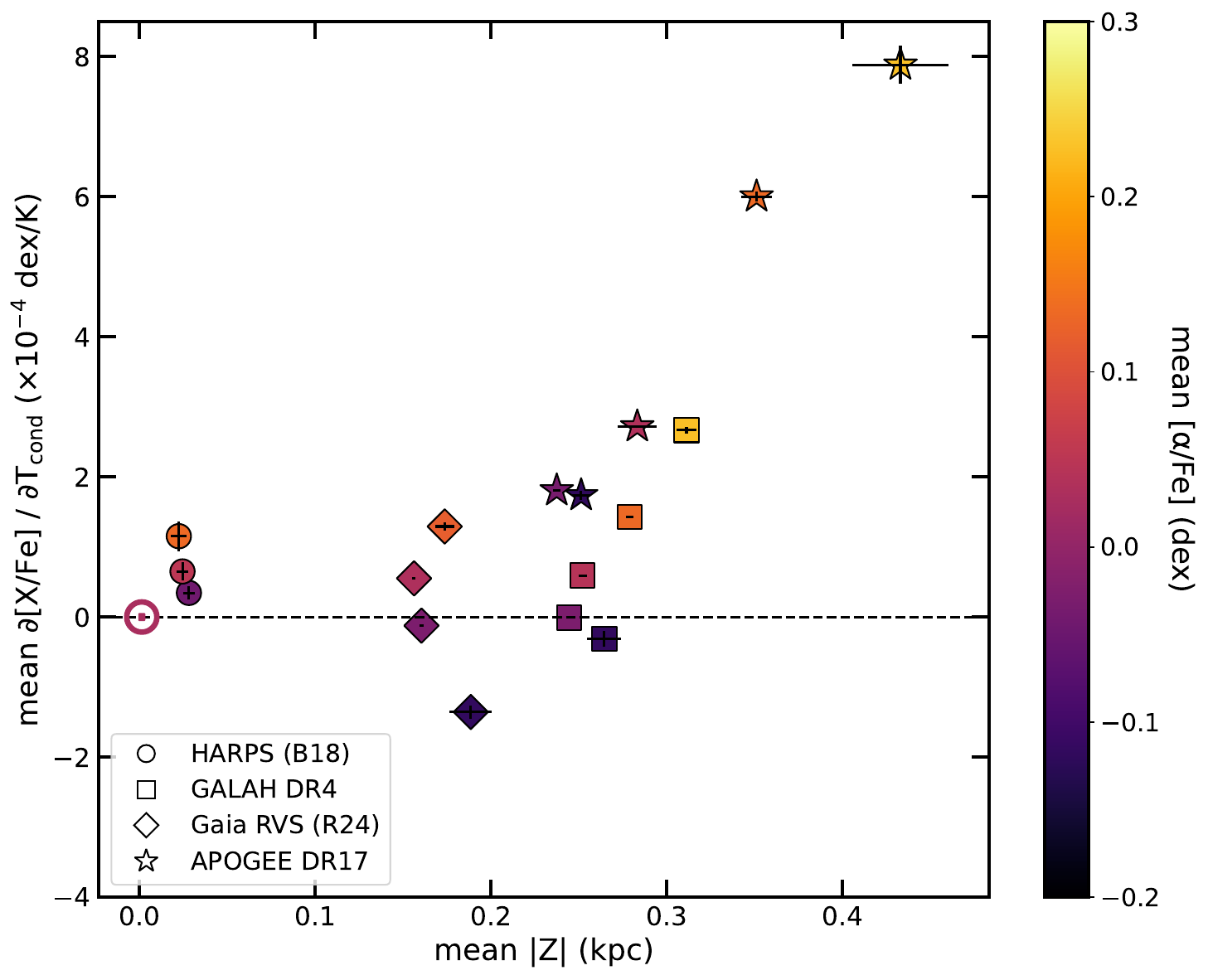}
    \caption{Mean \tcond\ slope (\tcondslope) as a function of binned absolute vertical distance from the Galaxy's midplane, $|Z|$, colored by mean \alphafe, shown for each survey: \cite{bedell18} as circles, GALAH as squares, \citetalias{Rampalli24} as diamonds, and APOGEE as stars. APOGEE shows systematically higher \tcond\ slopes and occupies larger $|Z|$  on average. While the dominant trend is that higher $|Z|$  corresponds to steeper \tcond\ slopes, APOGEE stars exhibit elevated \tcond\ slopes even compared to other survey bins with similar \alphafe, suggesting a residual dependence on vertical kinematics or correlated properties such as age. Bins with $<10$ stars are omitted from this figure.}
    \label{fig:z_tcond}
\end{figure}

Stars from APOGEE exhibit systematically higher \tcond\ slopes, which we attribute in part to its distinct survey selection: APOGEE samples more stars at larger vertical heights from the Galaxy's midplane ($|Z|$ ~$\sim 0.3$--$0.5$~kpc) compared to the other surveys ($|Z|$ ~$\lesssim 0.3$~kpc). As discussed by \cite{Rampalli25} in their Figure 6 and Section 5.1, APOGEE’s deeper, near-infrared targeting yields a kinematically hotter sample that is likely older, more $\alpha$-enhanced, and metal-poor. The latter two properties are correlated with steeper \tcond\ slopes as shown in Figure \ref{fig:fehalphafe}. However, as shown in Figure~\ref{fig:z_tcond}, we observe a weak trend between vertical height and \tcond\ slope for each sample. Across samples, even at similar \alphafe\ values (as indicated by color), APOGEE stars tend to lie at larger $|Z|$ and \tcond\ slopes. This persistent offset suggests that vertical height, or a correlated property such as age, may introduce additional structure in \tcond\ slope. Disentangling these effects is beyond the scope of this work. Nonetheless, the dominant trend across all samples is that \tcond\ slope increases with \alphafe\ and decreases with \feh.

\begin{figure}
    \centering
    \includegraphics[width=0.5\textwidth]{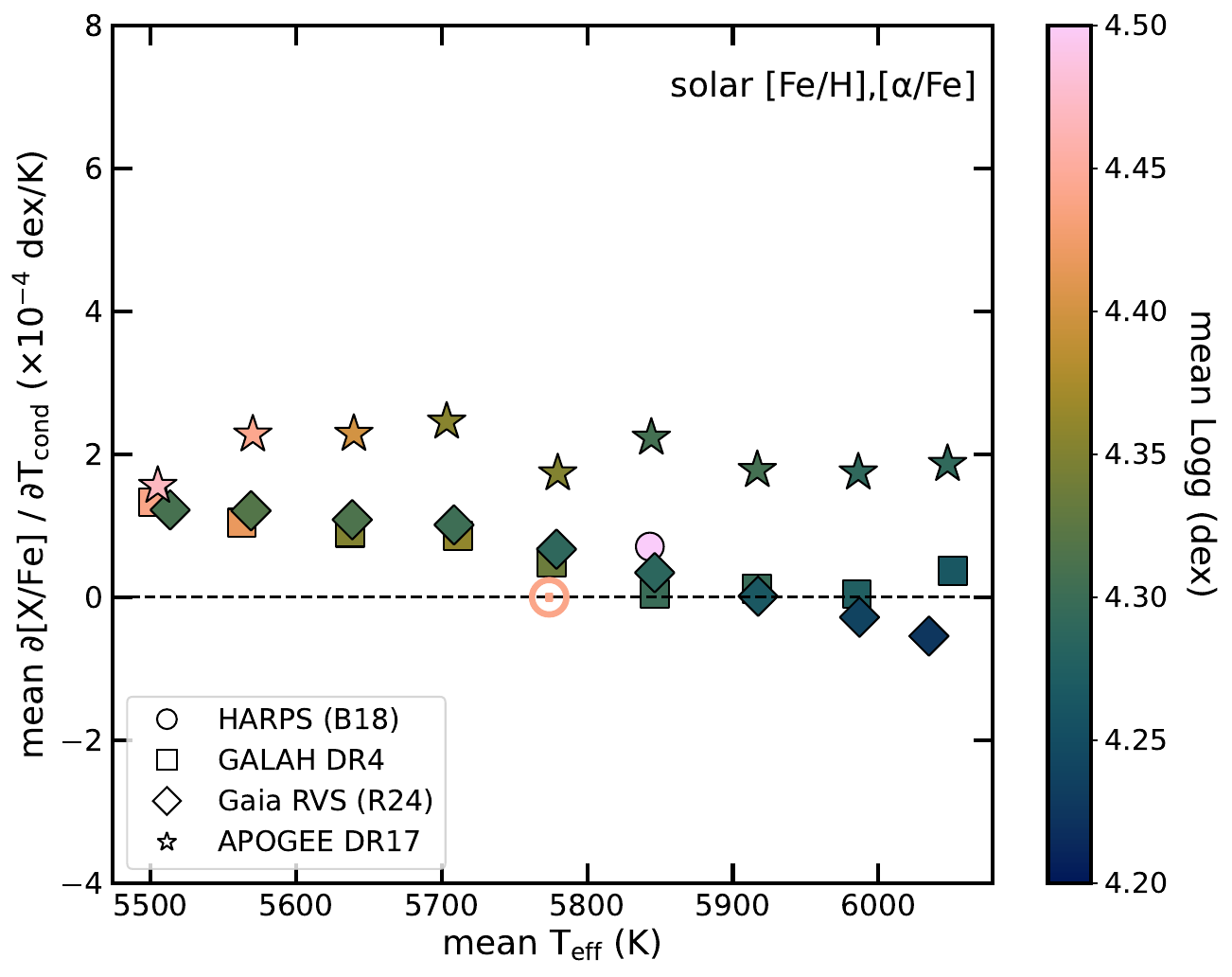}
    \caption{Mean \tcond\ slope (\tcondslope) as a function of mean \teff\ and colored by mean \logg\ for stars with the same \feh\ and \alphafe\ consistent with the Sun with \cite{bedell18} as circles (only one symbol shown since these stars occupy one bin), GALAH as squares, \citetalias{Rampalli24} as diamonds, and APOGEE as stars. There is a slight negative gradient in \tcond\ slope with \teff\ for each dataset (-3$\pm1$E-7 dex/K$^{2}$) and with \logg\ (-1$\pm3$E-4 K$^{2}$). Hotter and lower \logg\ stars tend to share \tcondslope\ = 0 than stars with similar \teff\ and \logg\ as the Sun. The APOGEE stars are systemically offset from the rest of the surveys given their difference in kinematic selection as shown in Figure \ref{fig:fehalphafe}. 
    }
    \label{fig:tefflogg}
\end{figure}

Given the strong influence of \feh\ and \alphafe\ on \tcond\ slopes, we restrict our dataset to stars with solar \feh\ and \alphafe\  within their 1$\sigma$ uncertainties to understand the influences of \teff\ and \logg\ on \tcond\ slope. In Figure \ref{fig:tefflogg}, we measure the gradients of
 \tcond\ slope as a function of \teff\ and \logg\ using linear regression. We find a negative gradient of -3$\pm1$E-7 dex/K$^{2}$ and -1$\pm3$E-4 K$^{-1}$ respectively. The Sun looks like an outlier in this visualization, with a \tcond\ slope that resembles slightly hotter stars, lower in \logg. However, we are only showing the mean trend, and the Sun sits well within the distribution when considering the entire dataset. APOGEE stars differ systematically from the rest of the surveys, which is a consequence of kinematic selection (see discussion and Figure \ref{fig:z_tcond} above). The broader range of \tcond\ slopes spanned by \feh\ and \alphafe\ compared to \teff\ and \logg\ suggests that stellar chemistry has a stronger influence on \tcond\ slope variance. We calculate \rsq\ values of $9\pm5\%$ for \teff\ and $13\pm16\%$ for \logg, confirming their weaker predictive power relative to \feh\ and \alphafe\ (each $>15\%$). 
When we run a linear regression of \tcond\ slope as a function of all of the parameters \teff, \logg, \feh, and \alphafe, we find an \rsq\ value of $49\pm18$ \%, indicating that ultimately all of these parameters have an effect in shaping \tcond\ slopes. The stronger correlation with \feh\ and \alphafe\ potentially indicate that \tcond\ slopes are shaped by nucleosynthesis, which we explore further in the next section.

\subsection{Nucleosynthetic Signatures in [X/Mg]–[Mg/Fe] Trends} 

\begin{figure*}
    \centering
    \includegraphics[width=\linewidth]{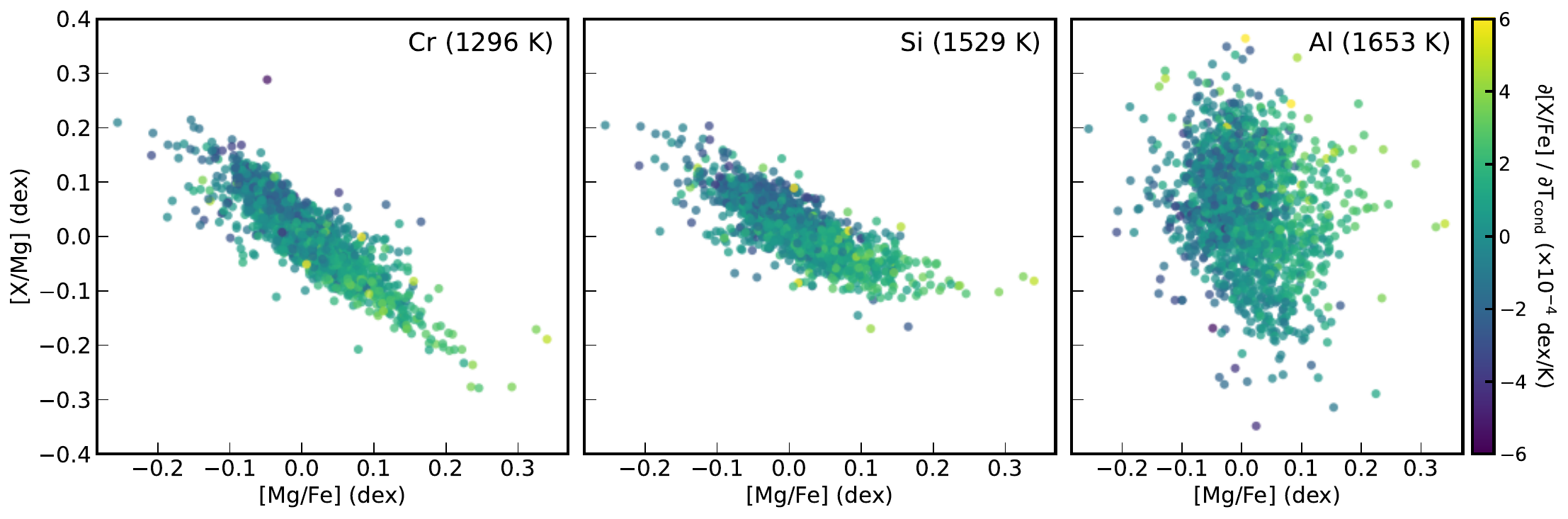}
    \caption{Observed [X/Mg]-[Mg/Fe] for Cr, Si, and Al for solar analogs with uncertainties $< 0.03$ dex, shown in order of their $T_{\rm cond}$ and colored by their observed $T_{\rm cond}$ slope. Low-\tcond\ elements like Cr show declining [X/Mg] with [Mg/H], while high-\tcond\ elements like Al track Mg more closely, resulting in flatter trends. This progressive flattening with increasing \tcond\ suggests that \tcond\ correlates with stellar yields and reflects systematic differences in nucleosynthetic origin and the relative contributions from core-collapse and Type Ia supernovae. {Data are not separated by survey here for visual clarity.}
}
    \label{fig:ex_xmg}
\end{figure*}

Given that solar analogs more metal-rich and $\alpha$-depleted than the Sun share its same abundance pattern, we investigate how GCE contributes to these trends. In Figure \ref{fig:ex_xmg}, we show [X/Mg] as a function of [Mg/Fe] for three elements, Cr, Si, and Al, that have a range of \tcond. Mg is predominantly produced in core-collapse supernovae, while Fe is produced in Type Ia supernovae; similarly to \alphafe, [Mg/Fe] traces the enrichment from core-collapse supernovae. [X/Mg] indicates whether an element shares Mg’s origin in core-collapse supernovae or diverges due to additional nucleosynthetic sources. For example, the declining [Cr/Mg] trend in the left panel reflects that Type Ia supernovae are an additional contributor to Cr enrichment, since Type Ia supernovae produce both Cr and Fe but not Mg. We observe that [X/Mg] trends progressively flatten with increasing \tcond, indicating that the relative contribution from core-collapse supernovae dominates for higher-\tcond\ elements. This result supports the idea that \tcond\ is correlated with stellar yields, or the element-by-element outputs of nucleosynthetic processes, as discussed in the previous section, and motivates our use of the data-driven KPM (Section~\ref{sec:kpm}) to model these abundance patterns and disentangle their nucleosynthetic origins.

\section{Interpreting \tcond\ slopes in a Physically Motivated Framework} \label{sec:kpm_results}

Building on the observed correlations between \tcond\ slopes and stellar parameters linked to nucleosynthetic enrichment in the previous section, we now turn to the data-driven KPM (as described in Section \ref{sec:kpm}) to interpret the trends in Figures \ref{fig:fehalphafe} and \ref{fig:ex_xmg}. By modeling stellar abundances as a combination of core-collapse and Type Ia supernovae contributions, the KPM offers a physically motivated framework to contextualize the processes that shape a star's \tcond\ slope.

\subsection{\acc\ and \aia: The Sun in Context}\label{sec:acc_aia_results}

\begin{figure*}
    \centering
    \includegraphics[width=0.75\textwidth]{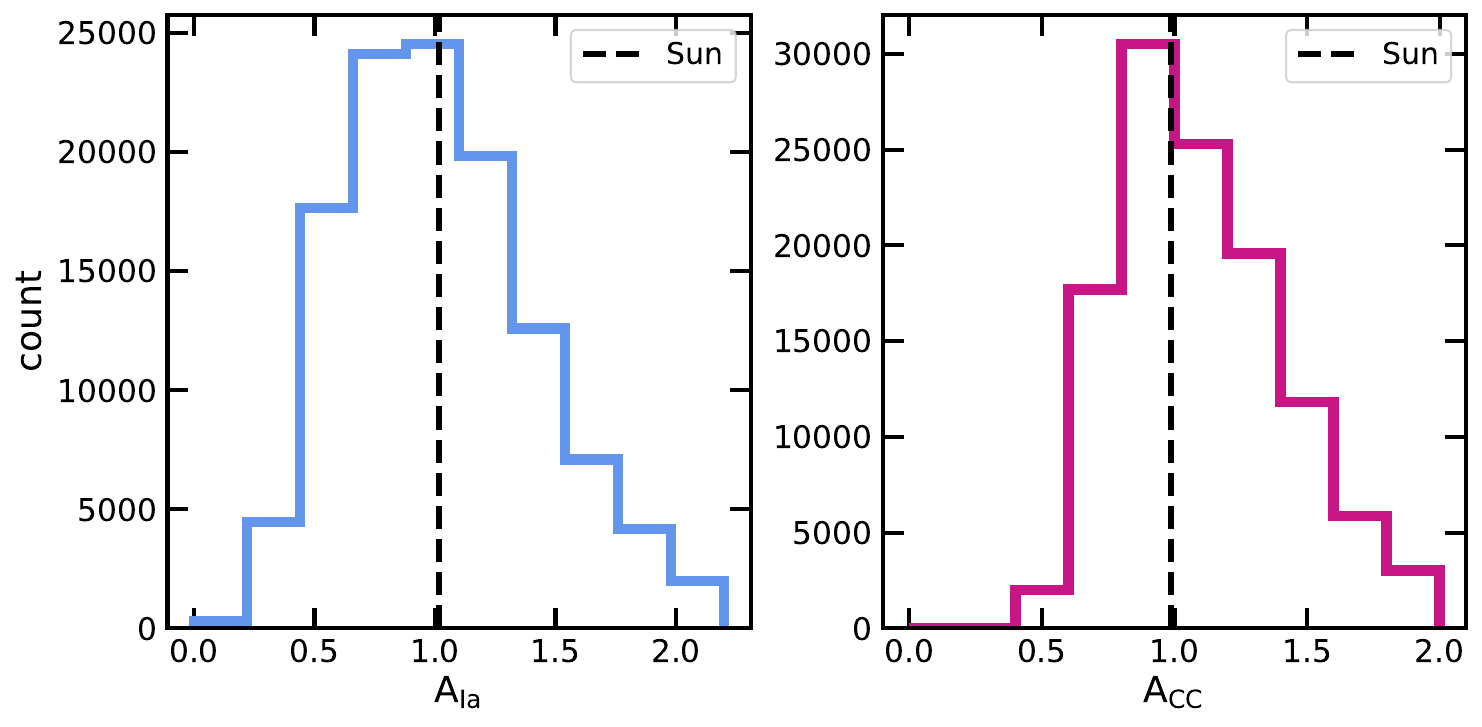}
    \caption{Distribution of relative contribution of Type Ia supernovae material, \aia\ (blue, left), and Type II/CC supernovae material, \acc\ (red, right) for the solar analog population. The Sun is within 1$\sigma$ in \aia\ and \acc\ compared to the rest of the distribution indicating it is chemically ordinary. {Data are not separated by survey here for visual clarity.}}
    \label{fig:ahists}
\end{figure*}

To place the Sun’s abundance pattern in a broader context, we examine its inferred enrichment history using the KPM parameters \acc\ and \aia, which quantify the relative contributions of core-collapse and Type Ia supernovae, respectively, to a star’s chemical makeup. Figure \ref{fig:ahists} shows the \aia\ (left, blue) and \acc\ distributions (right, red) for the entire solar analog dataset. The Sun is marked by a dashed, vertical line and has a similar \aia\ value (1) and \acc\ value (0.99) compared to the means of each distribution ($1\pm0.4$, $1.1\pm0.33$) respectively.

\label{sec:accaia}

\begin{figure*}
    \centering
    \includegraphics[width=\textwidth]{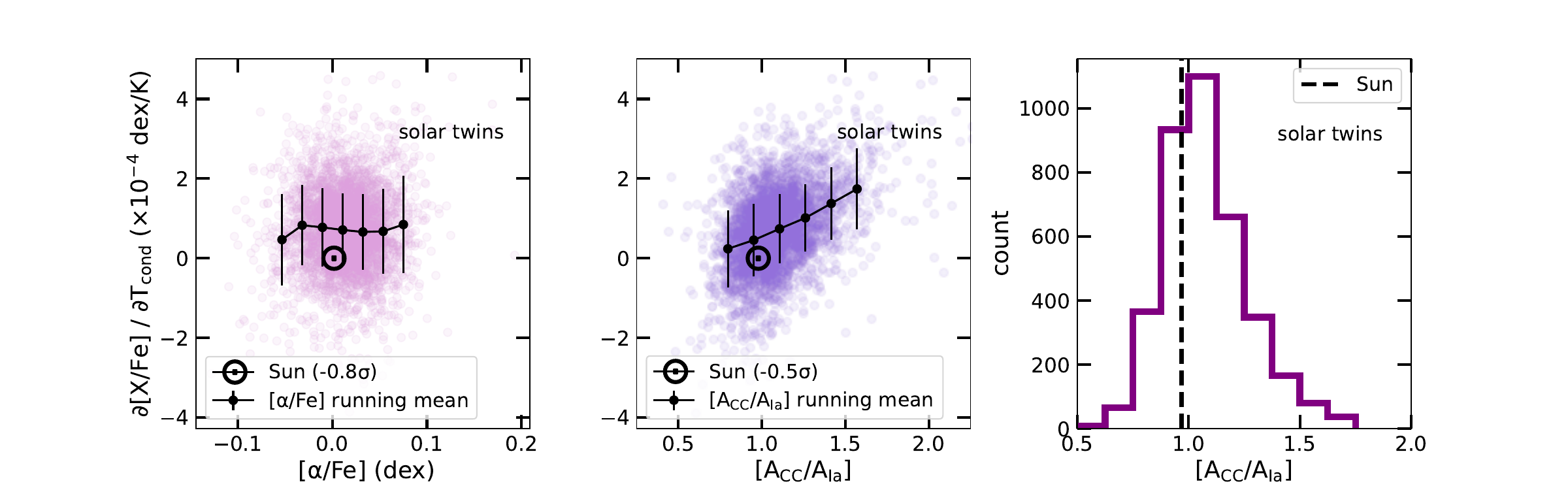}
    
  \caption{Comparison of \kpmalpha\ and \alphafe\ as predictors of \tcond\ slope for stars with with the \teff, \logg, and \feh, as the Sun within 1$\sigma$ uncertainties. {This includes the high-resolution HARPS solar-twin sample from \citet{bedell18} as well as stars from the other surveys meeting the same parameter criteria.} \textit{Left:} \tcond\ slope values as a function of \alphafe\ with the running mean and 1$\sigma$ standard deviation plotted. The Sun is 0.8$\sigma$ from the running mean. \textit{Middle:} \tcond\ slope values as a function of KPM-proxy of \alphafe, \kpmalpha\ with the running mean plotted. The Sun is 0.5$\sigma$ from the running mean. \textit{Right:} Histogram of \kpmalpha. The Sun sits within 1$\sigma$ in \kpmalpha\ space.  While the Sun sits slightly below the mean relation, it remains well within the intrinsic population scatter about the KPM-predicted trend. Data are not separated by survey here for visual clarity.}
    \label{fig:alphacomp}
\end{figure*}

We previously showed that both \feh\ and \alphafe\ correlate with \tcond\ trends across the full solar analog sample (Figure~\ref{fig:fehalphafe}). \feh\ traces overall metal content, while \alphafe\ (measured from the $\alpha$-process elements) reflects the relative contributions of core-collapse and Type Ia supernovae. In solar twin studies, \feh\ is fixed by design, making \alphafe\ the dominant tracer of chemical evolution and a natural benchmark for comparison with the KPM-derived parameters, \acc\ and \aia. Choosing the 3,814 stars with solar \teff, \logg, and \feh\ within their 1$\sigma$ uncertainties, or ``solar twins", we show that their \alphafe\ is not correlated with their \tcond\ slopes in the left-hand panel of Figure \ref{fig:alphacomp}; this differs from what we saw in Figure \ref{fig:fehalphafe} when considering the entire solar analog population. The Sun is an unremarkable member of the distribution in this space at 0.8$\sigma$ from the mean. Since there is no correlation between \alphafe\ and \tcond\ slope, this nominal offset does not carry physical significance. 

We calculate the KPM proxy for \alphafe\ by dividing the KPM process vectors, \acc\ and \aia, to get \kpmalpha. 
In this framework, \kpmalpha\ is analogous to \alphafe, while \acc\ + \aia\ captures the total enrichment and can be viewed as a physically motivated analog to \feh. Plotting \tcond\ slope as a function of \kpmalpha\ shown in the middle panel reveals a much stronger trend. The Sun is only $-0.5\sigma$ from the running mean and appears even more chemically ordinary than in the \alphafe\ plot on the left. We confirm that \kpmalpha\ is more strongly correlated to \tcond\ slope than \alphafe\ using a linear regression analysis. We find \kpmalpha\ is a much stronger predictor of solar twins' \tcond\ slopes than \alphafe, given their \rsq\ values of 22\% and 0.1\% respectively. We note that in the \cite{bedell18} sample specifically, however, \alphafe\ and \kpmalpha\ are comparable predictors of \tcond\ slopes. This likely reflects that in the high precision and uniformity of that sample, traditional parameters suffice. In contrast, \kpmalpha\ is more effective for survey-resolution data. Because \kpmalpha\ is inferred from the full suite of abundances, it is better able to reveal correlations in element abundance space when they exist, and to recover chemical trends that would otherwise be masked by measurement noise and abundance offsets.

In the right-hand plot of Figure \ref{fig:alphacomp}, the Sun lies within 0.5$\sigma$ of the \kpmalpha\ distribution. This placement is broadly consistent with its mild refractory depletion, which places it at the $\sim$20th percentile ($\sim$0.84$\sigma$) among solar twins \citep{bedell18}. Thus, the Sun’s refractory element depletion can be understood as a natural consequence of its nucleosynthetic enrichment history, as reflected by its \kpmalpha\ value. That is, the Sun’s relative refractory depletion does not necessarily require a special process for its explanation and instead appears consistent with being part of a distribution due to Galactic chemical evolution.

\subsection{Understanding the Origin of the \tcond\ Slope with KPM Parameters} \label{sec:fcc-tcond}

\begin{figure*}
    \centering
    \includegraphics[width=\textwidth]{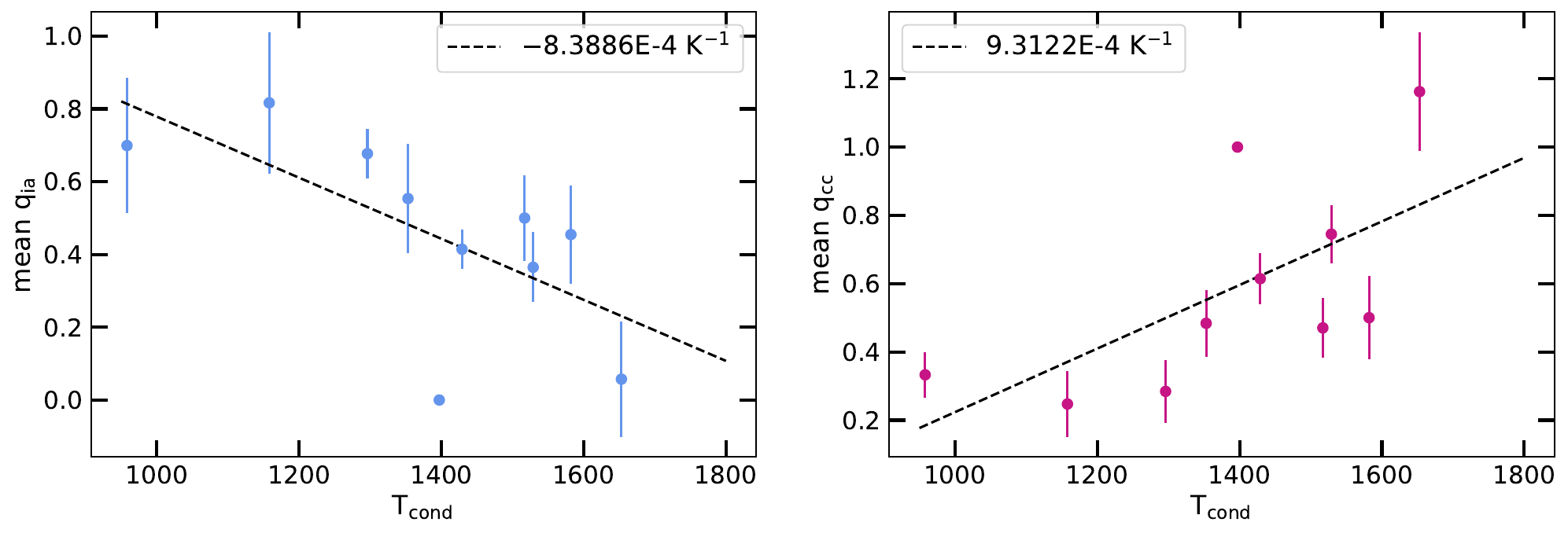}
    \includegraphics[width=\textwidth]{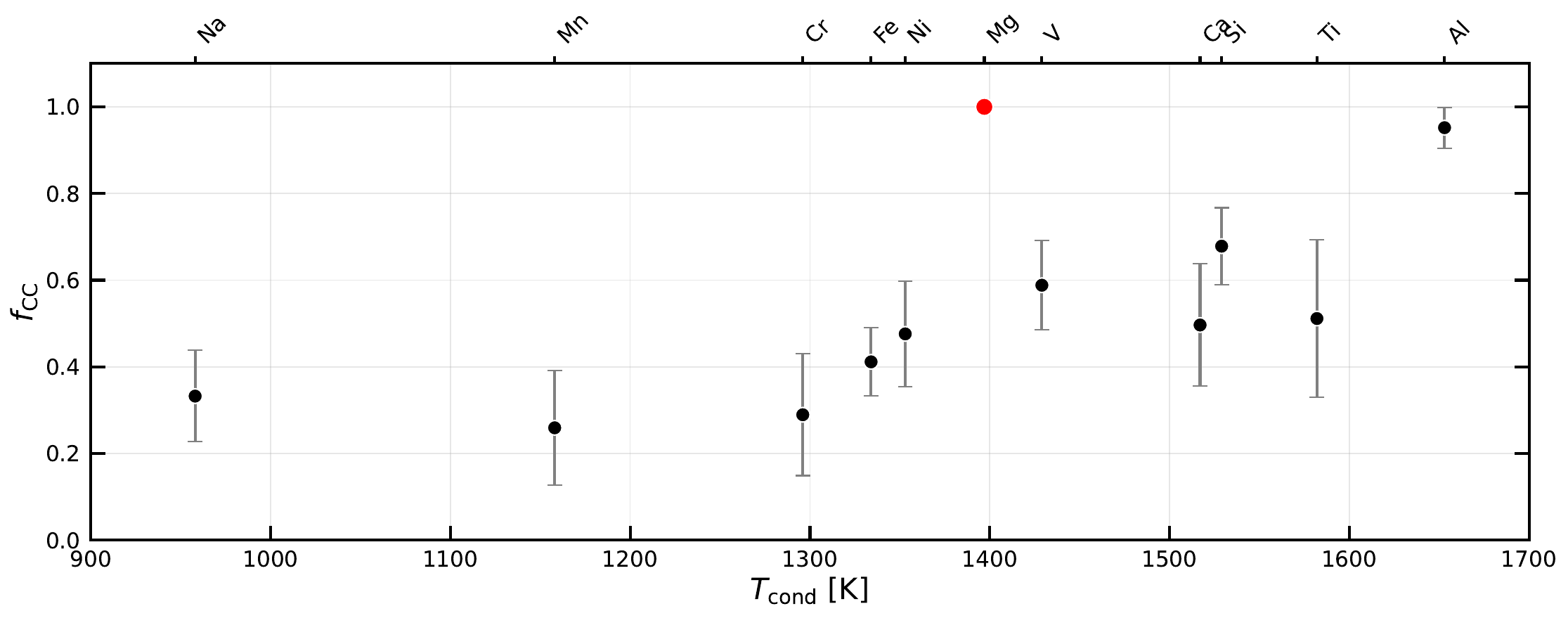}
    \caption{\textit{Top:} Mean $q_{\mathrm{ia}}$ (in blue, left) and $q_{\mathrm{CC}}$ (in red, right) or Type Ia and core-collapse ``yields" as a function of \tcond\ for each input refractory element, [X/H], in the KPM. \textit{Bottom:} \fcc\ computed from the $q$-values using Equation~\ref{eqn:fcc}, as a function of \tcond. We find a strong positive trend between \fcc\ and \tcond\, consistent with underlying nucleosynthetic yield patterns shown in the top panel. [Mg/H] appears as an outlier but is fixed at $f_{\mathrm{CC}} = 1$ in the KPM. {Data are not separated by survey here for visual clarity.} 
} 
    \label{fig:fcc}
\end{figure*}

To understand the link between chemical enrichment and \tcond\ trends, we calculate the average fraction of core-collapse supernovae enrichment (\fcc) for each refractory element, where \fcc\ (Equation~\ref{eqn:fcc}) is derived from the KPM amplitudes \acc\ and \aia\ and expresses the per-element contribution from core-collapse supernovae rather than per-star (like \acc\ and \aia). We provide the context for this trend by showing the KPM-inferred process vectors, $q_{\mathrm{CC}}$ and $q_{\mathrm{Ia}}$, or the yields from core-collapse and Type Ia supernovae, respectively, as a function of \tcond\ for each input refractory element, [X/H], in the KPM (top panel of Figure~\ref{fig:fcc}).

We plot the mean \fcc\ and standard deviation across the dataset as a function of \tcond\ for each refractory element in Figure \ref{fig:fcc}. We find a strong positive correlation between \fcc\ and \tcond. The more refractory elements tend to have higher \fcc\ values, as previously suggested by the abundance trends in Figure \ref{fig:ex_xmg}. 
[Mg/H] appears to be an outlier with an \fcc\ = 1 with no standard deviation, but this value reflects the simplifying assumption built into the KPM (see Section \ref{sec:kpm}).  The increase in \fcc\ with \tcond\ directly reflects the opposing trends in $q_{\mathrm{CC}}$ and $q_{\mathrm{Ia}}$: more refractory elements tend to have higher core-collapse yields and lower Type Ia yields.

Many of the refractory elements analyzed here are canonically classified as iron-peak elements. The modest increase in \fcc\ with \tcond\ likely reflects  the dominance of this group and drives the observed \tcond\ slope trends relative to other element families. Appendix~\ref{sec:fcc-explore} examines these trends as a function of different element families in more detail.

\begin{figure*}
    \centering
    \includegraphics[width=\linewidth]{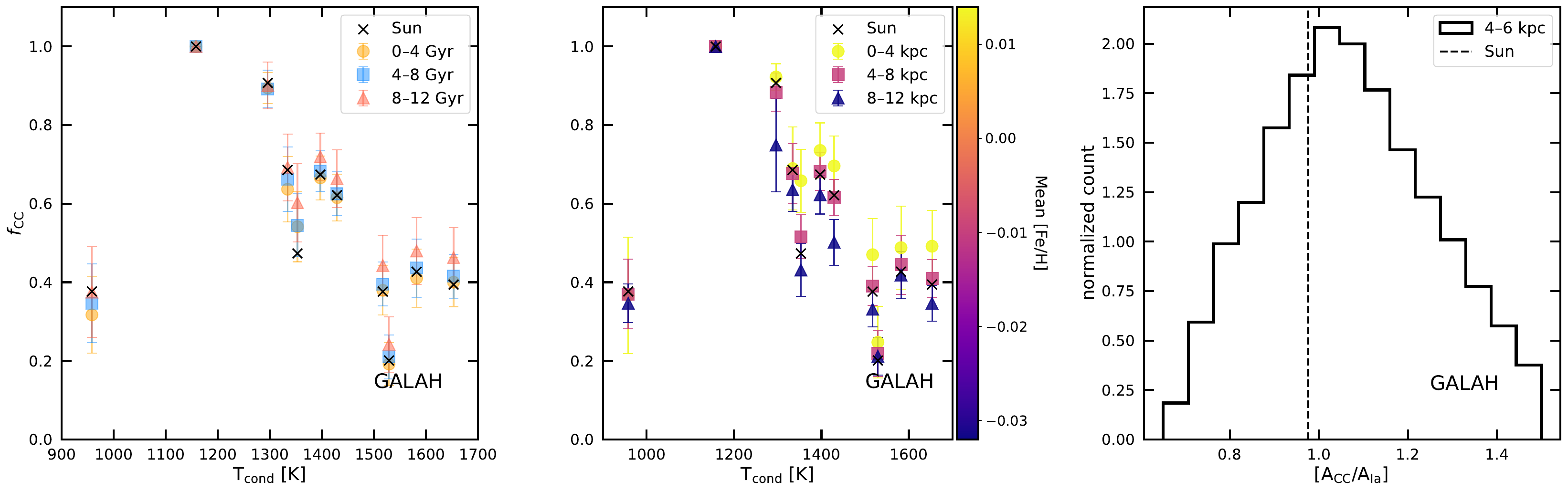}
    \caption{Testing age and Galactic \rbirth- related degeneracies with \fcc\ as a function of \tcond\ in the GALAH sample. The Sun is shown as inferred from GALAH-only abundances to ensure consistency with the rest of the sample. \textit{Left:} Breaking the sample into age bins (0--4, 4--8, and 8--12 Gyr), we find that the trend of \fcc\ with \tcond\ persists across age bins. Older stars show systematically higher \fcc, reflecting the delayed enrichment from Type Ia supernovae. \textit{Middle:} Breaking the sample into inner ($<4$ kpc), intermediate ($4$–$8$ kpc), and outer ($>8$ kpc) \rbirth\ bins and coloring the points by their mean \feh, we again find the trend with \tcond\ holds, but the absolute values of \fcc\ decrease with increasing \rbirth. This could reflect an increased relative contribution of Type Ia supernovae in the outer disk of the Galaxy and enhanced Type Ia supernovae production in low-metallicity environments \citep[e.g.,][]{Li11,Brown19,Wiseman21,Gandhi22,Johnson23}. \textit{Right:} Using the Sun’s estimated \rbirth\ of $\sim$5 kpc from \citet{Minchev18,Frankel19,Lu22}, and adopting a typical 1 kpc uncertainty in \rbirth\ estimates, we show the distribution of \kpmalpha\ for stars with \rbirth\ between 4 and 6 kpc. The Sun lies within $1\sigma$ of this distribution, and we note substantial spread in \kpmalpha\ even within narrow \rbirth\ bins. While \fcc\ varies with both age and \rbirth, the underlying trend with \tcond\ remains robust.
}
    \label{fig:test_age}
\end{figure*}

The observed relationship between \tcond\ and \fcc\ could reflect GCE trends driven by stellar age or birth location, rather than stellar yields. To test this, we examine how \fcc\ varies with reported isochrone ages and inferred Galactic birth radius (\rbirth) in the GALAH sample (Figure~\ref{fig:test_age}). We divide stars into age bins (young: $<$4 Gyr, intermediate: 4–8 Gyr, old: $>$8 Gyr; left panel) and by \rbirth\ (inner: $<$4 kpc, intermediate: 4–8 kpc, outer: $>$8 kpc; middle panel), adopting the \rbirth\ calibration from \cite{Wang2024}. While \fcc\ shows mild variation across these bins, the \fcc–\tcond\ trend persists. Specifically, \fcc\ increases with age due to the delayed onset of Type Ia supernovae enrichment \citep{Feuillet18,Feuillet19}, and decreases with \rbirth, potentially consistent with more efficient Type Ia production in low-metallicity outer disk environments driven by higher binary fractions \citep[e.g.,][]{Li11,Brown19,Badenes18,Moe19,Wiseman21,Gandhi22,Johnson23}.

Because the \fcc–\tcond\ trend holds across all age and \rbirth\ bins, we conclude that neither parameter alone drives the observed abundance patterns. In the right panel, we show that stars with \rbirth\ $\sim$5 kpc, similar to the Sun’s theorized birth radius \citep[e.g.,][]{Minchev18,Frankel19,Lu22}, span a wide range in \kpmalpha, and the Sun lies well within 1$\sigma$ of this distribution, again underscoring its chemical typicality. To assess whether the observed spread in \acc\ and \aia\ could be explained by measurement uncertainty, we generated 10 bootstrapped abundance realizations per star. The intrinsic scatter in the data consistently exceeded the bootstrap spread, confirming that these differences reflect real variation in nucleosynthetic enrichment.


\subsection{Observed \tcond\ Slopes Are Largely Explained by KPM's Treatment of Chemical Enrichment} \label{sec:kpm_predict}

\begin{figure*}
    \centering
    \includegraphics[width=0.9\linewidth]{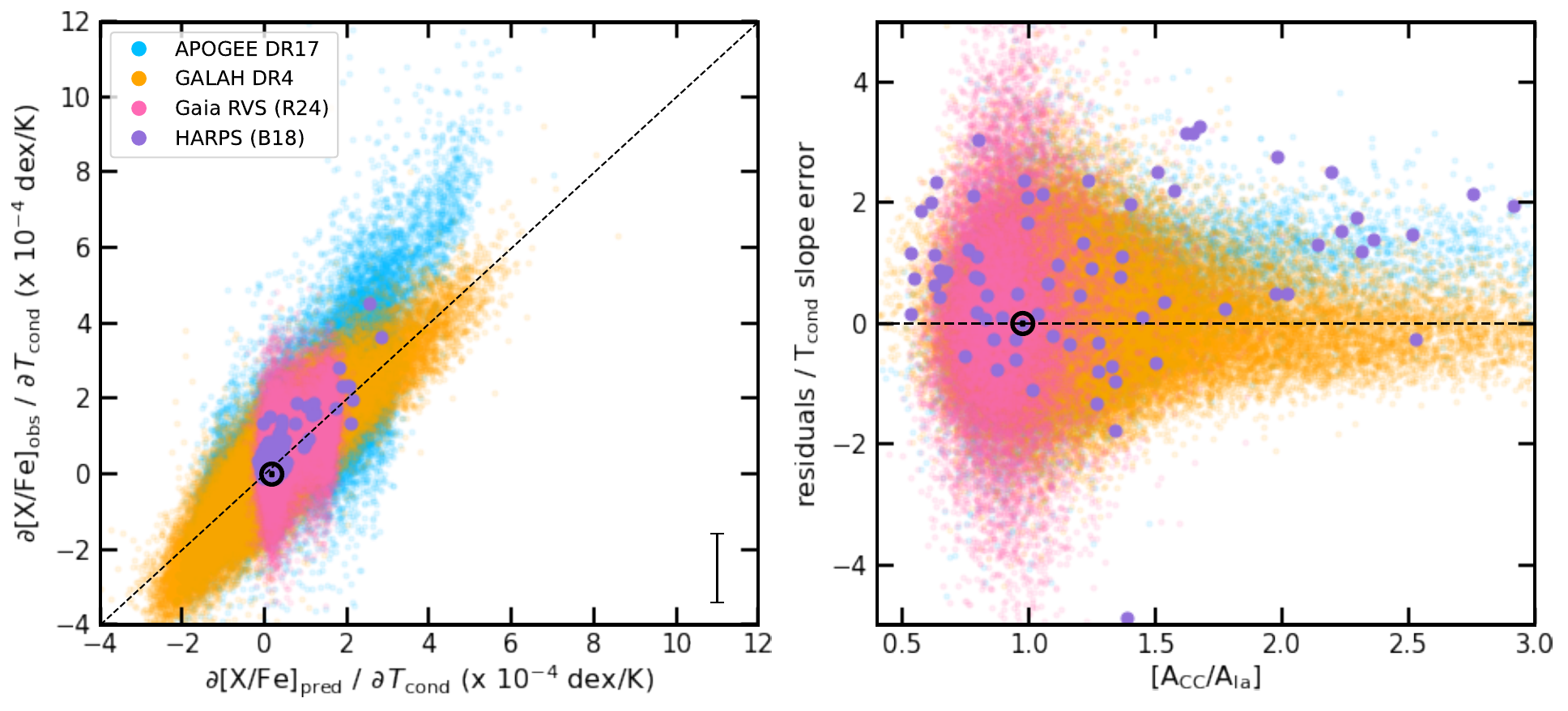}
    \caption{Comparison between KPM-predicted and observed \tcond\ slopes of \cite{bedell18} in purple, GALAH DR4 \citep{GALAHDR4} in orange, APOGEE DR17 \citep{apogeedr17} in blue, and Gaia RVS \citep{Rampalli24} in pink. \textit{Left:} Observed \tcond\ slopes versus predicted \tcond\ slopes, with a dashed 1:1 line in black showing reasonable agreement. Typical mean \tcond\ slope uncertainty is shown in the bottom right hand corner and shows that the scatter along the 1:1 line is largely due to this uncertainty. \textit{Right:} Residuals of \tcond\ slope (observed – predicted) normalized by their corresponding observational uncertainties as a function of \kpmalpha. The model-predicted slopes reproduce 83\% of observed slopes within 1$\sigma$ and 96\% within 2$\sigma$ uncertainties, indicating broad consistency across the sample and that chemical enrichment does indeed explain a majority of the observed \tcond\ slopes.
    }
    \label{fig:prediction}
\end{figure*}

Given the relationship between \fcc\ and \tcond, we test how much of the observed \tcond\ slopes can be accounted for by chemical enrichment alone. For each solar analog, we use the abundances predicted by the KPM and fit a linear trend with \tcond, following Equation~\ref{eqn:tcond} and the same procedure used to fit the observed \tcond\ slopes in Section~\ref{sec:tcond_slope}. At fixed total enrichment (solar \acc\ + \aia, analogous to solar \feh), the KPM predicts a broad spread of \tcond\ slopes as a function of \kpmalpha. This broad range is due to genuine nucleosynthetic diversity rather than model imprecision as discussed at the end of the previous section. 93\% of stars lie within $1\sigma$ of the model's prediction, indicating the observed scatter (including survey-dependent offsets) is consistent with the KPM expectation.

Figure~\ref{fig:prediction} compares the predicted and observed \tcond\ slopes for each survey. The left panel shows excellent agreement, with an average offset of $1 \times 10^{-4}$ dex/K, nearly identical to the typical measurement uncertainty. The apparent offset for APOGEE reflects its abundance systematics (discussed in Section \ref{sec:kpm}), but its stars still fall within 1$\sigma$ given their observational uncertainties. The right panel shows residuals (observed minus KPM-predicted \tcond\ slope), normalized by the observational uncertainties and plotted as a function of \kpmalpha. Predicted slopes match 83\% of observed values within 1$\sigma$ and 96\% within 2$\sigma$, suggesting most scatter is due to measurement uncertainty. The agreement is strongest in APOGEE and GALAH, with $\sim$75\% within 1$\sigma$ and $>$99\% within 3$\sigma$; Gaia RVS and \cite{bedell18} show lower 1$\sigma$ agreement ($\sim$50\%) but still $>$90\% within 3$\sigma$.

{Nucleosynthetic enrichment appears sufficient to explain the observed \tcond\ slopes.} Color-coding by survey shows that the residual scatter is present across all datasets. While the distributions are broadly consistent with Gaussian scatter, at low \kpmalpha, we note a mild broadening  and a small excess of large deviations. {These outliers could reflect additional variance from unmodeled enrichment channels or star-specific processes such as those related to planet formation.} This residual structure motivates the test presented in Appendix~\ref{sec:dimensionality}, where we assess the dimensionality of the abundance data. We use a 6-label regression model of stellar parameters and 4 representative abundances from different enrichment sources, to predict other elemental abundances and the \tcond\ slope for GALAH solar analogs. The dimensionality test confirms that enrichment parameters like \acc\ and \aia\ capture a majority (but not all) of the chemical diversity seen in stars via the \tcond\ slope metric.

\section{Discussion}\label{sec:discussion}

\subsection{The Sun is Statistically Ordinary Among its Peers: Comparison to Literature Results}

Previous analyses have suggested that the Sun's mild refractory depletion (\tcond\ slope $\leq 0$, where \tcond(Sun) is 0 by definition) is uncommon and found in only $\sim$20\% of other Sun-like stars \citep[e.g.,][]{Melendez09,Ramirez09,bedell18,Rampalli24}. However, \citet{Nibauer21} modeled the \tcond\ slope distribution as a mixture of refractory-depleted and refractory-enriched populations. Their analysis placed the Sun near the center of the refractory-depleted component, which in fact comprises the majority of Sun-like stars. Thus, considering the underlying population structure rather than the \tcond\ slope relative to zero, reframes the Sun as chemically typical. In our analysis, 5–21\% of Sun-like stars across the four surveys have \tcond\ slopes $\leq 0$, but 74–100\% are classified as refractory-depleted using a similar Gaussian mixture model as \citet{Nibauer21}. These results are consistent with both the view that the Sun’s \tcond\ slope is in the minority relative to zero, \textit{and} with the \citet{Nibauer21} interpretation that it is chemically typical when considering the underlying population structure. 

The KPM framework further explains this distribution, showing that abundance pattern differences are largely driven by variation in core-collapse and Type Ia supernovae enrichment. Therefore, the Sun’s chemistry is not anomalous, but consistent with expectations from its nucleosynthetic history. {We emphasize that this conclusion does not rely on broadening the comparison distribution by using loosely defined solar analogs. Figure~\ref{fig:alphacomp} is restricted to the high-resolution HARPS solar-twin sample from \citet{bedell18} and to stars consistent with solar parameters within reported 1$\sigma$ uncertainties. In this higher-precision regime, the Sun does lie slightly below the mean relation by eye, but the offset remains well within the intrinsic population scatter about the KPM-predicted trend. {Throughout this work, we use ``statistically ordinary'' to mean that, given its nucleosynthetic enrichment history as quantified by the KPM, the Sun's refractory--\tcond\ slope is consistent with expectation and does not require an additional, uniquely solar depletion mechanism beyond the dominant enrichment-driven contribution. While enrichment-driven trends dominate the observed \tcond\ behavior, we do note our sample is not optimized to isolate subtle abundance signatures at the precision required to distinguish additional contributions from stellar evolution or planet formation on a star-by-star basis.}

\subsection{Limitations of Traditional Galactic Chemical Evolution \& Enrichment Descriptors} \label{sec:limitations}

Previous studies of solar twins have attempted to correct for the effects of GCE by using trends in traditional stellar parameters to isolate intrinsic stellar abundance signatures \citep[e.g.,][]{Spina18,bedell18,Cowley22,Sun24}. These approaches apply first-order corrections for chemical evolution by using average age–abundance trends (despite often imprecise stellar ages, \citealt{soderblom}) and coarse tracers like \feh\ and \alphafe\ to summarize enrichment history. Even when we rerun the KPM on the GCE-corrected abundances from \citet{bedell18}, the correlation between \kpmalpha\ and \tcond\ slope persists, though flatter, {suggesting that proxy-based corrections do not fully parameterize enrichment history at the precision needed to describe \tcond\ slopes.} While such metrics and techniques are widely available and often sufficient for describing major trends in the Galaxy's history \citep[e.g.,][]{Edvardsson93,Bensby2003,Haywood13,Ness19,Manea23,Wang2024}, they rely on simplifying assumptions. In particular, \alphafe\ is an intuitive but coarse approximation of nucleosynthesis that treats elements as resulting from singular nucleosynthetic channels (e.g., Ca, Ti, Si, and Mg are \textit{solely} made from core-collapse supernovae). However, elements’ nucleosynthetic origins are, in reality, heterogeneous \citep{Burbidge57,Cameron57,Johnson19,Kobayashi20}.

Furthermore, corrections based on smooth GCE trends assume a chemically well-mixed ISM and neglect small-scale, local variation in chemical enrichment. Broad trends in star formation history and Galactic structure do shape average chemical abundances \citep[e.g.,][]{Spina18,Adibekyan16,Baba23}, but the persistent \fcc--\tcond\ trend across stellar age and the wide range of \kpmalpha\ at fixed \rbirth\ in Figure \ref{fig:test_age} suggest the presence of additional processes shaping \tcond\ slopes. Given the trends with \tcond\ are at the $< 0.1$ dex level, a likely candidate for this additional process is inhomogeneity present in the ISM at the time of star formation. The ISM only appears chemically homogeneous when averaged over spatial scales of ten to hundreds of parsecs
. As a result, even stars that are co-natal are shown to have chemical differences at the level of a few percent due to incomplete mixing during the star formation process \citep[e.g.,][]{BlandHawthorn10,Armillotta18}.

Thus, subtle abundance features, such as the Sun’s mild refractory depletion, are probably not purely the result of smooth, large-scale GCE. Interpreting the Sun's relative refractory depletion therefore necessitates a more detailed treatment of global \textit{and} local chemical evolution, as recent works have reiterated \citep[e.g.,][]{Magrini17,Blancato2019,Horta22,Ratcliffe23,Griffith24}. Our results demonstrate that the KPM successfully captures this complexity: by modeling the full suite of elemental abundances as a mixture of core-collapse and Type Ia supernovae contributions, it enables a more interpretable and chemically meaningful context for understanding the Sun. 

\subsection{KPM Recontextualizes the Sun's Abundance Distribution} \label{sec:kpm_sun}

Traditional stellar parameters like \feh\ and \alphafe\ summarize broad features of GCE in stars but lack the resolution to explain finer abundance patterns. In contrast, the KPM from \cite{Griffith24} models the full elemental abundance vector of each star and captures each element's mixed origin from core-collapse and Type Ia supernovae enrichment as \acc\ and \aia. Thus, \kpmalpha\ is likely a better predictor of \tcond\ slopes than \alphafe\ because it is a more physically motivated, interpretable metric that represent stars' detailed nucleosynthetic histories. While theoretical supernovae yield tables also attempt to model the Galaxy's evolution by invoking stellar physics, they are often in disagreement with each other and struggle to match observed abundances \citep[e.g.,][]{Rybizki17,Griffith21,Hegedus25}. The KPM empirically constrains enrichment histories from data, offering a more consistent and observationally grounded framework.

Our analysis shows the Sun was never truly an outlier in its mild depletion of refractory elements compared to stars matched in \feh, \alphafe, and age, when reexamined as a function of KPM metrics that more accurately model chemical enrichment. The Sun's position in \kpmalpha–\tcond\ slope space in Figure \ref{fig:alphacomp} is consistent with the bulk of the solar analog population. Therefore, its relative refractory depletion signature can be understood as a natural consequence of both large-scale GCE and smaller-scale variations in ISM enrichment history. {In this framework, enrichment-driven Galactic chemical evolution accounts for the dominant contribution to the observed \tcond\ slope variation, while not ruling out smaller additional processes (e.g., planet formation or localized effects such as binary mass transfer) that may contribute at a sub-dominant level.} While the KPM framework enables us to capture global and local enrichment signals, purely stochastic, uncorrelated local variations in element abundances will instead show up as residuals (not captured by the model, see Figure \ref{fig:prediction} and Table \ref{tab:resid_errors_condT}). Only if those ``stochastic” processes leave correlated signatures across abundances, the KPM can and will capture them.

The chemical differences between solar analogs are almost entirely captured by the two-process KPM and described by the variation in core-collapse and Type Ia supernovae enrichment: $> 83\%$ of the observed abundance diversity can be explained by differences in nucleosynthetic pathways alone as shown in Figure \ref{fig:prediction}. For the Sun, the average residuals are $-0.003\pm0.009$ dex {across all 4 survey KPM fits as shown in Table \ref{t:suns} and invoking Equation \ref{eqn:xh}}. Here, we do not suggest that each star has a distinct abundance pattern. Rather, across the solar analog population, variations in local ISM conditions drive subtle differences between stars, which correlate with \tcond\ given its relation to the relative yields of core-collapse and Type Ia supernovae (Figure \ref{fig:fcc}). These subtle variations are captured in the data-driven KPM but are missed by simpler proxies such as age, \feh, or \alphafe\ that are more useful for modeling global GCE trends. 
{We note that \tcond\ does not enter the KPM in any way. The model is fit solely to elemental abundances and decomposes them into core-collapse and Type Ia enrichment components without reference to refractory or volatile classification. There is therefore no a priori reason for the inferred KPM parameters to correlate with \tcond\ gradients. Importantly, systematic \tcond\ variations are visible directly in the data prior to any KPM modeling. In Figure~\ref{fig:ex_xmg} presents purely observational [X/Mg]–[Mg/Fe] sequences, with no KPM analysis applied, and shows clear, monotonic changes with \tcond. The presence of these structured \tcond\ trends in the raw abundance data demonstrates that the correlations are not imposed by the KPM, but instead reflect underlying nucleosynthetic yield physics.} Using this more detailed, physically-motivated, and data-driven framework, we find that the Sun’s abundance trend with \tcond\ is not chemically exceptional but lies within a statistically expected distribution for solar analogs.

{We therefore interpret our results as demonstrating that Galactic chemical evolution accounts for the dominant contribution to \tcond\ slope variation among solar analogs and is sufficient to explain the Sun’s abundance pattern within the observed scatter. Additional processes may contribute at a lower level but are not required by the data and are not resolvable with the current heterogeneous survey sample. Survey-level systematics may also broaden or distort the detailed \tcond\ slope distributions, reinforcing that our conclusions are limited to robust, population-level trends that are consistent across independently analyzed datasets.}

\subsection{Revisiting Planet-Driven Abundance Scenarios}\label{sec:planets}

\paragraph{Abundances of planet hosts} Given that our catalog includes 144 unique confirmed planet hosts, we examine whether these stars show distinct patterns in \acc\ and \aia. We find that the planet hosts do not exhibit distributions in these parameters that differ significantly {from the broader solar analog population within the precision of the current dataset}, which is consistent with the findings from \cite{Behmard23} and \citetalias{Rampalli24}. Many of the stars in our dataset that are not confirmed planet hosts may still host undetected planets \citep{Mulders18,Zink19}. The unknown planet status for a majority of stars makes it difficult to identify a clear chemical signature of planet formation in population-level studies. 

\paragraph{Engulfment enrichment patterns} Using a sample of wide binaries, \citet{Liu24} showed that planet engulfment can leave a distinctive element-by-element differential abundance profile when plotted as a function of \tcond, and observed this signature in one in 12 stars. We search for this signature in our solar analog\footnote{Our dataset, by definition, includes stars with nearly identical \teff, \logg, and \feh, such that their abundance patterns occupy the same correlated chemical space, so most elements can be predicted from a small set of others (Appendix~\ref{sec:dimensionality}; \citealt{Rampalli21,Ness22,Manea23,Mead25}). Even if these stars are not co-natal, differences in their abundance patterns remain physically meaningful to compare.} dataset, which contains $109,500 \times 109,499$ ordered star pairs, each consisting of the 12 reported elemental abundances. We count how often the abundance differences match the \citet{Liu24} engulfment signature within $\pm$0.05 dex across all overlapping elements. For comparison, we generate random abundance offset patterns and count how often abundance differences match these random patterns. Engulfment-like matches occur as frequently as random ones ($\sim$10$^{-5}$\%), suggesting that engulfment-like patterns are indistinguishable from the chemical diversity expected from Galactic enrichment alone {at the precision of the current sample}.

\paragraph{Refractory depletion in new solar analog samples} We find similar results looking at the recently published \citet{Martos25} sample, which includes 99 high signal-to-noise HARPS spectra of solar twins and analogs. \citeauthor{Martos25} argue that the Sun is unusually refractory-depleted compared to non-planet-hosts, with a significance of 9.5$\sigma$, and suggest this difference reflects planet formation. Applying the KPM to their sample, restricted to the elements in common with our analysis, we find that 64 out of 99 stars, including 9 out of 10 confirmed planet hosts, have \tcond\ slopes that agree with KPM-based predictions within 1$\sigma$, and 80 out of 99 agree within 2$\sigma$. The abundance patterns of these stars, including those with planets, are fully consistent with variations in nucleosynthetic enrichment alone {within the uncertainties of the present analysis}. 

{These results indicate that while planet-related chemical signatures may exist, they are difficult to isolate in population-level samples where enrichment-driven abundance diversity dominates.}  



\section{Conclusions}\label{sec:conclusion}

In this work, we explore how the local and global chemical evolution from the ISM and the Milky Way contributes to the Sun's refractory element depletion relative to other Sun-like stars. This is typically quantified by the slope of refractory element abundances (e.g., Na, Mn, Cr, Si, Fe, Ni, Mg, V, Ca, Ti, Al used in this work) as a function of their condensation temperatures, which we refer to here as the \tcond\ slope. Previous works in this space have corrected for GCE by applying  age-abundance trends to measured elemental abundances \citep[e.g.,][]{Nissen15,Nissen16,Spina18,bedell18}, which yield results that depend sensitively on techniques used \citep{bedell18,Cowley22,Sun24,Carlos25,Martos25}. We compile 109,500 unique solar analogs (144 of which are known planet hosts) and their abundances from four samples: Gaia RVS \citep{Rampalli24}, GALAH DR4 \citep{GALAHDR4}, APOGEE DR17 \citep{apogeedr17}, and the solar twin sample from \cite{bedell18}. We start with traditional stellar parameters (stellar \teff, \logg, \feh, and \alphafe) and then turn to more physically-motivated, and granular descriptors of chemical evolution. We use the data-driven KPM \citep{Griffith24} to explore how GCE and local ISM hetereogeneities impact \tcond\ slopes, and to assess whether the Sun is in fact chemically unique.

\begin{itemize}
    \item \textit{The Sun's chemistry connects to broader GCE trends:} \tcond\ slopes are correlated with \feh\ and \alphafe\ (Figure \ref{fig:fehalphafe}), suggesting a connection between GCE and abundance patterns. Solar analogs with \tcond\ slopes consistent with the Sun's are on average slightly more metal-rich and $\alpha$-depleted.  \teff\ and \logg\ also impact \tcond\ slopes but to a lesser degree (Figure \ref{fig:tefflogg}). Using a series of linear regressions to assess their predictive power for \tcond\ slopes, we compute \rsq\ values of 9\%, 13\%, 15\%, and 23\% for \teff, \logg, \feh\ and \alphafe\ respectively (larger values indicate stronger predictive power).
    
    \item \textit{Nucleosynthetic pathways correlate with refractory elements' \tcond:} Stars' \tcond\ slopes are shaped most strongly by \alphafe, which traces chemical enrichment from core-collapse supernovae versus Type Ia supernovae. We demonstrate this by looking [X/Mg] versus [Mg/Fe] (using Mg as a representative $\alpha$-process element, originating from core-collapse supernovae). We find that correlations in abundance ratios weaken with increasing \tcond, which indicates a weakening contribution from other enrichment sources (Figure \ref{fig:ex_xmg}). 
    
    \item \textit{The Sun is chemically ordinary among stars with similar chemical enrichment:} We apply the K-process model (KPM), a data-driven framework from \cite{Griffith24}, to model the enrichment contributions from Type Ia and core collapse supernovae (\aia\ and \acc). We do this for each solar analog and examine trends with \tcond\ slopes. {In this framework, the Sun appears chemically ordinary, falling within the intrinsic population scatter as shown in Figure \ref{fig:ahists}. Enrichment-driven Galactic chemical evolution accounts for the dominant contribution to the observed \tcond\ slope variation, while not excluding smaller additional processes (e.g., planet formation or localized effects). These additional contributions are not required by the data and are not resolvable at the precision of the present heterogeneous survey sample.}
    
    \item \textit{Data-driven chemical enrichment parameters improve \tcond\ slope predictions:} We then limit our dataset to the solar twins, which we define as any star with the same \teff, \logg, and \feh\ as the Sun within their 1$\sigma$ uncertainties. In Figure \ref{fig:alphacomp}, we show that \alphafe\ is no longer correlated with these stars' \tcond\ slopes, but the KPM's \alphafe\ proxy, \kpmalpha\ remains strongly correlated. This difference is due to how these two parameters are defined: \alphafe\ is measured by designating a few elements as entirely originating from core-collapse supernovae. \kpmalpha\ is derived by modeling the full set of elemental abundances to more accurately reflect each element’s mixed origin from both Type Ia and core-collapse supernovae. 

    \item \textit{Per-element nucleosynthetic enrichment fractions explain the \tcond\ slope's origin:} We use the KPM to calculate \fcc, the fraction of each element's abundance in a given star that orginated from core-collapse supernovae. Figure \ref{fig:fcc} shows that \fcc\ increases with \tcond, mirroring trends in the observed [X/Mg]–[Mg/Fe] correlations (Figure \ref{fig:ex_xmg}). In Appendix \ref{sec:fcc-explore}, we show that this trend appears to be primarily driven by iron-peak elements, which make up most of the refractory elements we study and exhibit a modest increase in \fcc\ with \tcond. The positive \fcc–\tcond\ correlation persists across stellar ages and Galactic birth radii, which only shift the relation vertically (Figure \ref{fig:test_age}). Stars born at similar \rbirth\ to the Sun span a wide range in \kpmalpha, likely reflecting chemical inhomogeneities in the ISM correlated in abundance space not captured by age, \feh, or \alphafe\ alone, but which are resolved by the KPM.
    
   \item \textit{Predicted \tcond\ slopes from KPM match observed values:} Using KPM-inferred abundances to predict \tcond\ slopes (Section \ref{sec:kpm_predict}), we recover the observed values with residual scatter that is broadly consistent with measurement uncertainties: 83\% of stars fall within $1\sigma$ and 96\% within $2\sigma$ (Figure~\ref{fig:prediction}). Most of the residuals are consistent with Gaussian noise, but we note a small excess of large deviations in Figure \ref{fig:prediction} that may reflect additional enrichment pathways or in-situ processes potentially related to planet formation. The independent, data-driven analysis in Appendix~\ref{sec:dimensionality} supports this view, showing that parameters like \acc\ and \aia\ capture most, but not all, of the chemical diversity traced by the \tcond\ slope. 

    \item {\textit{Planet-related signatures are difficult to isolate in the presence of enrichment-driven abundance diversity:}} In Section \ref{sec:planets}, we discuss that there are no significant differences in \acc\ or \aia\ between confirmed planet hosts and the broader solar analog population. When comparing chemically similar but unassociated solar analogs, we find engulfment-like abundance patterns are exceedingly rare and no more frequent than random patterns. This is consistent with our finding that even stars with known planets show abundance patterns that can be explained by enrichment history alone. Identifying planet-related chemical signatures at the population scale is extremely challenging as they are easily masked by the broader diversity of stellar abundances due to chemical enrichment at the local and Galactic scale.  
    
\end{itemize}

Our work demonstrates that the Sun’s relative refractory depletion can be understood as a natural outcome of GCE shaped by both large-scale trends and local inhomogeneities in the ISM. We apply physically motivated, data-driven learning techniques, typically used to study stellar populations in the Galactic context, to reframe trends in \tcond\ slope. By connecting planetary and Galactic scales via stellar abundances, we place the detailed chemical patterns of individual stars in a broader population-level context. {We show that the Sun’s abundance pattern is unremarkable, arising naturally from its chemical enrichment history rather than requiring a uniquely planetary origin. Our conclusions are therefore limited to robust, population-level trends that remain consistent across independently analyzed surveys despite known survey-level systematics.}

Our results also empirically reveal a connection between supernova yields and elements' \tcond. The physical origin of this relationship remains unclear and beyond the scope of this work but may arise from the overlap between atomic and nuclear properties. We simply report the empirical trend and highlight its relevance to exoplanet research: understanding subtle abundance signatures from processes like planet formation requires accounting for incomplete mixing in the ISM and GCE. Our framework provides a more complete picture of stellar chemical diversity and emphasizes the importance of multi-element modeling to disentangle global and local chemical histories of stars before turning to planet-related explanations.

\acknowledgements
R.R. is supported by the NSF Graduate Research Fellowship (DGE-2236868). E.J.G. is supported by an NSF Astronomy and Astrophysics Postdoctoral Fellowship under award AST-2202135.

This work has made use of data from the European Space Agency (ESA) mission Gaia (https://www.cosmos.esa.int/gaia), processed by the Gaia Data Processing and Analysis Consortium (DPAC, https://www.cosmos.esa.int/web /gaia/dpac/consortium). Funding for the DPAC has been provided by national institutions, in particular the institutions participating in the Gaia Multilateral Agreement.This work made use of the Third Data Release of the GALAH Survey (Buder et al. 2021). 

The GALAH Survey is based on data acquired through the Australian Astronomical Observatory under programs: A/2013B/13 (The GALAH pilot survey); A/2014A/25, A/2015A/19, A2017A/18 (The GALAH survey phase 1); A2018A/18 (Open clusters with HERMES); A2019A/1 (Hierarchical star formation in Ori OB1); A2019A/15 (The GALAH survey phase 2); A/2015B/19, A/2016A/22, A/2016B/10, A/2017B/16, A/2018B/15 (The HERMES-TESS program); and A/2015A/3, A/2015B/1, A/2015B/19, A/2016A/22, A/2016B/12, A/2017A/14 (The HERMES K2-follow-up program). 
We acknowledge the traditional owners of the land on which the AAT stands, the Gamilaraay people, and pay our respects to elders past and present. This paper includes data that has been provided by AAO Data Central (datacentral.org.au).

Funding for the Sloan Digital Sky Survey V has been provided by the Alfred P. Sloan Foundation, the Heising-Simons Foundation, the National Science Foundation, and the Participating Institutions. SDSS acknowledges support and resources from the Center for High-Performance Computing at the University of Utah. The SDSS web site is \url{www.sdss.org}. SDSS is managed by the Astrophysical Research Consortium for the Participating Institutions of the SDSS Collaboration, including the Carnegie Institution for Science, Chilean National Time Allocation Committee (CNTAC) ratified researchers, the Gotham Participation Group, Harvard University, Heidelberg University, The Johns Hopkins University, L’Ecole polytechnique federale de Lausanne (EPFL), Leibniz-Institut fur Astrophysik Potsdam (AIP), Max-Planck-Institut fur Astronomie (MPIA Heidelberg), Max-Planck-Institut fur Extraterrestrische Physik (MPE), Nanjing University, National Astronomical Observatories of China (NAOC), New Mexico State University, The Ohio State University, Pennsylvania State University, Smithsonian Astrophysical Observatory, Space Telescope Science Institute (STScI), the Stellar Astrophysics Participation Group, Universidad Nacional Autonoma de Mexico, University of Arizona, University of Colorado Boulder, University of Illinois at Urbana-Champaign, University of Toronto, University of Utah, University of Virginia, Yale University, and Yunnan University.


\facilities{Gaia \citep{GaiaMission}, GALAH \citep{galah_mission}, APOGEE \citep{apogee}}
\software{astropy \citep{2013A&A...558A..33A,astropyii},  astroquery \citep{astroquery}, numpy \citep{numpy}, pandas \citep{reback2020pandas,mckinney-proc-scipy-2010}, topcat \citep{topcat}, matplotlib \citep{Hunter:2007}, sklearn \citep{sklearn}, scipy \citep{scipy}}, KPM \citep{zenodo_kpm}




\begin{thebibliography}{}
\expandafter\ifx\csname natexlab\endcsname\relax\def\natexlab#1{#1}\fi
\providecommand{\url}[1]{\href{#1}{#1}}
\providecommand{\dodoi}[1]{doi:~\href{http://doi.org/#1}{\nolinkurl{#1}}}
\providecommand{\doeprint}[1]{\href{http://ascl.net/#1}{\nolinkurl{http://ascl.net/#1}}}
\providecommand{\doarXiv}[1]{\href{https://arxiv.org/abs/#1}{\nolinkurl{https://arxiv.org/abs/#1}}}

\bibitem[{{13emilygriffith} \& {Hogg}(2023)}]{zenodo_kpm}
{13emilygriffith}, \& {Hogg}, D.~W. 2023, {13emilygriffith/KProcessModel: KPM: The K Process Model}, v1.0.0,  Zenodo, \dodoi{10.5281/zenodo.10411910}

\bibitem[{{Abdurro'uf} {et~al.}(2022){Abdurro'uf}, {Accetta}, {Aerts}, {Silva Aguirre}, {Ahumada}, {Ajgaonkar}, {Filiz Ak}, {Alam}, {Allende Prieto}, {Almeida}, {Anders}, {Anderson}, {Andrews}, {Anguiano}, {Aquino-Ort{\'\i}z}, {Arag{\'o}n-Salamanca}, {Argudo-Fern{\'a}ndez}, {Ata}, {Aubert}, {Avila-Reese}, {Badenes}, {Barb{\'a}}, {Barger}, {Barrera-Ballesteros}, {Beaton}, {Beers}, {Belfiore}, {Bender}, {Bernardi}, {Bershady}, {Beutler}, {Bidin}, {Bird}, {Bizyaev}, {Blanc}, {Blanton}, {Boardman}, {Bolton}, {Boquien}, {Borissova}, {Bovy}, {Brandt}, {Brown}, {Brownstein}, {Brusa}, {Buchner}, {Bundy}, {Burchett}, {Bureau}, {Burgasser}, {Cabang}, {Campbell}, {Cappellari}, {Carlberg}, {Wanderley}, {Carrera}, {Cash}, {Chen}, {Chen}, {Cherinka}, {Chiappini}, {Choi}, {Chojnowski}, {Chung}, {Clerc}, {Cohen}, {Comerford}, {Comparat}, {da Costa}, {Covey}, {Crane}, {Cruz-Gonzalez}, {Culhane}, {Cunha}, {Dai}, {Damke}, {Darling}, {Davidson}, {Davies}, {Dawson}, {De Lee}, {Diamond-Stanic}, {Cano-D{\'\i}az}, {S{\'a}nchez},
  {Donor}, {Duckworth}, {Dwelly}, {Eisenstein}, {Elsworth}, {Emsellem}, {Eracleous}, {Escoffier}, {Fan}, {Farr}, {Feng}, {Fern{\'a}ndez-Trincado}, {Feuillet}, {Filipp}, {Fillingham}, {Frinchaboy}, {Fromenteau}, {Galbany}, {Garc{\'\i}a}, {Garc{\'\i}a-Hern{\'a}ndez}, {Ge}, {Geisler}, {Gelfand}, {G{\'e}ron}, {Gibson}, {Goddy}, {Godoy-Rivera}, {Grabowski}, {Green}, {Greener}, {Grier}, {Griffith}, {Guo}, {Guy}, {Hadjara}, {Harding}, {Hasselquist}, {Hayes}, {Hearty}, {Hern{\'a}ndez}, {Hill}, {Hogg}, {Holtzman}, {Horta}, {Hsieh}, {Hsu}, {Hsu}, {Huber}, {Huertas-Company}, {Hutchinson}, {Hwang}, {Ibarra-Medel}, {Chitham}, {Ilha}, {Imig}, {Jaekle}, {Jayasinghe}, {Ji}, {Johnson}, {Jones}, {J{\"o}nsson}, {Katkov}, {Khalatyan}, {Kinemuchi}, {Kisku}, {Knapen}, {Kneib}, {Kollmeier}, {Kong}, {Kounkel}, {Kreckel}, {Krishnarao}, {Lacerna}, {Lane}, {Langgin}, {Lavender}, {Law}, {Lazarz}, {Leung}, {Leung}, {Lewis}, {Li}, {Li}, {Lian}, {Liang}, {Lin}, {Lin}, {Lin}, {Lintott}, {Long}, {Longa-Pe{\~n}a}, {L{\'o}pez-Cob{\'a}}, {Lu},
  {Lundgren}, {Luo}, {Mackereth}, {de la Macorra}, {Mahadevan}, {Majewski}, {Manchado}, {Mandeville}, {Maraston}, {Margalef-Bentabol}, {Masseron}, {Masters}, {Mathur}, {McDermid}, {Mckay}, {Merloni}, {Merrifield}, {Meszaros}, {Miglio}, {Di Mille}, {Minniti}, {Minsley}, {Monachesi}, {Moon}, {Mosser}, {Mulchaey}, {Muna}, {Mu{\~n}oz}, {Myers}, {Myers}, {Nadathur}, {Nair}, {Nandra}, {Neumann}, {Newman}, {Nidever}, {Nikakhtar}, {Nitschelm}, {O'Connell}, {Garma-Oehmichen}, {Luan Souza de Oliveira}, {Olney}, {Oravetz}, {Ortigoza-Urdaneta}, {Osorio}, {Otter}, {Pace}, {Padilla}, {Pan}, {Pan}, {Parikh}, {Parker}, {Peirani}, {Pe{\~n}a Ram{\'\i}rez}, {Penny}, {Percival}, {Perez-Fournon}, {Pinsonneault}, {Poidevin}, {Poovelil}, {Price-Whelan}, {B{\'a}rbara de Andrade Queiroz}, {Raddick}, {Ray}, {Rembold}, {Riddle}, {Riffel}, {Riffel}, {Rix}, {Robin}, {Rodr{\'\i}guez-Puebla}, {Roman-Lopes}, {Rom{\'a}n-Z{\'u}{\~n}iga}, {Rose}, {Ross}, {Rossi}, {Rubin}, {Salvato}, {S{\'a}nchez}, {S{\'a}nchez-Gallego}, {Sanderson}, {Santana
  Rojas}, {Sarceno}, {Sarmiento}, {Sayres}, {Sazonova}, {Schaefer}, {Schiavon}, {Schlegel}, {Schneider}, {Schultheis}, {Schwope}, {Serenelli}, {Serna}, {Shao}, {Shapiro}, {Sharma}, {Shen}, {Shetrone}, {Shu}, {Simon}, {Skrutskie}, {Smethurst}, {Smith}, {Sobeck}, {Spoo}, {Sprague}, {Stark}, {Stassun}, {Steinmetz}, {Stello}, {Stone-Martinez}, {Storchi-Bergmann}, {Stringfellow}, {Stutz}, {Su}, {Taghizadeh-Popp}, {Talbot}, {Tayar}, {Telles}, {Teske}, {Thakar}, {Theissen}, {Tkachenko}, {Thomas}, {Tojeiro}, {Hernandez Toledo}, {Troup}, {Trump}, {Trussler}, {Turner}, {Tuttle}, {Unda-Sanzana}, {V{\'a}zquez-Mata}, {Valentini}, {Valenzuela}, {Vargas-Gonz{\'a}lez}, {Vargas-Maga{\~n}a}, {Alfaro}, {Villanova}, {Vincenzo}, {Wake}, {Warfield}, {Washington}, {Weaver}, {Weijmans}, {Weinberg}, {Weiss}, {Westfall}, {Wild}, {Wilde}, {Wilson}, {Wilson}, {Wilson}, {Wolf}, {Wood-Vasey}, {Yan}, {Zamora}, {Zasowski}, {Zhang}, {Zhao}, {Zheng}, {Zheng}, \& {Zhu}}]{apogeedr17}
{Abdurro'uf}, {Accetta}, K., {Aerts}, C., {et~al.} 2022, \apjs, 259, 35, \dodoi{10.3847/1538-4365/ac4414}

\bibitem[{{Adibekyan} {et~al.}(2016){Adibekyan}, {Delgado-Mena}, {Figueira}, {Sousa}, {Santos}, {Gonz{\'a}lez Hern{\'a}ndez}, {Minchev}, {Faria}, {Israelian}, {Harutyunyan}, {Su{\'a}rez-Andr{\'e}s}, \& {Hakobyan}}]{Adibekyan16}
{Adibekyan}, V., {Delgado-Mena}, E., {Figueira}, P., {et~al.} 2016, \aap, 592, A87, \dodoi{10.1051/0004-6361/201628883}

\bibitem[{{Adibekyan} {et~al.}(2014){Adibekyan}, {Gonz{\'a}lez Hern{\'a}ndez}, {Delgado Mena}, {Sousa}, {Santos}, {Israelian}, {Figueira}, \& {Bertran de Lis}}]{Adibekyan14}
{Adibekyan}, V.~Z., {Gonz{\'a}lez Hern{\'a}ndez}, J.~I., {Delgado Mena}, E., {et~al.} 2014, \aap, 564, L15, \dodoi{10.1051/0004-6361/201423435}

\bibitem[{{Anderson}(2019)}]{Anderson19}
{Anderson}, J.~P. 2019, \aap, 628, A7, \dodoi{10.1051/0004-6361/201935027}

\bibitem[{{Andrews} {et~al.}(2017){Andrews}, {Weinberg}, {Sch{\"o}nrich}, \& {Johnson}}]{Andrews17}
{Andrews}, B.~H., {Weinberg}, D.~H., {Sch{\"o}nrich}, R., \& {Johnson}, J.~A. 2017, \apj, 835, 224, \dodoi{10.3847/1538-4357/835/2/224}

\bibitem[{{Armillotta} {et~al.}(2018){Armillotta}, {Krumholz}, \& {Fujimoto}}]{Armillotta18}
{Armillotta}, L., {Krumholz}, M.~R., \& {Fujimoto}, Y. 2018, \mnras, 481, 5000, \dodoi{10.1093/mnras/sty2625}

\bibitem[{{Arnett}(1996)}]{Arnett96}
{Arnett}, D. 1996, {Supernovae and Nucleosynthesis: An Investigation of the History of Matter from the Big Bang to the Present}

\bibitem[{{Astropy Collaboration} {et~al.}(2013){Astropy Collaboration}, {Robitaille}, {Tollerud}, {Greenfield}, {Droettboom}, {Bray}, {Aldcroft}, {Davis}, {Ginsburg}, {Price-Whelan}, {Kerzendorf}, {Conley}, {Crighton}, {Barbary}, {Muna}, {Ferguson}, {Grollier}, {Parikh}, {Nair}, {Unther}, {Deil}, {Woillez}, {Conseil}, {Kramer}, {Turner}, {Singer}, {Fox}, {Weaver}, {Zabalza}, {Edwards}, {Azalee Bostroem}, {Burke}, {Casey}, {Crawford}, {Dencheva}, {Ely}, {Jenness}, {Labrie}, {Lim}, {Pierfederici}, {Pontzen}, {Ptak}, {Refsdal}, {Servillat}, \& {Streicher}}]{2013A&A...558A..33A}
{Astropy Collaboration}, {Robitaille}, T.~P., {Tollerud}, E.~J., {et~al.} 2013, \aap, 558, A33, \dodoi{10.1051/0004-6361/201322068}

\bibitem[{{Astropy Collaboration} {et~al.}(2018){Astropy Collaboration}, {Price-Whelan}, {Sip{\H{o}}cz}, {G{\"u}nther}, {Lim}, {Crawford}, {Conseil}, {Shupe}, {Craig}, {Dencheva}, {Ginsburg}, {VanderPlas}, {Bradley}, {P{\'e}rez-Su{\'a}rez}, {de Val-Borro}, {Aldcroft}, {Cruz}, {Robitaille}, {Tollerud}, {Ardelean}, {Babej}, {Bach}, {Bachetti}, {Bakanov}, {Bamford}, {Barentsen}, {Barmby}, {Baumbach}, {Berry}, {Biscani}, {Boquien}, {Bostroem}, {Bouma}, {Brammer}, {Bray}, {Breytenbach}, {Buddelmeijer}, {Burke}, {Calderone}, {Cano Rodr{\'\i}guez}, {Cara}, {Cardoso}, {Cheedella}, {Copin}, {Corrales}, {Crichton}, {D'Avella}, {Deil}, {Depagne}, {Dietrich}, {Donath}, {Droettboom}, {Earl}, {Erben}, {Fabbro}, {Ferreira}, {Finethy}, {Fox}, {Garrison}, {Gibbons}, {Goldstein}, {Gommers}, {Greco}, {Greenfield}, {Groener}, {Grollier}, {Hagen}, {Hirst}, {Homeier}, {Horton}, {Hosseinzadeh}, {Hu}, {Hunkeler}, {Ivezi{\'c}}, {Jain}, {Jenness}, {Kanarek}, {Kendrew}, {Kern}, {Kerzendorf}, {Khvalko}, {King}, {Kirkby}, {Kulkarni},
  {Kumar}, {Lee}, {Lenz}, {Littlefair}, {Ma}, {Macleod}, {Mastropietro}, {McCully}, {Montagnac}, {Morris}, {Mueller}, {Mumford}, {Muna}, {Murphy}, {Nelson}, {Nguyen}, {Ninan}, {N{\"o}the}, {Ogaz}, {Oh}, {Parejko}, {Parley}, {Pascual}, {Patil}, {Patil}, {Plunkett}, {Prochaska}, {Rastogi}, {Reddy Janga}, {Sabater}, {Sakurikar}, {Seifert}, {Sherbert}, {Sherwood-Taylor}, {Shih}, {Sick}, {Silbiger}, {Singanamalla}, {Singer}, {Sladen}, {Sooley}, {Sornarajah}, {Streicher}, {Teuben}, {Thomas}, {Tremblay}, {Turner}, {Terr{\'o}n}, {van Kerkwijk}, {de la Vega}, {Watkins}, {Weaver}, {Whitmore}, {Woillez}, {Zabalza}, \& {Astropy Contributors}}]{astropyii}
{Astropy Collaboration}, {Price-Whelan}, A.~M., {Sip{\H{o}}cz}, B.~M., {et~al.} 2018, \aj, 156, 123, \dodoi{10.3847/1538-3881/aabc4f}

\bibitem[{{Baba} {et~al.}(2023){Baba}, {Saitoh}, \& {Tsujimoto}}]{Baba23}
{Baba}, J., {Saitoh}, T.~R., \& {Tsujimoto}, T. 2023, \mnras, 526, 6088, \dodoi{10.1093/mnras/stad3188}

\bibitem[{{Badenes} {et~al.}(2018){Badenes}, {Mazzola}, {Thompson}, {Covey}, {Freeman}, {Walker}, {Moe}, {Troup}, {Nidever}, {Allende Prieto}, {Andrews}, {Barb{\'a}}, {Beers}, {Bovy}, {Carlberg}, {De Lee}, {Johnson}, {Lewis}, {Majewski}, {Pinsonneault}, {Sobeck}, {Stassun}, {Stringfellow}, \& {Zasowski}}]{Badenes18}
{Badenes}, C., {Mazzola}, C., {Thompson}, T.~A., {et~al.} 2018, \apj, 854, 147, \dodoi{10.3847/1538-4357/aaa765}

\bibitem[{{Bedell} {et~al.}(2018){Bedell}, {Bean}, {Mel{\'e}ndez}, {Spina}, {Ram{\'\i}rez}, {Asplund}, {Alves-Brito}, {dos Santos}, {Dreizler}, {Yong}, {Monroe}, \& {Casagrande}}]{bedell18}
{Bedell}, M., {Bean}, J.~L., {Mel{\'e}ndez}, J., {et~al.} 2018, \apj, 865, 68, \dodoi{10.3847/1538-4357/aad908}

\bibitem[{{Behmard} {et~al.}(2023){Behmard}, {Ness}, {Cunningham}, \& {Bedell}}]{Behmard23}
{Behmard}, A., {Ness}, M.~K., {Cunningham}, E.~C., \& {Bedell}, M. 2023, \aj, 165, 178, \dodoi{10.3847/1538-3881/acc32a}

\bibitem[{{Bensby} {et~al.}(2003){Bensby}, {Feltzing}, \& {Lundstr{\"o}m}}]{Bensby2003}
{Bensby}, T., {Feltzing}, S., \& {Lundstr{\"o}m}, I. 2003, \aap, 410, 527, \dodoi{10.1051/0004-6361:20031213}

\bibitem[{{Berke} {et~al.}(2023){Berke}, {Murphy}, {Flynn}, \& {Liu}}]{Berke23}
{Berke}, D.~A., {Murphy}, M.~T., {Flynn}, C., \& {Liu}, F. 2023, \mnras, 519, 1221, \dodoi{10.1093/mnras/stac2037}

\bibitem[{{Blancato} {et~al.}(2019){Blancato}, {Ness}, {Johnston}, {Rybizki}, \& {Bedell}}]{Blancato2019}
{Blancato}, K., {Ness}, M., {Johnston}, K.~V., {Rybizki}, J., \& {Bedell}, M. 2019, \apj, 883, 34, \dodoi{10.3847/1538-4357/ab39e5}

\bibitem[{{Bland-Hawthorn} {et~al.}(2010){Bland-Hawthorn}, {Krumholz}, \& {Freeman}}]{BlandHawthorn10}
{Bland-Hawthorn}, J., {Krumholz}, M.~R., \& {Freeman}, K. 2010, \apj, 713, 166, \dodoi{10.1088/0004-637X/713/1/166}

\bibitem[{{Booth} \& {Owen}(2020)}]{Booth20}
{Booth}, R.~A., \& {Owen}, J.~E. 2020, \mnras, 493, 5079, \dodoi{10.1093/mnras/staa578}

\bibitem[{{Brown} {et~al.}(2019){Brown}, {Stanek}, {Holoien}, {Kochanek}, {Shappee}, {Prieto}, {Dong}, {Chen}, {Thompson}, {Beacom}, {Stritzinger}, {Bersier}, \& {Brimacombe}}]{Brown19}
{Brown}, J.~S., {Stanek}, K.~Z., {Holoien}, T.~W.~S., {et~al.} 2019, \mnras, 484, 3785, \dodoi{10.1093/mnras/stz258}

\bibitem[{{Buder} {et~al.}(2024){Buder}, {Kos}, {Wang}, {McKenzie}, {Howell}, {Martell}, {Hayden}, {Zucker}, {Nordlander}, {Montet}, {Traven}, {Bland-Hawthorn}, {De Silva}, {Freeman}, {Lewis}, {Lind}, {Sharma}, {Simpson}, {Stello}, {Zwitter}, {Amarsi}, {Armstrong}, {Banks}, {Beavis}, {Beeson}, {Chen}, {Ciuc{\u{a}}}, {Da Costa}, {de Grijs}, {Martin}, {Nataf}, {Ness}, {Rains}, {Scarr}, {Vogrin{\v{c}}i{\v{c}}}, {Wang}, {Wittenmyer}, {Xie}, \& {The GALAH Collaboration}}]{GALAHDR4}
{Buder}, S., {Kos}, J., {Wang}, E.~X., {et~al.} 2024, arXiv e-prints, arXiv:2409.19858, \dodoi{10.48550/arXiv.2409.19858}

\bibitem[{{Burbidge} {et~al.}(1957){Burbidge}, {Burbidge}, {Fowler}, \& {Hoyle}}]{Burbidge57}
{Burbidge}, E.~M., {Burbidge}, G.~R., {Fowler}, W.~A., \& {Hoyle}, F. 1957, Reviews of Modern Physics, 29, 547, \dodoi{10.1103/RevModPhys.29.547}

\bibitem[{{Cameron}(1957)}]{Cameron57}
{Cameron}, A.~G.~W. 1957, \pasp, 69, 201, \dodoi{10.1086/127051}

\bibitem[{{Carlos} {et~al.}(2025){Carlos}, {Amarsi}, {Nissen}, \& {Canocchi}}]{Carlos25}
{Carlos}, M., {Amarsi}, A.~M., {Nissen}, P.~E., \& {Canocchi}, G. 2025, arXiv e-prints, arXiv:2505.22615.
\newblock \doarXiv{2505.22615}

\bibitem[{{Cowley} \& {Y{\"u}ce}(2022)}]{Cowley22}
{Cowley}, C.~R., \& {Y{\"u}ce}, K. 2022, \mnras, 512, 3684, \dodoi{10.1093/mnras/stac637}

\bibitem[{{De Silva} {et~al.}(2015){De Silva}, {Freeman}, {Bland-Hawthorn}, {Martell}, {de Boer}, {Asplund}, {Keller}, {Sharma}, {Zucker}, {Zwitter}, {Anguiano}, {Bacigalupo}, {Bayliss}, {Beavis}, {Bergemann}, {Campbell}, {Cannon}, {Carollo}, {Casagrande}, {Casey}, {Da Costa}, {D'Orazi}, {Dotter}, {Duong}, {Heger}, {Ireland}, {Kafle}, {Kos}, {Lattanzio}, {Lewis}, {Lin}, {Lind}, {Munari}, {Nataf}, {O'Toole}, {Parker}, {Reid}, {Schlesinger}, {Sheinis}, {Simpson}, {Stello}, {Ting}, {Traven}, {Watson}, {Wittenmyer}, {Yong}, \& {{\v{Z}}erjal}}]{galah_mission}
{De Silva}, G.~M., {Freeman}, K.~C., {Bland-Hawthorn}, J., {et~al.} 2015, \mnras, 449, 2604, \dodoi{10.1093/mnras/stv327}

\bibitem[{{Di Matteo} {et~al.}(2013){Di Matteo}, {Haywood}, {Combes}, {Semelin}, \& {Snaith}}]{DiMatteo13}
{Di Matteo}, P., {Haywood}, M., {Combes}, F., {Semelin}, B., \& {Snaith}, O.~N. 2013, \aap, 553, A102, \dodoi{10.1051/0004-6361/201220539}

\bibitem[{{Edvardsson} {et~al.}(1993){Edvardsson}, {Andersen}, {Gustafsson}, {Lambert}, {Nissen}, \& {Tomkin}}]{Edvardsson93}
{Edvardsson}, B., {Andersen}, J., {Gustafsson}, B., {et~al.} 1993, \aap, 275, 101

\bibitem[{{Feuillet} {et~al.}(2019){Feuillet}, {Frankel}, {Lind}, {Frinchaboy}, {Garc{\'\i}a-Hern{\'a}ndez}, {Lane}, {Nitschelm}, \& {Roman-Lopes}}]{Feuillet19}
{Feuillet}, D.~K., {Frankel}, N., {Lind}, K., {et~al.} 2019, \mnras, 489, 1742, \dodoi{10.1093/mnras/stz2221}

\bibitem[{{Feuillet} {et~al.}(2018){Feuillet}, {Bovy}, {Holtzman}, {Weinberg}, {Garc{\'\i}a-Hern{\'a}ndez}, {Hearty}, {Majewski}, {Roman-Lopes}, {Rybizki}, \& {Zamora}}]{Feuillet18}
{Feuillet}, D.~K., {Bovy}, J., {Holtzman}, J., {et~al.} 2018, \mnras, 477, 2326, \dodoi{10.1093/mnras/sty779}

\bibitem[{{Flores} {et~al.}(2023){Flores}, {Galarza}, {Miquelarena}, {Saffe}, {Jaque Arancibia}, {Iba{\~n}ez Bustos}, {Jofr{\'e}}, {Alacoria1}, \& {Gunella}}]{Flores23}
{Flores}, M., {Galarza}, J.~Y., {Miquelarena}, P., {et~al.} 2023, arXiv e-prints, arXiv:2311.17272, \dodoi{10.48550/arXiv.2311.17272}

\bibitem[{{Frankel} {et~al.}(2019){Frankel}, {Sanders}, {Rix}, {Ting}, \& {Ness}}]{Frankel19}
{Frankel}, N., {Sanders}, J., {Rix}, H.-W., {Ting}, Y.-S., \& {Ness}, M. 2019, \apj, 884, 99, \dodoi{10.3847/1538-4357/ab4254}

\bibitem[{{Gaia Collaboration} {et~al.}(2016){Gaia Collaboration}, {Prusti}, {de Bruijne}, {Brown}, {Vallenari}, {Babusiaux}, {Bailer-Jones}, {Bastian}, {Biermann}, {Evans}, {Eyer}, {Jansen}, {Jordi}, {Klioner}, {Lammers}, {Lindegren}, {Luri}, {Mignard}, {Milligan}, {Panem}, {Poinsignon}, {Pourbaix}, {Randich}, {Sarri}, {Sartoretti}, {Siddiqui}, {Soubiran}, {Valette}, {van Leeuwen}, {Walton}, {Aerts}, {Arenou}, {Cropper}, {Drimmel}, {H{\o}g}, {Katz}, {Lattanzi}, {O'Mullane}, {Grebel}, {Holland}, {Huc}, {Passot}, {Bramante}, {Cacciari}, {Casta{\~n}eda}, {Chaoul}, {Cheek}, {De Angeli}, {Fabricius}, {Guerra}, {Hern{\'a}ndez}, {Jean-Antoine-Piccolo}, {Masana}, {Messineo}, {Mowlavi}, {Nienartowicz}, {Ord{\'o}{\~n}ez-Blanco}, {Panuzzo}, {Portell}, {Richards}, {Riello}, {Seabroke}, {Tanga}, {Th{\'e}venin}, {Torra}, {Els}, {Gracia-Abril}, {Comoretto}, {Garcia-Reinaldos}, {Lock}, {Mercier}, {Altmann}, {Andrae}, {Astraatmadja}, {Bellas-Velidis}, {Benson}, {Berthier}, {Blomme}, {Busso}, {Carry}, {Cellino}, {Clementini},
  {Cowell}, {Creevey}, {Cuypers}, {Davidson}, {De Ridder}, {de Torres}, {Delchambre}, {Dell'Oro}, {Ducourant}, {Fr{\'e}mat}, {Garc{\'\i}a-Torres}, {Gosset}, {Halbwachs}, {Hambly}, {Harrison}, {Hauser}, {Hestroffer}, {Hodgkin}, {Huckle}, {Hutton}, {Jasniewicz}, {Jordan}, {Kontizas}, {Korn}, {Lanzafame}, {Manteiga}, {Moitinho}, {Muinonen}, {Osinde}, {Pancino}, {Pauwels}, {Petit}, {Recio-Blanco}, {Robin}, {Sarro}, {Siopis}, {Smith}, {Smith}, {Sozzetti}, {Thuillot}, {van Reeven}, {Viala}, {Abbas}, {Abreu Aramburu}, {Accart}, {Aguado}, {Allan}, {Allasia}, {Altavilla}, {{\'A}lvarez}, {Alves}, {Anderson}, {Andrei}, {Anglada Varela}, {Antiche}, {Antoja}, {Ant{\'o}n}, {Arcay}, {Atzei}, {Ayache}, {Bach}, {Baker}, {Balaguer-N{\'u}{\~n}ez}, {Barache}, {Barata}, {Barbier}, {Barblan}, {Baroni}, {Barrado y Navascu{\'e}s}, {Barros}, {Barstow}, {Becciani}, {Bellazzini}, {Bellei}, {Bello Garc{\'\i}a}, {Belokurov}, {Bendjoya}, {Berihuete}, {Bianchi}, {Bienaym{\'e}}, {Billebaud}, {Blagorodnova}, {Blanco-Cuaresma}, {Boch},
  {Bombrun}, {Borrachero}, {Bouquillon}, {Bourda}, {Bouy}, {Bragaglia}, {Breddels}, {Brouillet}, {Br{\"u}semeister}, {Bucciarelli}, {Budnik}, {Burgess}, {Burgon}, {Burlacu}, {Busonero}, {Buzzi}, {Caffau}, {Cambras}, {Campbell}, {Cancelliere}, {Cantat-Gaudin}, {Carlucci}, {Carrasco}, {Castellani}, {Charlot}, {Charnas}, {Charvet}, {Chassat}, {Chiavassa}, {Clotet}, {Cocozza}, {Collins}, {Collins}, {Costigan}, {Crifo}, {Cross}, {Crosta}, {Crowley}, {Dafonte}, {Damerdji}, {Dapergolas}, {David}, {David}, {De Cat}, {de Felice}, {de Laverny}, {De Luise}, {De March}, {de Martino}, {de Souza}, {Debosscher}, {del Pozo}, {Delbo}, {Delgado}, {Delgado}, {di Marco}, {Di Matteo}, {Diakite}, {Distefano}, {Dolding}, {Dos Anjos}, {Drazinos}, {Dur{\'a}n}, {Dzigan}, {Ecale}, {Edvardsson}, {Enke}, {Erdmann}, {Escolar}, {Espina}, {Evans}, {Eynard Bontemps}, {Fabre}, {Fabrizio}, {Faigler}, {Falc{\~a}o}, {Farr{\`a}s Casas}, {Faye}, {Federici}, {Fedorets}, {Fern{\'a}ndez-Hern{\'a}ndez}, {Fernique}, {Fienga}, {Figueras}, {Filippi},
  {Findeisen}, {Fonti}, {Fouesneau}, {Fraile}, {Fraser}, {Fuchs}, {Furnell}, {Gai}, {Galleti}, {Galluccio}, {Garabato}, {Garc{\'\i}a-Sedano}, {Gar{\'e}}, {Garofalo}, {Garralda}, {Gavras}, {Gerssen}, {Geyer}, {Gilmore}, {Girona}, {Giuffrida}, {Gomes}, {Gonz{\'a}lez-Marcos}, {Gonz{\'a}lez-N{\'u}{\~n}ez}, {Gonz{\'a}lez-Vidal}, {Granvik}, {Guerrier}, {Guillout}, {Guiraud}, {G{\'u}rpide}, {Guti{\'e}rrez-S{\'a}nchez}, {Guy}, {Haigron}, {Hatzidimitriou}, {Haywood}, {Heiter}, {Helmi}, {Hobbs}, {Hofmann}, {Holl}, {Holland}, {Hunt}, {Hypki}, {Icardi}, {Irwin}, {Jevardat de Fombelle}, {Jofr{\'e}}, {Jonker}, {Jorissen}, {Julbe}, {Karampelas}, {Kochoska}, {Kohley}, {Kolenberg}, {Kontizas}, {Koposov}, {Kordopatis}, {Koubsky}, {Kowalczyk}, {Krone-Martins}, {Kudryashova}, {Kull}, {Bachchan}, {Lacoste-Seris}, {Lanza}, {Lavigne}, {Le Poncin-Lafitte}, {Lebreton}, {Lebzelter}, {Leccia}, {Leclerc}, {Lecoeur-Taibi}, {Lemaitre}, {Lenhardt}, {Leroux}, {Liao}, {Licata}, {Lindstr{\o}m}, {Lister}, {Livanou}, {Lobel}, {L{\"o}ffler},
  {L{\'o}pez}, {Lopez-Lozano}, {Lorenz}, {Loureiro}, {MacDonald}, {Magalh{\~a}es Fernandes}, {Managau}, {Mann}, {Mantelet}, {Marchal}, {Marchant}, {Marconi}, {Marie}, {Marinoni}, {Marrese}, {Marschalk{\'o}}, {Marshall}, {Mart{\'\i}n-Fleitas}, {Martino}, {Mary}, {Matijevi{\v{c}}}, {Mazeh}, {McMillan}, {Messina}, {Mestre}, {Michalik}, {Millar}, {Miranda}, {Molina}, {Molinaro}, {Molinaro}, {Moln{\'a}r}, {Moniez}, {Montegriffo}, {Monteiro}, {Mor}, {Mora}, {Morbidelli}, {Morel}, {Morgenthaler}, {Morley}, {Morris}, {Mulone}, {Muraveva}, {Musella}, {Narbonne}, {Nelemans}, {Nicastro}, {Noval}, {Ord{\'e}novic}, {Ordieres-Mer{\'e}}, {Osborne}, {Pagani}, {Pagano}, {Pailler}, {Palacin}, {Palaversa}, {Parsons}, {Paulsen}, {Pecoraro}, {Pedrosa}, {Pentik{\"a}inen}, {Pereira}, {Pichon}, {Piersimoni}, {Pineau}, {Plachy}, {Plum}, {Poujoulet}, {Pr{\v{s}}a}, {Pulone}, {Ragaini}, {Rago}, {Rambaux}, {Ramos-Lerate}, {Ranalli}, {Rauw}, {Read}, {Regibo}, {Renk}, {Reyl{\'e}}, {Ribeiro}, {Rimoldini}, {Ripepi}, {Riva}, {Rixon},
  {Roelens}, {Romero-G{\'o}mez}, {Rowell}, {Royer}, {Rudolph}, {Ruiz-Dern}, {Sadowski}, {Sagrist{\`a} Sell{\'e}s}, {Sahlmann}, {Salgado}, {Salguero}, {Sarasso}, {Savietto}, {Schnorhk}, {Schultheis}, {Sciacca}, {Segol}, {Segovia}, {Segransan}, {Serpell}, {Shih}, {Smareglia}, {Smart}, {Smith}, {Solano}, {Solitro}, {Sordo}, {Soria Nieto}, {Souchay}, {Spagna}, {Spoto}, {Stampa}, {Steele}, {Steidelm{\"u}ller}, {Stephenson}, {Stoev}, {Suess}, {S{\"u}veges}, {Surdej}, {Szabados}, {Szegedi-Elek}, {Tapiador}, {Taris}, {Tauran}, {Taylor}, {Teixeira}, {Terrett}, {Tingley}, {Trager}, {Turon}, {Ulla}, {Utrilla}, {Valentini}, {van Elteren}, {Van Hemelryck}, {van Leeuwen}, {Varadi}, {Vecchiato}, {Veljanoski}, {Via}, {Vicente}, {Vogt}, {Voss}, {Votruba}, {Voutsinas}, {Walmsley}, {Weiler}, {Weingrill}, {Werner}, {Wevers}, {Whitehead}, {Wyrzykowski}, {Yoldas}, {{\v{Z}}erjal}, {Zucker}, {Zurbach}, {Zwitter}, {Alecu}, {Allen}, {Allende Prieto}, {Amorim}, {Anglada-Escud{\'e}}, {Arsenijevic}, {Azaz}, {Balm}, {Beck}, {Bernstein},
  {Bigot}, {Bijaoui}, {Blasco}, {Bonfigli}, {Bono}, {Boudreault}, {Bressan}, {Brown}, {Brunet}, {Bunclark}, {Buonanno}, {Butkevich}, {Carret}, {Carrion}, {Chemin}, {Ch{\'e}reau}, {Corcione}, {Darmigny}, {de Boer}, {de Teodoro}, {de Zeeuw}, {Delle Luche}, {Domingues}, {Dubath}, {Fodor}, {Fr{\'e}zouls}, {Fries}, {Fustes}, {Fyfe}, {Gallardo}, {Gallegos}, {Gardiol}, {Gebran}, {Gomboc}, {G{\'o}mez}, {Grux}, {Gueguen}, {Heyrovsky}, {Hoar}, {Iannicola}, {Isasi Parache}, {Janotto}, {Joliet}, {Jonckheere}, {Keil}, {Kim}, {Klagyivik}, {Klar}, {Knude}, {Kochukhov}, {Kolka}, {Kos}, {Kutka}, {Lainey}, {LeBouquin}, {Liu}, {Loreggia}, {Makarov}, {Marseille}, {Martayan}, {Martinez-Rubi}, {Massart}, {Meynadier}, {Mignot}, {Munari}, {Nguyen}, {Nordlander}, {Ocvirk}, {O'Flaherty}, {Olias Sanz}, {Ortiz}, {Osorio}, {Oszkiewicz}, {Ouzounis}, {Palmer}, {Park}, {Pasquato}, {Peltzer}, {Peralta}, {P{\'e}turaud}, {Pieniluoma}, {Pigozzi}, {Poels}, {Prat}, {Prod'homme}, {Raison}, {Rebordao}, {Risquez}, {Rocca-Volmerange}, {Rosen},
  {Ruiz-Fuertes}, {Russo}, {Sembay}, {Serraller Vizcaino}, {Short}, {Siebert}, {Silva}, {Sinachopoulos}, {Slezak}, {Soffel}, {Sosnowska}, {Strai{\v{z}}ys}, {ter Linden}, {Terrell}, {Theil}, {Tiede}, {Troisi}, {Tsalmantza}, {Tur}, {Vaccari}, {Vachier}, {Valles}, {Van Hamme}, {Veltz}, {Virtanen}, {Wallut}, {Wichmann}, {Wilkinson}, {Ziaeepour}, \& {Zschocke}}]{GaiaMission}
{Gaia Collaboration}, {Prusti}, T., {de Bruijne}, J.~H.~J., {et~al.} 2016, \aap, 595, A1, \dodoi{10.1051/0004-6361/201629272}

\bibitem[{{Gaia Collaboration} {et~al.}(2023){Gaia Collaboration}, {Vallenari}, {Brown}, {Prusti}, {de Bruijne}, {Arenou}, {Babusiaux}, {Biermann}, {Creevey}, {Ducourant}, {Evans}, {Eyer}, {Guerra}, {Hutton}, {Jordi}, {Klioner}, {Lammers}, {Lindegren}, {Luri}, {Mignard}, {Panem}, {Pourbaix}, {Randich}, {Sartoretti}, {Soubiran}, {Tanga}, {Walton}, {Bailer-Jones}, {Bastian}, {Drimmel}, {Jansen}, {Katz}, {Lattanzi}, {van Leeuwen}, {Bakker}, {Cacciari}, {Casta{\~n}eda}, {De Angeli}, {Fabricius}, {Fouesneau}, {Fr{\'e}mat}, {Galluccio}, {Guerrier}, {Heiter}, {Masana}, {Messineo}, {Mowlavi}, {Nicolas}, {Nienartowicz}, {Pailler}, {Panuzzo}, {Riclet}, {Roux}, {Seabroke}, {Sordo}, {Th{\'e}venin}, {Gracia-Abril}, {Portell}, {Teyssier}, {Altmann}, {Andrae}, {Audard}, {Bellas-Velidis}, {Benson}, {Berthier}, {Blomme}, {Burgess}, {Busonero}, {Busso}, {C{\'a}novas}, {Carry}, {Cellino}, {Cheek}, {Clementini}, {Damerdji}, {Davidson}, {de Teodoro}, {Nu{\~n}ez Campos}, {Delchambre}, {Dell'Oro}, {Esquej},
  {Fern{\'a}ndez-Hern{\'a}ndez}, {Fraile}, {Garabato}, {Garc{\'\i}a-Lario}, {Gosset}, {Haigron}, {Halbwachs}, {Hambly}, {Harrison}, {Hern{\'a}ndez}, {Hestroffer}, {Hodgkin}, {Holl}, {Jan{\ss}en}, {Jevardat de Fombelle}, {Jordan}, {Krone-Martins}, {Lanzafame}, {L{\"o}ffler}, {Marchal}, {Marrese}, {Moitinho}, {Muinonen}, {Osborne}, {Pancino}, {Pauwels}, {Recio-Blanco}, {Reyl{\'e}}, {Riello}, {Rimoldini}, {Roegiers}, {Rybizki}, {Sarro}, {Siopis}, {Smith}, {Sozzetti}, {Utrilla}, {van Leeuwen}, {Abbas}, {{\'A}brah{\'a}m}, {Abreu Aramburu}, {Aerts}, {Aguado}, {Ajaj}, {Aldea-Montero}, {Altavilla}, {{\'A}lvarez}, {Alves}, {Anders}, {Anderson}, {Anglada Varela}, {Antoja}, {Baines}, {Baker}, {Balaguer-N{\'u}{\~n}ez}, {Balbinot}, {Balog}, {Barache}, {Barbato}, {Barros}, {Barstow}, {Bartolom{\'e}}, {Bassilana}, {Bauchet}, {Becciani}, {Bellazzini}, {Berihuete}, {Bernet}, {Bertone}, {Bianchi}, {Binnenfeld}, {Blanco-Cuaresma}, {Blazere}, {Boch}, {Bombrun}, {Bossini}, {Bouquillon}, {Bragaglia}, {Bramante}, {Breedt},
  {Bressan}, {Brouillet}, {Brugaletta}, {Bucciarelli}, {Burlacu}, {Butkevich}, {Buzzi}, {Caffau}, {Cancelliere}, {Cantat-Gaudin}, {Carballo}, {Carlucci}, {Carnerero}, {Carrasco}, {Casamiquela}, {Castellani}, {Castro-Ginard}, {Chaoul}, {Charlot}, {Chemin}, {Chiaramida}, {Chiavassa}, {Chornay}, {Comoretto}, {Contursi}, {Cooper}, {Cornez}, {Cowell}, {Crifo}, {Cropper}, {Crosta}, {Crowley}, {Dafonte}, {Dapergolas}, {David}, {David}, {de Laverny}, {De Luise}, {De March}, {De Ridder}, {de Souza}, {de Torres}, {del Peloso}, {del Pozo}, {Delbo}, {Delgado}, {Delisle}, {Demouchy}, {Dharmawardena}, {Di Matteo}, {Diakite}, {Diener}, {Distefano}, {Dolding}, {Edvardsson}, {Enke}, {Fabre}, {Fabrizio}, {Faigler}, {Fedorets}, {Fernique}, {Fienga}, {Figueras}, {Fournier}, {Fouron}, {Fragkoudi}, {Gai}, {Garcia-Gutierrez}, {Garcia-Reinaldos}, {Garc{\'\i}a-Torres}, {Garofalo}, {Gavel}, {Gavras}, {Gerlach}, {Geyer}, {Giacobbe}, {Gilmore}, {Girona}, {Giuffrida}, {Gomel}, {Gomez}, {Gonz{\'a}lez-N{\'u}{\~n}ez},
  {Gonz{\'a}lez-Santamar{\'\i}a}, {Gonz{\'a}lez-Vidal}, {Granvik}, {Guillout}, {Guiraud}, {Guti{\'e}rrez-S{\'a}nchez}, {Guy}, {Hatzidimitriou}, {Hauser}, {Haywood}, {Helmer}, {Helmi}, {Sarmiento}, {Hidalgo}, {Hilger}, {H{\l}adczuk}, {Hobbs}, {Holland}, {Huckle}, {Jardine}, {Jasniewicz}, {Jean-Antoine Piccolo}, {Jim{\'e}nez-Arranz}, {Jorissen}, {Juaristi Campillo}, {Julbe}, {Karbevska}, {Kervella}, {Khanna}, {Kontizas}, {Kordopatis}, {Korn}, {K{\'o}sp{\'a}l}, {Kostrzewa-Rutkowska}, {Kruszy{\'n}ska}, {Kun}, {Laizeau}, {Lambert}, {Lanza}, {Lasne}, {Le Campion}, {Lebreton}, {Lebzelter}, {Leccia}, {Leclerc}, {Lecoeur-Taibi}, {Liao}, {Licata}, {Lindstr{\o}m}, {Lister}, {Livanou}, {Lobel}, {Lorca}, {Loup}, {Madrero Pardo}, {Magdaleno Romeo}, {Managau}, {Mann}, {Manteiga}, {Marchant}, {Marconi}, {Marcos}, {Marcos Santos}, {Mar{\'\i}n Pina}, {Marinoni}, {Marocco}, {Marshall}, {Martin Polo}, {Mart{\'\i}n-Fleitas}, {Marton}, {Mary}, {Masip}, {Massari}, {Mastrobuono-Battisti}, {Mazeh}, {McMillan}, {Messina}, {Michalik},
  {Millar}, {Mints}, {Molina}, {Molinaro}, {Moln{\'a}r}, {Monari}, {Mongui{\'o}}, {Montegriffo}, {Montero}, {Mor}, {Mora}, {Morbidelli}, {Morel}, {Morris}, {Muraveva}, {Murphy}, {Musella}, {Nagy}, {Noval}, {Oca{\~n}a}, {Ogden}, {Ordenovic}, {Osinde}, {Pagani}, {Pagano}, {Palaversa}, {Palicio}, {Pallas-Quintela}, {Panahi}, {Payne-Wardenaar}, {Pe{\~n}alosa Esteller}, {Penttil{\"a}}, {Pichon}, {Piersimoni}, {Pineau}, {Plachy}, {Plum}, {Poggio}, {Pr{\v{s}}a}, {Pulone}, {Racero}, {Ragaini}, {Rainer}, {Raiteri}, {Rambaux}, {Ramos}, {Ramos-Lerate}, {Re Fiorentin}, {Regibo}, {Richards}, {Rios Diaz}, {Ripepi}, {Riva}, {Rix}, {Rixon}, {Robichon}, {Robin}, {Robin}, {Roelens}, {Rogues}, {Rohrbasser}, {Romero-G{\'o}mez}, {Rowell}, {Royer}, {Ruz Mieres}, {Rybicki}, {Sadowski}, {S{\'a}ez N{\'u}{\~n}ez}, {Sagrist{\`a} Sell{\'e}s}, {Sahlmann}, {Salguero}, {Samaras}, {Sanchez Gimenez}, {Sanna}, {Santove{\~n}a}, {Sarasso}, {Schultheis}, {Sciacca}, {Segol}, {Segovia}, {S{\'e}gransan}, {Semeux}, {Shahaf}, {Siddiqui}, {Siebert},
  {Siltala}, {Silvelo}, {Slezak}, {Slezak}, {Smart}, {Snaith}, {Solano}, {Solitro}, {Souami}, {Souchay}, {Spagna}, {Spina}, {Spoto}, {Steele}, {Steidelm{\"u}ller}, {Stephenson}, {S{\"u}veges}, {Surdej}, {Szabados}, {Szegedi-Elek}, {Taris}, {Taylor}, {Teixeira}, {Tolomei}, {Tonello}, {Torra}, {Torra}, {Torralba Elipe}, {Trabucchi}, {Tsounis}, {Turon}, {Ulla}, {Unger}, {Vaillant}, {van Dillen}, {van Reeven}, {Vanel}, {Vecchiato}, {Viala}, {Vicente}, {Voutsinas}, {Weiler}, {Wevers}, {Wyrzykowski}, {Yoldas}, {Yvard}, {Zhao}, {Zorec}, {Zucker}, \& {Zwitter}}]{GaiaDr3}
{Gaia Collaboration}, {Vallenari}, A., {Brown}, A.~G.~A., {et~al.} 2023, \aap, 674, A1, \dodoi{10.1051/0004-6361/202243940}

\bibitem[{{Gaidos}(2015)}]{Gaidos15}
{Gaidos}, E. 2015, \apj, 804, 40, \dodoi{10.1088/0004-637X/804/1/40}

\bibitem[{{Gandhi} {et~al.}(2022){Gandhi}, {Wetzel}, {Hopkins}, {Shappee}, {Wheeler}, \& {Faucher-Gigu{\`e}re}}]{Gandhi22}
{Gandhi}, P.~J., {Wetzel}, A., {Hopkins}, P.~F., {et~al.} 2022, \mnras, 516, 1941, \dodoi{10.1093/mnras/stac2228}

\bibitem[{{Garc{\'\i}a P{\'e}rez} {et~al.}(2016){Garc{\'\i}a P{\'e}rez}, {Allende Prieto}, {Holtzman}, {Shetrone}, {M{\'e}sz{\'a}ros}, {Bizyaev}, {Carrera}, {Cunha}, {Garc{\'\i}a-Hern{\'a}ndez}, {Johnson}, {Majewski}, {Nidever}, {Schiavon}, {Shane}, {Smith}, {Sobeck}, {Troup}, {Zamora}, {Weinberg}, {Bovy}, {Eisenstein}, {Feuillet}, {Frinchaboy}, {Hayden}, {Hearty}, {Nguyen}, {O'Connell}, {Pinsonneault}, {Wilson}, \& {Zasowski}}]{aspcap}
{Garc{\'\i}a P{\'e}rez}, A.~E., {Allende Prieto}, C., {Holtzman}, J.~A., {et~al.} 2016, \aj, 151, 144, \dodoi{10.3847/0004-6256/151/6/144}

\bibitem[{{Ginsburg} {et~al.}(2019){Ginsburg}, {Sip{\H{o}}cz}, {Brasseur}, {Cowperthwaite}, {Craig}, {Deil}, {Guillochon}, {Guzman}, {Liedtke}, {Lian Lim}, {Lockhart}, {Mommert}, {Morris}, {Norman}, {Parikh}, {Persson}, {Robitaille}, {Segovia}, {Singer}, {Tollerud}, {de Val-Borro}, {Valtchanov}, {Woillez}, {Astroquery Collaboration}, \& {a subset of astropy Collaboration}}]{astroquery}
{Ginsburg}, A., {Sip{\H{o}}cz}, B.~M., {Brasseur}, C.~E., {et~al.} 2019, \aj, 157, 98, \dodoi{10.3847/1538-3881/aafc33}

\bibitem[{{Griffith} {et~al.}(2019){Griffith}, {Johnson}, \& {Weinberg}}]{Griffith19}
{Griffith}, E., {Johnson}, J.~A., \& {Weinberg}, D.~H. 2019, \apj, 886, 84, \dodoi{10.3847/1538-4357/ab4b5d}

\bibitem[{{Griffith} {et~al.}(2021){Griffith}, {Weinberg}, {Johnson}, {Beaton}, {Garc{\'\i}a-Hern{\'a}ndez}, {Hasselquist}, {Holtzman}, {Johnson}, {J{\"o}nsson}, {Lane}, {Nataf}, \& {Roman-Lopes}}]{Griffith21}
{Griffith}, E., {Weinberg}, D.~H., {Johnson}, J.~A., {et~al.} 2021, \apj, 909, 77, \dodoi{10.3847/1538-4357/abd6be}

\bibitem[{{Griffith} {et~al.}(2024){Griffith}, {Hogg}, {Dalcanton}, {Hasselquist}, {Ratcliffe}, {Ness}, \& {Weinberg}}]{Griffith24}
{Griffith}, E.~J., {Hogg}, D.~W., {Dalcanton}, J.~J., {et~al.} 2024, \aj, 167, 98, \dodoi{10.3847/1538-3881/ad19c7}

\bibitem[{{Griffith} {et~al.}(2022){Griffith}, {Weinberg}, {Buder}, {Johnson}, {Johnson}, \& {Vincenzo}}]{Griffith22}
{Griffith}, E.~J., {Weinberg}, D.~H., {Buder}, S., {et~al.} 2022, \apj, 931, 23, \dodoi{10.3847/1538-4357/ac5826}

\bibitem[{{Gustafsson} {et~al.}(2008){Gustafsson}, {Edvardsson}, {Eriksson}, {J{\o}rgensen}, {Nordlund}, \& {Plez}}]{MARCSatmospheres}
{Gustafsson}, B., {Edvardsson}, B., {Eriksson}, K., {et~al.} 2008, \aap, 486, 951, \dodoi{10.1051/0004-6361:200809724}

\bibitem[{Harris {et~al.}(2020)Harris, Millman, van~der Walt, Gommers, Virtanen, Cournapeau, Wieser, Taylor, Berg, Smith, Kern, Picus, Hoyer, van Kerkwijk, Brett, Haldane, del R{\'{i}}o, Wiebe, Peterson, G{\'{e}}rard-Marchant, Sheppard, Reddy, Weckesser, Abbasi, Gohlke, \& Oliphant}]{numpy}
Harris, C.~R., Millman, K.~J., van~der Walt, S.~J., {et~al.} 2020, Nature, 585, 357, \dodoi{10.1038/s41586-020-2649-2}

\bibitem[{{Haywood} {et~al.}(2013){Haywood}, {Di Matteo}, {Lehnert}, {Katz}, \& {G{\'o}mez}}]{Haywood13}
{Haywood}, M., {Di Matteo}, P., {Lehnert}, M.~D., {Katz}, D., \& {G{\'o}mez}, A. 2013, \aap, 560, A109, \dodoi{10.1051/0004-6361/201321397}

\bibitem[{{Heged{\H{u}}s} {et~al.}(2025){Heged{\H{u}}s}, {M{\'e}sz{\'a}ros}, {Vil{\'a}gos}, {Pignatari}, {Griffith}, {Souto}, \& {Lugaro}}]{Hegedus25}
{Heged{\H{u}}s}, V., {M{\'e}sz{\'a}ros}, S., {Vil{\'a}gos}, B., {et~al.} 2025, arXiv e-prints, arXiv:2506.00503, \dodoi{10.48550/arXiv.2506.00503}

\bibitem[{{Hinkel} {et~al.}(2016){Hinkel}, {Young}, {Pagano}, {Desch}, {Anbar}, {Adibekyan}, {Blanco-Cuaresma}, {Carlberg}, {Delgado Mena}, {Liu}, {Nordlander}, {Sousa}, {Korn}, {Gruyters}, {Heiter}, {Jofr{\'e}}, {Santos}, \& {Soubiran}}]{Hinkel16}
{Hinkel}, N.~R., {Young}, P.~A., {Pagano}, M.~D., {et~al.} 2016, \apjs, 226, 4, \dodoi{10.3847/0067-0049/226/1/4}

\bibitem[{{Horta} {et~al.}(2022){Horta}, {Ness}, {Rybizki}, {Schiavon}, \& {Buder}}]{Horta22}
{Horta}, D., {Ness}, M.~K., {Rybizki}, J., {Schiavon}, R.~P., \& {Buder}, S. 2022, \mnras, 513, 5477, \dodoi{10.1093/mnras/stac953}

\bibitem[{Hunter(2007)}]{Hunter:2007}
Hunter, J.~D. 2007, Computing in Science \& Engineering, 9, 90, \dodoi{10.1109/MCSE.2007.55}

\bibitem[{{Jofr{\'e}} {et~al.}(2019){Jofr{\'e}}, {Heiter}, \& {Soubiran}}]{Jofre19}
{Jofr{\'e}}, P., {Heiter}, U., \& {Soubiran}, C. 2019, \araa, 57, 571, \dodoi{10.1146/annurev-astro-091918-104509}

\bibitem[{{Johnson}(2019)}]{Johnson19}
{Johnson}, J.~A. 2019, Science, 363, 474, \dodoi{10.1126/science.aau9540}

\bibitem[{{Johnson} {et~al.}(2023){Johnson}, {Kochanek}, \& {Stanek}}]{Johnson23}
{Johnson}, J.~W., {Kochanek}, C.~S., \& {Stanek}, K.~Z. 2023, \mnras, 526, 5911, \dodoi{10.1093/mnras/stad3019}

\bibitem[{{J{\"o}nsson} {et~al.}(2020){J{\"o}nsson}, {Holtzman}, {Allende Prieto}, {Cunha}, {Garc{\'\i}a-Hern{\'a}ndez}, {Hasselquist}, {Masseron}, {Osorio}, {Shetrone}, {Smith}, {Stringfellow}, {Bizyaev}, {Edvardsson}, {Majewski}, {M{\'e}sz{\'a}ros}, {Souto}, {Zamora}, {Beaton}, {Bovy}, {Donor}, {Pinsonneault}, {Poovelil}, \& {Sobeck}}]{Jonsson2020}
{J{\"o}nsson}, H., {Holtzman}, J.~A., {Allende Prieto}, C., {et~al.} 2020, \aj, 160, 120, \dodoi{10.3847/1538-3881/aba592}

\bibitem[{{Kobayashi} {et~al.}(2020){Kobayashi}, {Karakas}, \& {Lugaro}}]{Kobayashi20}
{Kobayashi}, C., {Karakas}, A.~I., \& {Lugaro}, M. 2020, \apj, 900, 179, \dodoi{10.3847/1538-4357/abae65}

\bibitem[{{Kurucz}(1979)}]{Kurucz}
{Kurucz}, R.~L. 1979, \apjs, 40, 1, \dodoi{10.1086/190589}

\bibitem[{{Li} {et~al.}(2011){Li}, {Chornock}, {Leaman}, {Filippenko}, {Poznanski}, {Wang}, {Ganeshalingam}, \& {Mannucci}}]{Li11}
{Li}, W., {Chornock}, R., {Leaman}, J., {et~al.} 2011, \mnras, 412, 1473, \dodoi{10.1111/j.1365-2966.2011.18162.x}

\bibitem[{{Liu} {et~al.}(2024){Liu}, {Ting}, {Yong}, {Bitsch}, {Karakas}, {Murphy}, {Joyce}, {Dotter}, \& {Dai}}]{Liu24}
{Liu}, F., {Ting}, Y.-S., {Yong}, D., {et~al.} 2024, \nat, 627, 501, \dodoi{10.1038/s41586-024-07091-y}

\bibitem[{{Lodders}(2003)}]{Lodders03}
{Lodders}, K. 2003, \apj, 591, 1220, \dodoi{10.1086/375492}

\bibitem[{{Lu} {et~al.}(2022){Lu}, {Minchev}, {Buck}, {Khoperskov}, {Steinmetz}, {Libeskind}, {Cescutti}, {Freeman}, \& {Ratcliffe}}]{Lu22}
{Lu}, Y., {Minchev}, I., {Buck}, T., {et~al.} 2022, arXiv e-prints, arXiv:2212.04515, \dodoi{10.48550/arXiv.2212.04515}

\bibitem[{{Magrini} {et~al.}(2017){Magrini}, {Randich}, {Kordopatis}, {Prantzos}, {Romano}, {Chieffi}, {Limongi}, {Fran{\c{c}}ois}, {Pancino}, {Friel}, {Bragaglia}, {Tautvai{\v{s}}ien{\.{e}}}, {Spina}, {Overbeek}, {Cantat-Gaudin}, {Donati}, {Vallenari}, {Sordo}, {Jim{\'e}nez-Esteban}, {Tang}, {Drazdauskas}, {Sousa}, {Duffau}, {Jofr{\'e}}, {Gilmore}, {Feltzing}, {Alfaro}, {Bensby}, {Flaccomio}, {Koposov}, {Lanzafame}, {Smiljanic}, {Bayo}, {Carraro}, {Casey}, {Costado}, {Damiani}, {Franciosini}, {Hourihane}, {Lardo}, {Lewis}, {Monaco}, {Morbidelli}, {Sacco}, {Sbordone}, {Worley}, \& {Zaggia}}]{Magrini17}
{Magrini}, L., {Randich}, S., {Kordopatis}, G., {et~al.} 2017, \aap, 603, A2, \dodoi{10.1051/0004-6361/201630294}

\bibitem[{{Majewski} {et~al.}(2017){Majewski}, {Schiavon}, {Frinchaboy}, {Allende Prieto}, {Barkhouser}, {Bizyaev}, {Blank}, {Brunner}, {Burton}, {Carrera}, {Chojnowski}, {Cunha}, {Epstein}, {Fitzgerald}, {Garc{\'\i}a P{\'e}rez}, {Hearty}, {Henderson}, {Holtzman}, {Johnson}, {Lam}, {Lawler}, {Maseman}, {M{\'e}sz{\'a}ros}, {Nelson}, {Nguyen}, {Nidever}, {Pinsonneault}, {Shetrone}, {Smee}, {Smith}, {Stolberg}, {Skrutskie}, {Walker}, {Wilson}, {Zasowski}, {Anders}, {Basu}, {Beland}, {Blanton}, {Bovy}, {Brownstein}, {Carlberg}, {Chaplin}, {Chiappini}, {Eisenstein}, {Elsworth}, {Feuillet}, {Fleming}, {Galbraith-Frew}, {Garc{\'\i}a}, {Garc{\'\i}a-Hern{\'a}ndez}, {Gillespie}, {Girardi}, {Gunn}, {Hasselquist}, {Hayden}, {Hekker}, {Ivans}, {Kinemuchi}, {Klaene}, {Mahadevan}, {Mathur}, {Mosser}, {Muna}, {Munn}, {Nichol}, {O'Connell}, {Parejko}, {Robin}, {Rocha-Pinto}, {Schultheis}, {Serenelli}, {Shane}, {Silva Aguirre}, {Sobeck}, {Thompson}, {Troup}, {Weinberg}, \& {Zamora}}]{apogee}
{Majewski}, S.~R., {Schiavon}, R.~P., {Frinchaboy}, P.~M., {et~al.} 2017, \aj, 154, 94, \dodoi{10.3847/1538-3881/aa784d}

\bibitem[{{Manea} {et~al.}(2023){Manea}, {Hawkins}, {Ness}, {Buder}, {Martell}, \& {Zucker}}]{Manea23}
{Manea}, C., {Hawkins}, K., {Ness}, M.~K., {et~al.} 2023, arXiv e-prints, arXiv:2310.15257, \dodoi{10.48550/arXiv.2310.15257}

\bibitem[{{Martos} {et~al.}(2025){Martos}, {Mel{\'e}ndez}, {Spina}, \& {Lucatello}}]{Martos25}
{Martos}, G., {Mel{\'e}ndez}, J., {Spina}, L., \& {Lucatello}, S. 2025, \aap, 699, A46, \dodoi{10.1051/0004-6361/202554675}

\bibitem[{{Mead} {et~al.}(2025){Mead}, {De La Garza}, \& {Ness}}]{Mead25}
{Mead}, J., {De La Garza}, R., \& {Ness}, M. 2025, arXiv e-prints, arXiv:2504.18532.
\newblock \doarXiv{2504.18532}

\bibitem[{{Mel{\'e}ndez} {et~al.}(2009){Mel{\'e}ndez}, {Asplund}, {Gustafsson}, \& {Yong}}]{Melendez09}
{Mel{\'e}ndez}, J., {Asplund}, M., {Gustafsson}, B., \& {Yong}, D. 2009, \apjl, 704, L66, \dodoi{10.1088/0004-637X/704/1/L66}

\bibitem[{{Minchev} \& {Famaey}(2010)}]{Minchev10}
{Minchev}, I., \& {Famaey}, B. 2010, \apj, 722, 112, \dodoi{10.1088/0004-637X/722/1/112}

\bibitem[{{Minchev} {et~al.}(2018){Minchev}, {Anders}, {Recio-Blanco}, {Chiappini}, {de Laverny}, {Queiroz}, {Steinmetz}, {Adibekyan}, {Carrillo}, {Cescutti}, {Guiglion}, {Hayden}, {de Jong}, {Kordopatis}, {Majewski}, {Martig}, \& {Santiago}}]{Minchev18}
{Minchev}, I., {Anders}, F., {Recio-Blanco}, A., {et~al.} 2018, \mnras, 481, 1645, \dodoi{10.1093/mnras/sty2033}

\bibitem[{{Moe} {et~al.}(2019){Moe}, {Kratter}, \& {Badenes}}]{Moe19}
{Moe}, M., {Kratter}, K.~M., \& {Badenes}, C. 2019, \apj, 875, 61, \dodoi{10.3847/1538-4357/ab0d88}

\bibitem[{{Mulders} {et~al.}(2018){Mulders}, {Pascucci}, {Apai}, \& {Ciesla}}]{Mulders18}
{Mulders}, G.~D., {Pascucci}, I., {Apai}, D., \& {Ciesla}, F.~J. 2018, \aj, 156, 24, \dodoi{10.3847/1538-3881/aac5ea}

\bibitem[{{NASA Exoplanet Archive}(2023)}]{ps}
{NASA Exoplanet Archive}. 2023, Planetary Systems, Version: 2024-09-23,  NExScI-Caltech/IPAC, \dodoi{10.26133/NEA12}

\bibitem[{{Ness} {et~al.}(2015){Ness}, {Hogg}, {Rix}, {Ho}, \& {Zasowski}}]{cannon}
{Ness}, M., {Hogg}, D.~W., {Rix}, H.~W., {Ho}, A. Y.~Q., \& {Zasowski}, G. 2015, \apj, 808, 16, \dodoi{10.1088/0004-637X/808/1/16}

\bibitem[{{Ness} {et~al.}(2019){Ness}, {Johnston}, {Blancato}, {Rix}, {Beane}, {Bird}, \& {Hawkins}}]{Ness19}
{Ness}, M.~K., {Johnston}, K.~V., {Blancato}, K., {et~al.} 2019, \apj, 883, 177, \dodoi{10.3847/1538-4357/ab3e3c}

\bibitem[{{Ness} {et~al.}(2022){Ness}, {Wheeler}, {McKinnon}, {Horta}, {Casey}, {Cunningham}, \& {Price-Whelan}}]{Ness22}
{Ness}, M.~K., {Wheeler}, A.~J., {McKinnon}, K., {et~al.} 2022, \apj, 926, 144, \dodoi{10.3847/1538-4357/ac4754}

\bibitem[{{Nibauer} {et~al.}(2021){Nibauer}, {Baxter}, {Jain}, {Van Saders}, {Beaton}, \& {Teske}}]{Nibauer21}
{Nibauer}, J., {Baxter}, E.~J., {Jain}, B., {et~al.} 2021, \apj, 907, 116, \dodoi{10.3847/1538-4357/abd0f1}

\bibitem[{{Nissen}(2015)}]{Nissen15}
{Nissen}, P.~E. 2015, \aap, 579, A52, \dodoi{10.1051/0004-6361/201526269}

\bibitem[{{Nissen}(2016)}]{Nissen16}
---. 2016, \aap, 593, A65, \dodoi{10.1051/0004-6361/201628888}

\bibitem[{pandas~development team(2020)}]{reback2020pandas}
pandas~development team, T. 2020, pandas-dev/pandas: Pandas, latest,  Zenodo, \dodoi{10.5281/zenodo.3509134}

\bibitem[{Pedregosa {et~al.}(2011)Pedregosa, Varoquaux, Gramfort, Michel, Thirion, Grisel, Blondel, Prettenhofer, Weiss, Dubourg, Vanderplas, Passos, Cournapeau, Brucher, Perrot, \& Duchesnay}]{sklearn}
Pedregosa, F., Varoquaux, G., Gramfort, A., {et~al.} 2011, Journal of Machine Learning Research, 12, 2825

\bibitem[{{Piskunov} \& {Valenti}(2017)}]{SME2}
{Piskunov}, N., \& {Valenti}, J.~A. 2017, \aap, 597, A16, \dodoi{10.1051/0004-6361/201629124}

\bibitem[{{Pr{\v{s}}a} {et~al.}(2016){Pr{\v{s}}a}, {Harmanec}, {Torres}, {Mamajek}, {Asplund}, {Capitaine}, {Christensen-Dalsgaard}, {Depagne}, {Haberreiter}, {Hekker}, {Hilton}, {Kopp}, {Kostov}, {Kurtz}, {Laskar}, {Mason}, {Milone}, {Montgomery}, {Richards}, {Schmutz}, {Schou}, \& {Stewart}}]{Prvsa16}
{Pr{\v{s}}a}, A., {Harmanec}, P., {Torres}, G., {et~al.} 2016, \aj, 152, 41, \dodoi{10.3847/0004-6256/152/2/41}

\bibitem[{{Ram{\'\i}rez} {et~al.}(2009){Ram{\'\i}rez}, {Mel{\'e}ndez}, \& {Asplund}}]{Ramirez09}
{Ram{\'\i}rez}, I., {Mel{\'e}ndez}, J., \& {Asplund}, M. 2009, \aap, 508, L17, \dodoi{10.1051/0004-6361/200913038}

\bibitem[{{Rampalli} {et~al.}(2021){Rampalli}, {Ness}, \& {Wylie}}]{Rampalli21}
{Rampalli}, R., {Ness}, M., \& {Wylie}, S. 2021, \apj, 921, 78, \dodoi{10.3847/1538-4357/ac1ac8}

\bibitem[{{Rampalli} {et~al.}(2024){Rampalli}, {Ness}, {Edwards}, {Newton}, \& {Bedell}}]{Rampalli24}
{Rampalli}, R., {Ness}, M.~K., {Edwards}, G.~H., {Newton}, E.~R., \& {Bedell}, M. 2024, \apj, 965, 176, \dodoi{10.3847/1538-4357/ad303e}

\bibitem[{{Rampalli} {et~al.}(2025){Rampalli}, {Ness}, {Newton}, {Vanderburg}, {Buck}, \& {Mills}}]{Rampalli25}
{Rampalli}, R., {Ness}, M.~K., {Newton}, E.~R., {et~al.} 2025, arXiv e-prints, arXiv:2506.16511, \dodoi{10.48550/arXiv.2506.16511}

\bibitem[{{Ratcliffe} {et~al.}(2023){Ratcliffe}, {Minchev}, {Anders}, {Khoperskov}, {Guiglion}, {Buck}, {Cunha}, {Queiroz}, {Nitschelm}, {Meszaros}, {Steinmetz}, {de Jong}, {Nepal}, {Lane}, \& {Sobeck}}]{Ratcliffe23}
{Ratcliffe}, B., {Minchev}, I., {Anders}, F., {et~al.} 2023, \mnras, 525, 2208, \dodoi{10.1093/mnras/stad1573}

\bibitem[{{Recio-Blanco} {et~al.}(2016){Recio-Blanco}, {de Laverny}, {Allende Prieto}, {Fustes}, {Manteiga}, {Arcay}, {Bijaoui}, {Dafonte}, {Ordenovic}, \& {Ordo{\~n}ez Blanco}}]{recio-blanco}
{Recio-Blanco}, A., {de Laverny}, P., {Allende Prieto}, C., {et~al.} 2016, \aap, 585, A93, \dodoi{10.1051/0004-6361/201425030}

\bibitem[{{Ro{\v{s}}kar} {et~al.}(2008){Ro{\v{s}}kar}, {Debattista}, {Quinn}, {Stinson}, \& {Wadsley}}]{Roskar08}
{Ro{\v{s}}kar}, R., {Debattista}, V.~P., {Quinn}, T.~R., {Stinson}, G.~S., \& {Wadsley}, J. 2008, \apjl, 684, L79, \dodoi{10.1086/592231}

\bibitem[{{Rybizki} {et~al.}(2017){Rybizki}, {Just}, \& {Rix}}]{Rybizki17}
{Rybizki}, J., {Just}, A., \& {Rix}, H.-W. 2017, \aap, 605, A59, \dodoi{10.1051/0004-6361/201730522}

\bibitem[{{Sellwood} \& {Binney}(2002)}]{SellwoodBinney}
{Sellwood}, J.~A., \& {Binney}, J.~J. 2002, \mnras, 336, 785, \dodoi{10.1046/j.1365-8711.2002.05806.x}

\bibitem[{{Sneden}(1973)}]{MOOG}
{Sneden}, C. 1973, \apj, 184, 839, \dodoi{10.1086/152374}

\bibitem[{{Soderblom}(2010)}]{soderblom}
{Soderblom}, D.~R. 2010, \araa, 48, 581, \dodoi{10.1146/annurev-astro-081309-130806}

\bibitem[{{Spina} {et~al.}(2018){Spina}, {Mel{\'e}ndez}, {Karakas}, {dos Santos}, {Bedell}, {Asplund}, {Ram{\'\i}rez}, {Yong}, {Alves-Brito}, {Bean}, \& {Dreizler}}]{Spina18}
{Spina}, L., {Mel{\'e}ndez}, J., {Karakas}, A.~I., {et~al.} 2018, \mnras, 474, 2580, \dodoi{10.1093/mnras/stx2938}

\bibitem[{{Sun} {et~al.}(2025){Sun}, {Ji}, {Xuesong Wang}, {Lin}, {Teske}, {Ting}, {Bedell}, \& {Liu}}]{Sun25}
{Sun}, Q., {Ji}, C., {Xuesong Wang}, S., {et~al.} 2025, arXiv e-prints, arXiv:2508.06976, \dodoi{10.48550/arXiv.2508.06976}

\bibitem[{{Sun} {et~al.}(2024){Sun}, {Xuesong Wang}, {Gan}, {Ji}, {Lin}, {Ting}, {Teske}, {Li}, {Liu}, {Hua}, {Tang}, {Yu}, {Zhang}, {Badenas-Agusti}, {Vanderburg}, {Ricker}, {Vanderspek}, {Latham}, {Seager}, {Jenkins}, {Schwarz}, {Guillot}, {Tan}, {Conti}, {Collins}, {Srdoc}, {Stockdale}, {Suarez}, {Zambelli}, {Radford}, {Barkaoui}, {Evans}, \& {Bieryla}}]{Sun24}
{Sun}, Q., {Xuesong Wang}, S., {Gan}, T., {et~al.} 2024, arXiv e-prints, arXiv:2411.13825, \dodoi{10.48550/arXiv.2411.13825}

\bibitem[{{Taylor}(2005)}]{topcat}
{Taylor}, M.~B. 2005, in Astronomical Society of the Pacific Conference Series, Vol. 347, Astronomical Data Analysis Software and Systems XIV, ed. P.~{Shopbell}, M.~{Britton}, \& R.~{Ebert}, 29

\bibitem[{{Thielemann} {et~al.}(2002){Thielemann}, {Argast}, {Brachwitz}, {Martinez-Pinedo}, {Rauscher}, {Liebend{\"o}rfer}, {Mezzacappa}, {H{\"o}flich}, \& {Nomoto}}]{Thielemann02}
{Thielemann}, F.~K., {Argast}, D., {Brachwitz}, F., {et~al.} 2002, \apss, 281, 25, \dodoi{10.1023/A:1019543110473}

\bibitem[{{Valenti} \& {Piskunov}(1996)}]{SME1}
{Valenti}, J.~A., \& {Piskunov}, N. 1996, \aaps, 118, 595

\bibitem[{Virtanen {et~al.}(2020)Virtanen, Gommers, Oliphant, Haberland, Reddy, Cournapeau, Burovski, Peterson, Weckesser, Bright, {van der Walt}, Brett, Wilson, Millman, Mayorov, Nelson, Jones, Kern, Larson, Carey, Polat, Feng, Moore, {VanderPlas}, Laxalde, Perktold, Cimrman, Henriksen, Quintero, Harris, Archibald, Ribeiro, Pedregosa, {van Mulbregt}, \& {SciPy 1.0 Contributors}}]{scipy}
Virtanen, P., Gommers, R., Oliphant, T.~E., {et~al.} 2020, Nature Methods, 17, 261, \dodoi{10.1038/s41592-019-0686-2}

\bibitem[{{Wang} {et~al.}(2024){Wang}, {Carrillo}, {Ness}, \& {Buck}}]{Wang2024}
{Wang}, K., {Carrillo}, A., {Ness}, M.~K., \& {Buck}, T. 2024, \mnras, 527, 321, \dodoi{10.1093/mnras/stad3182}

\bibitem[{{Weinberg} {et~al.}(2019){Weinberg}, {Holtzman}, {Hasselquist}, {Bird}, {Johnson}, {Shetrone}, {Sobeck}, {Allende Prieto}, {Bizyaev}, {Carrera}, {Cohen}, {Cunha}, {Ebelke}, {Fernandez-Trincado}, {Garc{\'\i}a-Hern{\'a}ndez}, {Hayes}, {J{\"o}nsson}, {Lane}, {Majewski}, {Malanushenko}, {M{\'e}sz{\'a}ros}, {Nidever}, {Nitschelm}, {Pan}, {Rix}, {Rybizki}, {Schiavon}, {Schneider}, {Wilson}, \& {Zamora}}]{Weinberg19}
{Weinberg}, D.~H., {Holtzman}, J.~A., {Hasselquist}, S., {et~al.} 2019, \apj, 874, 102, \dodoi{10.3847/1538-4357/ab07c7}

\bibitem[{{Weinberg} {et~al.}(2022){Weinberg}, {Holtzman}, {Johnson}, {Hayes}, {Hasselquist}, {Shetrone}, {Ting}, {Beaton}, {Beers}, {Bird}, {Bizyaev}, {Blanton}, {Cunha}, {Fern{\'a}ndez-Trincado}, {Frinchaboy}, {Garc{\'\i}a-Hern{\'a}ndez}, {Griffith}, {Johnson}, {J{\"o}nsson}, {Lane}, {Leung}, {Mackereth}, {Majewski}, {M{\'e}sz{\'a}ros}, {Nitschelm}, {Pan}, {Schiavon}, {Schneider}, {Schultheis}, {Smith}, {Sobeck}, {Stassun}, {Stringfellow}, {Vincenzo}, {Wilson}, \& {Zasowski}}]{Weinberg22}
{Weinberg}, D.~H., {Holtzman}, J.~A., {Johnson}, J.~A., {et~al.} 2022, \apjs, 260, 32, \dodoi{10.3847/1538-4365/ac6028}

\bibitem[{{W}es {M}c{K}inney(2010)}]{mckinney-proc-scipy-2010}
{W}es {M}c{K}inney. 2010, in {P}roceedings of the 9th {P}ython in {S}cience {C}onference, ed. {S}t\'efan van~der {W}alt \& {J}arrod {M}illman, 56 -- 61, \dodoi{10.25080/Majora-92bf1922-00a}

\bibitem[{{Wiseman} {et~al.}(2021){Wiseman}, {Sullivan}, {Smith}, {Frohmaier}, {Vincenzi}, {Graur}, {Popovic}, {Armstrong}, {Brout}, {Davis}, {Galbany}, {Hinton}, {Kelsey}, {Kessler}, {Lidman}, {M{\"o}ller}, {Nichol}, {Rose}, {Scolnic}, {Toy}, {Zontou}, {Asorey}, {Carollo}, {Glazebrook}, {Lewis}, {Tucker}, {Abbott}, {Aguena}, {Allam}, {Andrade-Oliveira}, {Annis}, {Bacon}, {Bertin}, {Brooks}, {Buckley-Geer}, {Burke}, {Carnero Rosell}, {Carrasco Kind}, {Carretero}, {Costanzi}, {da Costa}, {Pereira}, {Desai}, {Diehl}, {Doel}, {Everett}, {Ferrero}, {Flaugher}, {Fosalba}, {Frieman}, {Garc{\'\i}a-Bellido}, {Gaztanaga}, {Giannantonio}, {Gruen}, {Gruendl}, {Gschwend}, {Gutierrez}, {Hollowood}, {Honscheid}, {Hoyle}, {James}, {Krause}, {Kuehn}, {Kuropatkin}, {Maia}, {Marshall}, {Martini}, {Menanteau}, {Miquel}, {Morgan}, {Ogando}, {Palmese}, {Paz-Chinch{\'o}n}, {Petravick}, {Pieres}, {Plazas Malag{\'o}n}, {Romer}, {Sanchez}, {Scarpine}, {Schubnell}, {Serrano}, {Sevilla-Noarbe}, {Soares-Santos}, {Suchyta}, {Swanson},
  {Tarle}, {Thomas}, {To}, {Varga}, {Walker}, \& {DES Collaboration}}]{Wiseman21}
{Wiseman}, P., {Sullivan}, M., {Smith}, M., {et~al.} 2021, \mnras, 506, 3330, \dodoi{10.1093/mnras/stab1943}

\bibitem[{{Woosley} \& {Weaver}(1995)}]{Woosley95}
{Woosley}, S.~E., \& {Weaver}, T.~A. 1995, \apjs, 101, 181, \dodoi{10.1086/192237}

\bibitem[{{Zink} {et~al.}(2019){Zink}, {Christiansen}, \& {Hansen}}]{Zink19}
{Zink}, J.~K., {Christiansen}, J.~L., \& {Hansen}, B. M.~S. 2019, \mnras, 483, 4479, \dodoi{10.1093/mnras/sty3463}

\end{thebibliography}
\begin{longtable*}{lllcl}
\caption{Catalog Parameters for Solar Analogs Used in this Work} \label{t:sas} \\

\hline
\hline
\colhead{Column} & \colhead{Format} & \colhead{Units} & \colhead{Example} & \colhead{Description} \\
\hline
\endfirsthead

\multicolumn{5}{c}{{\tablename\ \thetable{} -- Continued}} \\
\hline
\colhead{Column} & \colhead{Format} & \colhead{Units} & \colhead{Example} & \colhead{Description} \\
\hline
\endhead

\hline
\multicolumn{5}{r}{{Continued on next page}} \\
\endfoot

\hline
\endlastfoot

\sidehead{\textit{Identifiers:}} 
ID    & string & \nodata & HIP10175 & Unique star identifier \\
prov    & string & \nodata & HARPS (B18) & Survey source \\
planet\_host? & integer & \nodata & 0 & Planet host flag (1=Yes, 0=No) \\

\sidehead{\textit{Kinematics:}} 
ra      & float & degrees & 32.717524 & Right ascension \\
dec     & float & degrees & 13.682958 & Declination \\
parallax     & float & mas & 23.3 & Parallax \\

\sidehead{\textit{Fundamental Stellar Parameters:}} 
teff      & float & K & 5719 & Effective temperature \\
e\_teff   & float & K & 3 & Uncertainty in Teff \\
logg     & float & dex & 4.49 & Surface gravity \\
e\_logg  & float & dex & 0.01 & Uncertainty in Logg \\
Fe\_h     & float & dex & -0.028 & Iron abundance [Fe/H] \\
e\_Fe\_h  & float & dex & 0.002 & Uncertainty in [Fe/H] \\
alpha\_fe  & float & dex & 0.039 & Alpha-element enhancement [\(\alpha\)/Fe] \\
alpha\_fe\_err  & float & dex & 0.017 & Uncertainty in [\(\alpha\)/Fe] \\

\sidehead{\textit{Abundance Data:}} 
O\_fe    & float & dex    & 0.052 & Oxygen abundance [O/Fe] \\
e\_O\_fe & float & dex    & 0.009 & Uncertainty in [O/Fe] \\
Na\_fe   & float & dex    & -0.023 & Sodium abundance [Na/Fe] \\
e\_Na\_fe & float & dex    & 0.022 & Uncertainty in [Na/Fe] \\
Mg\_fe   & float & dex    & 0.022 & Magnesium abundance [Mg/Fe] \\
e\_Mg\_fe & float & dex    & 0.007 & Uncertainty in [Mg/Fe] \\
Al\_fe   & float & dex    & -0.006 & Aluminum abundance [Al/Fe] \\
e\_Al\_fe & float & dex    & 0.008 & Uncertainty in [Al/Fe] \\
Si\_fe   & float & dex    & 0.018 & Silicon abundance [Si/Fe] \\
e\_Si\_fe & float & dex    & 0.004 & Uncertainty in [Si/Fe] \\
Ca\_fe   & float & dex    & 0.051 & Calcium abundance [Ca/Fe] \\
e\_Ca\_fe & float & dex    & 0.005 & Uncertainty in [Ca/Fe] \\
Ti\_fe   & float & dex    & 0.051 & Titanium abundance [Ti/Fe] \\
e\_Ti\_fe & float & dex    & 0.005 & Uncertainty in [Ti/Fe] \\
V\_fe    & float & dex    & 0.026 & Vanadium abundance [V/Fe] \\
e\_V\_fe & float & dex    & 0.004 & Uncertainty in [V/Fe] \\
Cr\_fe   & float & dex    & 0.041 & Chromium abundance [Cr/Fe] \\
e\_Cr\_fe & float & dex    & 0.005 & Uncertainty in [Cr/Fe] \\
Mn\_fe   & float & dex    & -0.001 & Manganese abundance [Mn/Fe] \\
e\_Mn\_fe & float & dex    & 0.005 & Uncertainty in [Mn/Fe] \\
Ni\_fe   & float & dex    & -0.013 & Nickel abundance [Ni/Fe] \\
e\_Ni\_fe & float & dex    & 0.003 & Uncertainty in [Ni/Fe] \\

\sidehead{\textit{Inferred Parameters:}} 
slope      & float & dex/K & 0.000073 & Tcond slope \\
slope\_err  & float & dex/K & 0.000055 & slope unc. \\
slope\_noalpha    & float & dex/K & 0.000073 & Tcond slope (no $\alpha$) \\
slope\_noalpha\_err & float & dex/K & 0.000055 & unc. (no $\alpha$) \\
intc  & float & dex & -0.086161 & intercept \\
intc\_err  & float & dex & 0.078398 & intercept unc. \\
intc\_noalpha  & float & dex & -0.003374 & intercept (no $\alpha$) \\
intc\_noalpha\_err  & float & dex & 0.113942 & unc. (no $\alpha$) \\
Acc        & float & \nodata & 0.89 & CC amp \\
AIa        & float & \nodata & 0.98 & Ia amp \\

fcc\_O     & float & \nodata & 0.72 & CC frac (O) \\
fcc\_Na    & float & \nodata & 0.37 & CC frac (Na) \\
fcc\_Mg    & float & \nodata & 0.99 & CC frac (Mg) \\
fcc\_Al    & float & \nodata & 0.97 & CC frac (Al) \\
fcc\_Si    & float & \nodata & 0.68 & CC frac (Si) \\
fcc\_Ca    & float & \nodata & 0.46 & CC frac (Ca) \\
fcc\_Ti    & float & \nodata & 0.47 & CC frac (Ti) \\
fcc\_V     & float & \nodata & 0.58 & CC frac (V) \\
fcc\_Cr    & float & \nodata & 0.24 & CC frac (Cr) \\
fcc\_Mn    & float & \nodata & 0.25 & CC frac (Mn) \\
fcc\_Ni    & float & \nodata & 0.47 & CC frac (Ni) \\
fcc\_Fe    & float & \nodata & 0.55 & CC frac (Fe) \\

qcc\_O     & float & \nodata & 0.9 & CC yield (O) \\
qcc\_Na    & float & \nodata & 0.4 & CC yield (Na) \\
qcc\_Mg    & float & \nodata & 1.0 & CC yield (Mg) \\
qcc\_Al    & float & \nodata & 1.0 & CC yield (Al) \\
qcc\_Si    & float & \nodata & 0.7 & CC yield (Si) \\
qcc\_Ca    & float & \nodata & 0.5 & CC yield (Ca) \\
qcc\_Ti    & float & \nodata & 0.5 & CC yield (Ti) \\
qcc\_V     & float & \nodata & 0.6 & CC yield (V) \\
qcc\_Cr    & float & \nodata & 0.3 & CC yield (Cr) \\
qcc\_Mn    & float & \nodata & 0.3 & CC yield (Mn) \\
qcc\_Ni    & float & \nodata & 0.5 & CC yield (Ni) \\
qcc\_Fe    & float & \nodata & 0.4 & CC yield (Fe) \\

qia\_O     & float & \nodata & 0.3 & Ia yield (O) \\
qia\_Na    & float & \nodata & 0.6 & Ia yield (Na) \\
qia\_Mg    & float & \nodata & 0.0001 & Ia yield (Mg) \\
qia\_Al    & float & \nodata & 0.005 & Ia yield (Al) \\
qia\_Si    & float & \nodata & 0.3 & Ia yield (Si) \\
qia\_Ca    & float & \nodata & 0.5 & Ia yield (Ca) \\
qia\_Ti    & float & \nodata & 0.5 & Ia yield (Ti) \\
qia\_V     & float & \nodata & 0.4 & Ia yield (V) \\
qia\_Cr    & float & \nodata & 0.7 & Ia yield (Cr) \\
qia\_Mn    & float & \nodata & 0.8 & Ia yield (Mn) \\
qia\_Ni    & float & \nodata & 0.5 & Ia yield (Ni) \\
qia\_Fe    & float & \nodata & 0.6 & Ia yield (Fe) \\

\end{longtable*}

\begin{longtable}{lllcl}
\caption{KPM Results for Sun by Survey} \label{t:suns} \\

\hline
\hline
\colhead{Column} & \colhead{Format} & \colhead{Units} & \colhead{Example} & \colhead{Description} \\
\hline
\endfirsthead

\multicolumn{5}{c}{{\tablename\ \thetable{} -- Continued}} \\
\hline
\colhead{Column} & \colhead{Format} & \colhead{Units} & \colhead{Example} & \colhead{Description} \\
\hline
\endhead

\hline
\multicolumn{5}{r}{{Continued on next page}} \\
\endfoot

\hline
\endlastfoot

\sidehead{\textit{Identifiers:}} 
prov    & string & \nodata & HARPS (B18) & survey \\

\sidehead{\textit{Inferred Parameters:}} 

Acc        & float & \nodata & 0.89 & CC amp \\
AIa        & float & \nodata & 0.98 & Ia amp \\
fcc\_O     & float & \nodata & 0.72 & CC frac (O) \\
fcc\_Na    & float & \nodata & 0.37 & CC frac (Na) \\
fcc\_Mg    & float & \nodata & 0.99 & CC frac (Mg) \\
fcc\_Al    & float & \nodata & 0.97 & CC frac (Al) \\
fcc\_Si    & float & \nodata & 0.68 & CC frac (Si) \\
fcc\_Ca    & float & \nodata & 0.46 & CC frac (Ca) \\
fcc\_Ti    & float & \nodata & 0.47 & CC frac (Ti) \\
fcc\_V     & float & \nodata & 0.58 & CC frac (V) \\
fcc\_Cr    & float & \nodata & 0.24 & CC frac (Cr) \\
fcc\_Mn    & float & \nodata & 0.25 & CC frac (Mn) \\
fcc\_Ni    & float & \nodata & 0.47 & CC frac (Ni) \\
fcc\_Fe    & float & \nodata & 0.55 & CC frac (Fe) \\

qcc\_O     & float & \nodata & 0.9 & CC yield (O) \\
qcc\_Na    & float & \nodata & 0.4 & CC yield (Na) \\
qcc\_Mg    & float & \nodata & 1.0 & CC yield (Mg) \\
qcc\_Al    & float & \nodata & 1.0 & CC yield (Al) \\
qcc\_Si    & float & \nodata & 0.7 & CC yield (Si) \\
qcc\_Ca    & float & \nodata & 0.5 & CC yield (Ca) \\
qcc\_Ti    & float & \nodata & 0.5 & CC yield (Ti) \\
qcc\_V     & float & \nodata & 0.6 & CC yield (V) \\
qcc\_Cr    & float & \nodata & 0.3 & CC yield (Cr) \\
qcc\_Mn    & float & \nodata & 0.3 & CC yield (Mn) \\
qcc\_Ni    & float & \nodata & 0.5 & CC yield (Ni) \\
qcc\_Fe    & float & \nodata & 0.4 & CC yield (Fe) \\

qia\_O     & float & \nodata & 0.3 & Ia yield (O) \\
qia\_Na    & float & \nodata & 0.6 & Ia yield (Na) \\
qia\_Mg    & float & \nodata & 0.0001 & Ia yield (Mg) \\
qia\_Al    & float & \nodata & 0.005 & Ia yield (Al) \\
qia\_Si    & float & \nodata & 0.3 & Ia yield (Si) \\
qia\_Ca    & float & \nodata & 0.5 & Ia yield (Ca) \\
qia\_Ti    & float & \nodata & 0.5 & Ia yield (Ti) \\
qia\_V     & float & \nodata & 0.4 & Ia yield (V) \\
qia\_Cr    & float & \nodata & 0.7 & Ia yield (Cr) \\
qia\_Mn    & float & \nodata & 0.8 & Ia yield (Mn) \\
qia\_Ni    & float & \nodata & 0.5 & Ia yield (Ni) \\
qia\_Fe    & float & \nodata & 0.6 & Ia yield (Fe) \\

\end{longtable}

\appendix
\renewcommand{\thesection}{\Alph{section}} 
\renewcommand{\thesubsection}{\thesection.\arabic{subsection}} 

\renewcommand{\thefigure}{\thesection.\arabic{figure}} 
\renewcommand{\thetable}{\thesection.\arabic{table}}  

\counterwithin{figure}{section}
\counterwithin{table}{section}

\section{Element Families' \fcc\ as a function of \tcond} \label{sec:fcc-explore}

Motivated by the positive correlation of elements' \fcc\ as a function of \tcond\ shown in Figure \ref{fig:fcc}, we see if elements with different nucleosynthetic origins behave differently from each other in this space. We use a subset of the solar analogs solely from GALAH with a signal-to-noise $> 100$ and consider additional elements beyond those used in the main text: O, K, Co, Ba, Ce, Y, and Sc (182, 1006, 1352, 1447, 1477, 1647, 1659 K respectively). Of the 11 refractory elements we consider  (Na, Mn, Cr, Si, Fe, Ni, Mg, V, Ca, Ti, and Al), 5 are iron-peak elements. Si, Ca, Ti, and Mg are associated with $\alpha$-process nucleosynthetic origins but have contributions from Type Ia supernovae, giving them abundance patterns resembling iron-peak elements. Na and Al are odd Z elements. As such, the linear trend observed in Figure~\ref{fig:fcc} may be primarily driven by the iron-peak elements. The additional elements we include are 2 iron-peak elements (Co, Sc), O as an $\alpha$-process element, and 3 s-process elements (Y - first peak, Ce and Ba - second peak).

In Figure~\ref{fig:fcc_explore}, we show the distribution of elemental \fcc\ values as a function of \tcond, grouped by nucleosynthetic family. Iron-peak elements are shown in blue circles, $\alpha$-process elements in magenta squares, s-process elements in green diamonds, and odd Z elements in gray triangles. The \fcc\ values shown here were inferred separately using the data-driven KPM applied only to the GALAH stars, and therefore do not exactly match those shown in Figure~\ref{fig:fcc}, which are averaged across all the samples. 

To test whether the iron-peak elements’ linear trend is statistically significant, we bootstrap-sampled synthetic realizations of \fcc\ using each element’s standard deviation and fit linear models to the resampled data. The resulting slope distribution for the iron-peak group yields a median value of $6\pm3$E$-4$ K$^{-1}$, consistent with a weak but clear linear increase in \fcc\ with \tcond. Since the general increase of \fcc\ with \tcond\ is what gives rise to the observed refractory-\tcond\ slopes in the Sun and other stars (shown in Figure \ref{fig:prediction}), this result suggests that such trends are primarily driven by the iron-peak elements rather than other nucleosynthetic families. 

However, these trends are more easily interpreted if one moves beyond such classifications. Specifically, whether or not an element follows this trend in \fcc\ with \tcond\ appears to loosely trace proton number. Light nuclei are dominated by massive star nucleosynthesis, sitting near \fcc\ = 1 by definition, and span a broad range of \tcond. Heavier nuclei have significant contributions from Type Ia supernovae and follow this correlation. The transition occurs around the proton number of Si. S is a noteworthy exception, which has two more protons than Si but sits near \fcc\ = 1 based on previous results \citep[e.g.,][]{Weinberg19} and at \tcond\ = 704 K. Several odd-Z elements with lighter nuclei also appear to follow this trend (Na, K, and Al), though we only have three of them, each of which has been identified as having peculiar abundance distributions, metallicity-dependent yields, or systematic issues in abundance determination. At yet heavier nuclei, Type Ia supernovae cease to be an important contributor, and neutron capture nucleosynthesis becomes signficant. These elements do not show a coherent trend in \fcc\ with \tcond\ in Figure \ref{fig:fcc_explore}, but we only have data for these elements.  
A deeper interpretation of these trends, including their connection to nucleosynthetic yield models and Galactic chemical evolution, is beyond the scope of this paper and is left for future work.

\begin{figure}
    \centering
    \includegraphics[width=\linewidth]{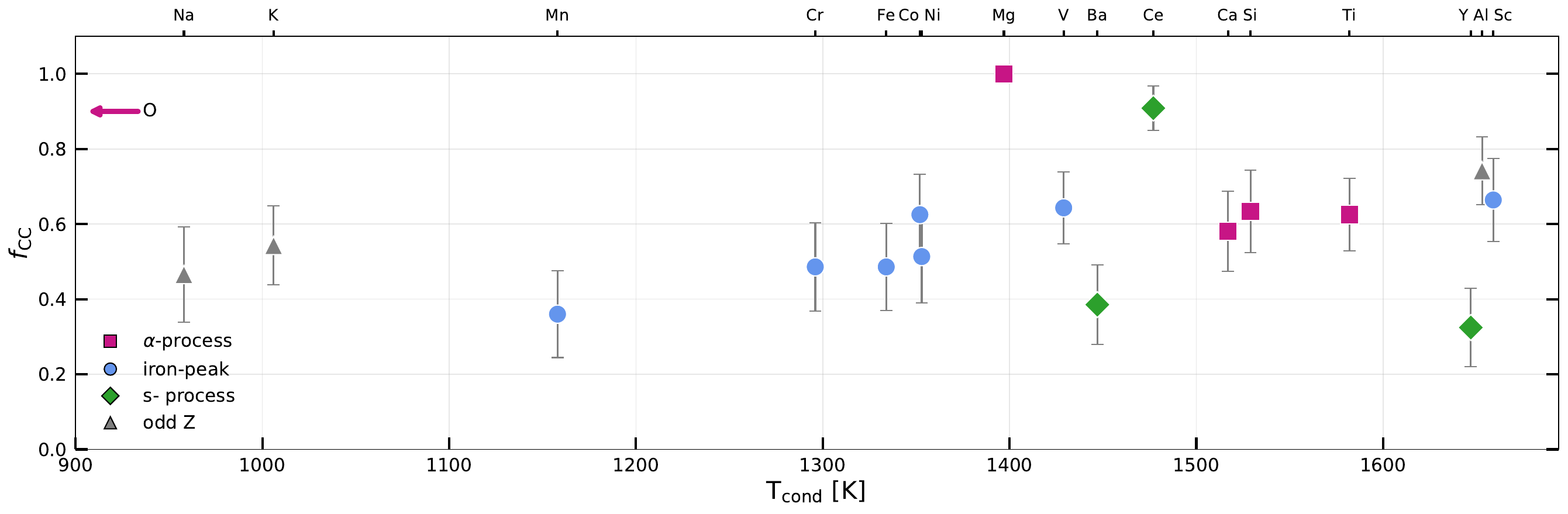}
    \caption{Mean values of \fcc\ for individual elements as a function of \tcond, color-coded by nucleosynthetic origin: iron-peak (blue circles), $\alpha$-process (magenta squares), s-process (green diamonds), and other odd-$Z$ elements (gray triangles). Error bars indicate the standard deviation of \fcc\ for each element across the GALAH solar analog sample. We fit linear models to the iron-peak group to test whether their trend underlies the observed linear refractory-\tcond\ slopes in stellar abundance patterns and find a weak but clear trend. The other groups contain too few elements to support robust trends, but show either weak decreases or element-to-element scatter that differs from the iron-peak group. We do note that the $\alpha$-process elements do also receive significant enrichment from Type Ia supernovae, like the iron-peak elements, and thus also ultimately contribute to the overall refractory–\tcond\ trend. The oxygen point is indicated by a leftward magenta arrow, reflecting its high \fcc\ value at low \tcond\ as an $\alpha$-process element beyond the plot range.}

    \label{fig:fcc_explore}
\end{figure}

\section{Contextualizing the Sun in the Low Dimensionality of Chemical Abundance Space} \label{sec:dimensionality}
We can also consider the relative uniqueness of the Sun’s chemistry in the context of abundance dimensionality. This concept explores how many stellar parameters (or “labels”) are needed to explain the variance seen across stars’ elemental abundances. Here we ask, when we model stars’ abundances using a 6-label regression model, does the Sun emerge as an outlier or fall within the expected variance of the population? Following Appendix \ref{sec:fcc-explore}, we solely use the GALAH solar analogs with a signal-to-noise $> 100$ and consider four additional elements beyond those used among all four surveys in the earlier analysis (O, Ba, Ce, Y). We build a set of regression models following Equation \ref{eqn:lr} using \teff, \logg, \feh, \mg, [Y/Fe], and [Ce/Fe] to predict the stars’ other 11 element abundances and \tcond\ slopes. These labels are motivated by \cite{Mead25}, who select a set of labels that capture the evolutionary state of the star (\teff, \logg) and key abundances that represent nucleosynthetic channels: iron-peak, alpha, and first- and second-peak s-process elements. We also use these models to predict the 11 other abundances and \tcond\ slope for the Sun. We show the observed versus model predicted abundances for each element ordered by its \tcond\ in Figure \ref{fig:regressions}.  The predicted versus observed abundances for all 11 elements show tight correlations, with residual scatter typically below 0.05 dex. The Sun’s element abundances are not always exactly perfectly predicted but fall consistently within the scatter of the solar analog population for each element. This once again shows that the Sun’s chemistry is not anomalous but falls within the statistically expected distribution in abundance space in addition to \acc\ and \aia\ space and \feh\ and \alphafe\ space as shown in Figures \ref{fig:fehalphafe} and \ref{fig:alphacomp}. 

We also use same six labels to predict  stars \tcond\ slope. We find we can perfectly predict Sun’s \tcond\ slope within uncertainties and find an overall \rsq\ of 73\% using these 6 labels of \teff, \logg, \feh, \mg, [Y/Fe], and [Ce/Fe] compared to \rsq\ of 49\% found when just using \teff, \logg, \feh, and \alphafe\ in Section \ref{sec:tcondtrends}. This suggests that the addition of first and second-peak s process elements (Y, Ce) are helpful in modeling the variance of \tcond\ slope because they represent additional nucleosynthetic enrichment sources.

When examining the residuals in Figure \ref{fig:regressions} as a function of \tcond\ slope, we find elements with higher \tcond, and thus larger core-collapse contributions (as shown in Figure \ref{fig:fcc}), like Ti, Al, and Ca, tend to be underpredicted in the refractory-enriched stars, or with high \tcond\ slopes. Elements with lower \tcond\ that are less core-collapse supernovae dominated, like Mn and Na, are underpredicted for the refractory-depleted solar analog population. We posit that \tcond\ slope may be correlated with an underlying enrichment axis that is not fully captured by the six input labels, despite their strong overall performance. To test whether this residual structure reflects nucleosynthetic origin, we compare the current regression model using representative abundances, Fe and Mg, to a regression model that uses \acc\ and \aia\, the more physically-motivated enrichment parameters that more fully model core-collapse and Type Ia supernova enrichment for reasons we discuss in Section \ref{sec:kpm_sun}. We find that the latter model yields a higher \rsq\ (84\%) than the standard abundance model (73\%), indicating that these enrichment parameters better capture the structure encoded in \tcond\ slope. This result reinforces the idea that \tcond\ slope carries signatures of stellar chemical evolution tied to nucleosynthetic history. The fact that the \rsq\ increases when using \acc\ and \aia\ suggests that these are more accurate descriptions of nucleosynthesis than traditional proxies. However, residual trends with \acc\ and \aia\ do persist for several elements, indicating that while they represent a more physical enrichment basis, they still do not fully encode the complexity of stellar nucleosynthesis.

\begin{figure}
    \centering
    \includegraphics[width=\linewidth]{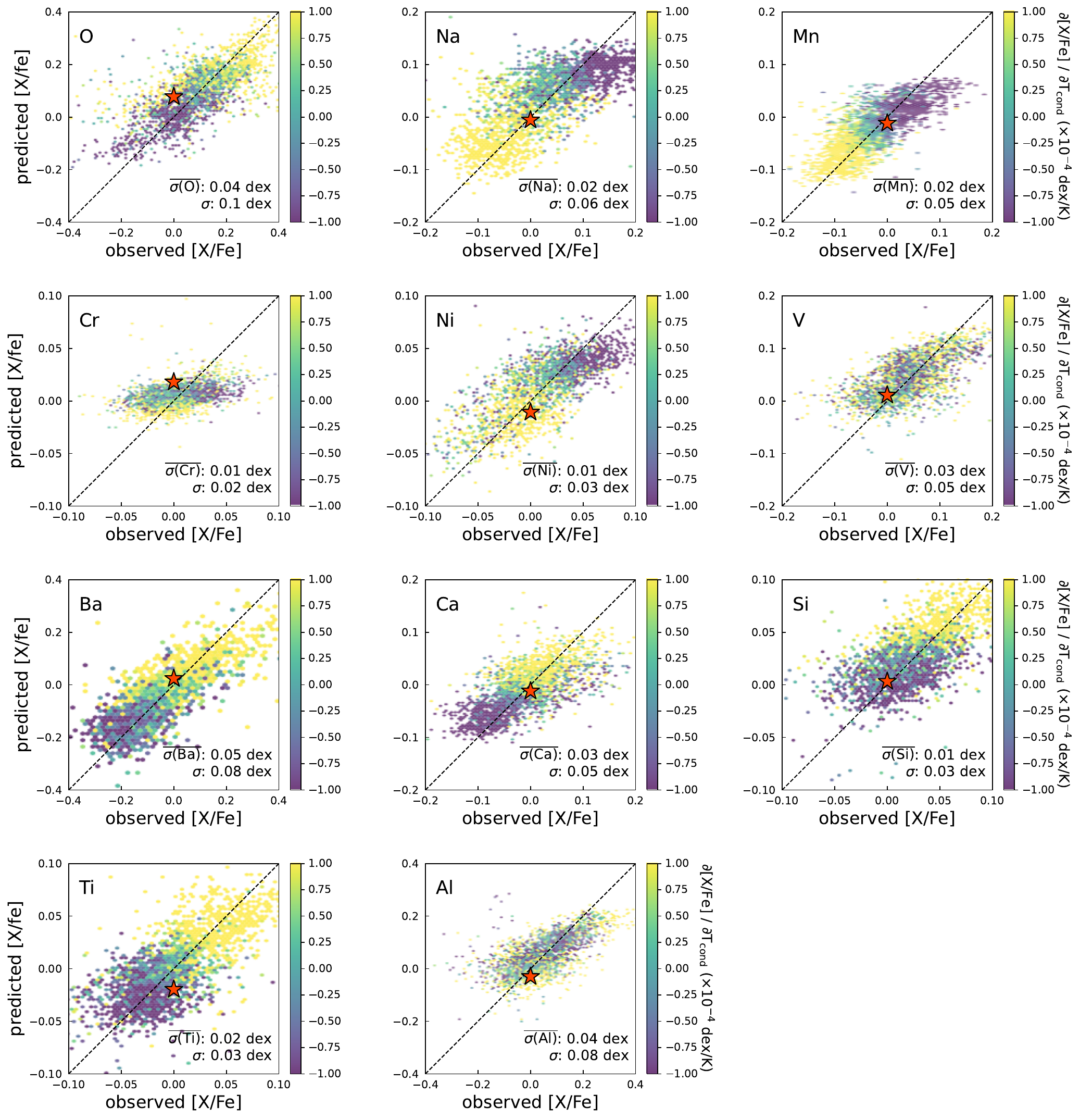}
    \caption{Observed [X/Fe] versus predicted [X/Fe] from 6 label regression model (\teff, \logg, [Fe/H], [Mg/Fe], [Y/Fe], [Ce/Fe]) for GALAH solar analogs. Panels are ordered by each element’s condensation temperature (\tcond) and colored by stars' \tcond\ slope. The 1:1 line is shown as a black dashed line, and the Sun is marked as a red star. The standard deviation of the residuals (observed – predicted) and the mean observational uncertainty are also reported in each panel. We find that as an element’s \tcond\ increases, it tends to yield more positive residuals for stars that are more refractory-enriched, or with more positive \tcond\ slopes.}
    \label{fig:regressions}
\end{figure}


\end{document}